\documentclass{article}
\usepackage{amsthm}

\usepackage{PRIMEarxiv}
\usepackage{natbib}
\usepackage[utf8]{inputenc} % allow utf-8 input
\usepackage[T1]{fontenc}    % use 8-bit T1 fonts
\usepackage{hyperref}       % hyperlinks
\usepackage{url}            % simple URL typesetting
\usepackage{booktabs}       % professional-quality tables
\usepackage{amsfonts}       % blackboard math symbols
\usepackage{nicefrac}       % compact symbols for 1/2, etc.
\usepackage{microtype}      % microtypography
\usepackage{lipsum}
\usepackage{fancyhdr}       % header
\usepackage{graphicx}       % graphics
\graphicspath{{media/}}     % organize your images and other figures under media/ folder
\usepackage{amsmath}
\usepackage{algpseudocode}
\usepackage{algorithm2e}
\SetKwInOut{Input}{Input}
\SetKwInOut{Output}{Output}
\SetKwInOut{Parameter}{Parameter}
\SetKwInOut{Return}{Return}
\DeclareMathOperator*{\argmax}{arg\,max}
\DeclareMathOperator*{\argmin}{arg\,min}

\usepackage[dvipsnames]{xcolor}
\usepackage{dutchcal}
\usepackage{comment}
\usepackage{ulem}
\usepackage{tikz}
\usepackage{comment}
\usepackage{subcaption}
\usepackage{adjustbox}
\usepackage{dcolumn,booktabs}
\newcolumntype{d}[1]{D{.}{.}{#1}}

\newtheorem{remark}{Remark}
\newtheorem{definition}{Definition}

 \usepackage{float}
  
\title{Generative Flexible Latent Structure Regression (GFLSR) model
\thanks{\textit{\underline{Citation}}: 
\textbf{To be continued}} 
}

\author{
   Clara Grazian \\
    The University of Sydney \\
    and \\
    ARC Training Centre in Data Analytics for Resources \& Environments \\
    Sydney \\
   \texttt{clara.grazian@sydney.edu.au} \\
   \And
  Qian Jin \\
  The University of New South Wales \\
  Sydney\\
  \texttt{qian.jin@unsw.edu.au} \\
   \And
  Pierre Lafaye De Micheaux \\
  The University of New South Wales \\
  Sydney \\
  \texttt{lafaye@unsw.edu.au}
}

\newtheorem{theorem}{Theorem}
\newtheorem{lemma}[theorem]{Lemma}
\newtheorem{prop}{Proposition}
\begin{document}
\maketitle

\begin{abstract}
Latent structure methods, specifically linear continuous latent structure methods, are a type of fundamental statistical learning strategy. They are widely used for dimension reduction, regression and prediction, in the fields of chemometrics, economics, social science and etc. However, due to the lack of model inference, generative form, and unidentifiable parameters, most of these methods are always used as an algorithm, instead of a model. This paper proposed a Generative Flexible Latent Structure Regression (GFLSR) model structure to address this problem. Moreover, we show that most linear continuous latent variable methods can be represented under the proposed framework. The recursive structure allows potential model inference and residual analysis. Then, the traditional Partial Least Squares (PLS) is focused; we show that the PLS can be specialised in the proposed model structure, named Generative-PLS. With a model structure, we analyse the convergence of the parameters and the latent variables. Under additional distribution assumptions, we show that the proposed model structure can lead to model inference without solving the probabilistic model. Additionally, we proposed a novel bootstrap algorithm that enables uncertainty on parameters and on prediction for new datasets. A simulation study and a Real-world dataset are used to verify the proposed Generative-PLS model structure. Although the traditional PLS is a special case, this proposed GFLSRM structure leads to a potential inference structure for all the linear continuous latent variable methods.
\end{abstract}

% keywords can be removed

\section{Introduction}

We introduce the \textit{Generative Flexible Latent Structure Regression (GFLSR) model}, a parametric framework that formalizes and extends the latent-variable foundations of regression methods such as Partial Least Squares Regression (GPLSR). Let $(\boldsymbol{X},\boldsymbol{Y})\in\mathbb{R}^p\times\mathbb{R}^q$ be a pair of observable random vectors, where $\boldsymbol{X}$ denotes a vector of $p$ predictors and $\boldsymbol{Y}$ a vector of $q$ responses. While PLSR is traditionally defined in algorithmic terms --- as a procedure for extracting predictive components from $X$ and $Y$ --- the FLSR model provides an explicit data-generating process (DGP) that incorporates both the observed variables and their latent structure. At its core, the model defines two sets of $H\geq2$ scalar random latent variables, 
$\boldsymbol{\xi}:=(\xi_1,\ldots, \xi_H)$ and $\boldsymbol{\omega}:=(\omega_1, \ldots, \omega_H)$, whose pairwise dependence is specified via parametrized, invertible, and possibly nonlinear transformations of shared latent randomness. This generalises the linear dependency implicitly assumed in PLSR. The explanatory variables $\boldsymbol{X}$ are constructed through a classical linear deflation mechanism (as in PLSR), while the conditional expectation 
$\mathbb{E}[\boldsymbol{Y}\mid\boldsymbol{\xi}]$ is modelled via a rich class of nonlinear functions.

All parameters involved in this generative construction --- including the latent coupling functions, nonlinear mappings to $\boldsymbol{Y}$, and variances of the noise components --- are made explicit and subject to estimation. This allows not only for direct inference and simulation, but also for methodological developments such as residual-based bootstrap algorithms. In particular, we demonstrate that PLS arises as a special case of the GFLSR model under specific structural constraints, thereby recasting it as a fully generative model without introducing additional distributional assumptions. This perspective enables a unified statistical foundation for a wide class of latent-variable regression methods, offering new insights and inference tools beyond what is available under the standard algorithmic view.

Latent Variable methods refer to the approaches where, rather than focusing on the explanatory variables observed, it is assumed that the observed response variables are generated by underlying latent variables that are not directly observed but inferred from the explanatory variables. These hidden latent variables are meaningful in the context of the study and are capable of representing the original observed data. Latent variable methods are widely used across various disciplines, including economics \cite{bollen2007socio}, social and behavioural science \cite{bollen2002latent}, chemometrics \cite{eriksson2014chemometrics}, medical data analysis \cite{yang1997latent}, and image processing \cite{gulrajani2016pixelvae}. When the direct measure of a concept is challenging, the latent variable method offers a powerful alternative. Furthermore, these methods enable users to uncover the underlying structure and provide insight into the potential interpretation of complex datasets.

Typically, the latent variable methods serve as dimensionality-reduction techniques, crucial for managing high-dimensional data and avoiding the curse of dimensionality when the number of predictive variables is large. Several widely-used examples include Principal Component Analysis (PCA) \cite{jolliffe2002principal}, Partial Least Squares (PLS) \cite{wold1968nonlinear}, Independent Component Analysis (ICA) \cite{lee1998independent}, Canonical Correlation Analysis (CCA) \cite{hotelling1992relations}, Factor Analysis (FA) \cite{spearman1961general}, kernel methods \cite{scholkopf1998nonlinear}, and autoencoders \cite{hinton2006reducing}. Although nonlinear or deep latent variable models offer flexibility, classical statistical learning-based classic latent variable methods remain preferred due to their clear interpretability, computational efficiency, and fewer required parameters. This paper concentrates specifically on linear latent variable factor methods, particularly emphasising the Partial Least Squares (PLS) approach.

PLS is a foundational technique in machine learning and statistical modelling, which is also known as Projection on Latent Structures (PLS) \cite{abdi2010partial}. The method was initially introduced as a latent variable method for regression and has since been extended to classification and other applications \cite{cha1994partial}. As data dimensions increase and multicollinearity among variables becomes prevalent, deriving meaningful inferences becomes increasingly challenging. PLS addresses this challenge by constructing latent variables that linearly represent the original dataset. These latent variables are straightforward to compute, highly interpretable, and preserve strong relationships with the original data. The weights of the components explicitly represent the linear relationships between the latent variables and the original predictors, making PLS a powerful tool for exploratory data analysis. In most applications, the first few components are sufficient to represent the dataset and deliver strong regression performance \cite{lafaye2019pls}, achieving dimensionality reduction while retaining essential information.

PLS has demonstrated exceptional performance across diverse fields, including chemometrics \cite{wold2001pls}, bioinformatics \cite{land2011partial}, genomics \cite{gavaghan2002physiological}, finance \cite{avkiran2018partial}, and economics \cite{mustofa2022exploring,purwanto2021partial}. Its ability to handle multicollinearity, combined with its versatility, accuracy, and interpretability, makes PLS particularly popular in industry.  However, traditional PLS is often represented as an algorithm rather than a statistical model because it lacks identifiability; the rotation of parameters does not affect the results \cite{wang2005interpretation}. In many applications, a formal model framework is often preferred over an algorithm for enhanced interpretability and statistical inference capabilities.

Methods to estimate uncertainty within PLS can be broadly categorised into probabilistic and resampling techniques. Probabilistic PLS models have been explored in previous studies \cite{Li2011, Li_2015, Zheng2016, Zheng2018,el_Bouhaddani_2018, Eti_vant_2022}. Within probabilistic modelling, the distribution of data is assumed, therefore, the unknown parameters, together with the distributions, can be estimated and naturally quantify uncertainty in parameter estimates. The idea of incorporating probability within the PLS model is introduced in \cite{gustafsson2001probabilistic}. Then Li et al. \cite{Li2011, Li_2015} proposed the initial probabilistic partial least squares regression model (PPLSR). Zheng et al. \cite{Zheng2016, Zheng2018} extended the model to a more general form. However, the models are different from the traditional PLS and are not identifiable. While \cite{el_Bouhaddani_2018} proposed a symmetric probabilistic model and proved its identifiability up to a sign, \cite{Eti_vant_2022} criticised it for being overly restrictive and proposed a more flexible PPLS-SVD model. However, the PPLS-SVD model is computationally complex and challenging to solve using the EM algorithm. 

Bayesian PLS models \cite{Vidaurre_2013, Urbas_2024} are another branch of the probabilistic model. Bayesian techniques provide prediction intervals for new datasets, however, the models also differ from traditional PLS and are not symmetric.  Additionally, most previous studies assume normal distributions with diagonal variance matrices, limiting their applicability to datasets with different distributional characteristics.

Alternatively, Bootstrap methods have been explored in various studies \cite{faber2002uncertainty, Odgers2023}. While \cite{faber2002uncertainty} employed bootstrap by residuals, \cite{Odgers2023} noted that this approach ignores critical variable information. Pair bootstrap methods are computationally intensive, but remain the most reliable methods for traditional PLS inference.

Beyond model inference, characterising the statistical properties of Partial Least Squares (PLS) remains challenging. Numerous studies have been conducted on investigating the statistical properties of PLS. For instance, Helland (1988) demonstrated equivalence between the original PLS algorithm proposed by Wold \cite{wold1984collinearity} and the NIPALS algorithm introduced by Martens \cite{martens1992multivariate}. Furthermore, Naik and Tsai \cite{naik2000partial} established consistency of PLS estimators up to a scaling constant within a single-index model framework. Kramer \cite{kramer2011degrees} provided an unbiased estimator for the degrees of freedom in regression settings, while Delaigle and Hall \cite{delaigle2012methodology} investigated convergence properties in functional PLS models within infinite-dimensional function spaces. Despite these advancements, comprehensive statistical theory regarding parameters and latent variables remains underdeveloped due to the absence of a formal structural model for PLS. Additionally, J. Henseler (2010) \cite{henseler2010convergence} highlighted convergence issues by presenting six specific cases where iterative algorithms in PLS path modelling failed to converge, emphasising the urgent need for a formally defined model with rigorous inference mechanisms.

In this paper, we propose the Generative Flexible Latent Structure Regression (GFLSR) model, a unified framework encompassing a broad class of linear latent variable models. This formulation enables both theoretical and practical extensions to nonlinear data structures, broadening the model’s applicability to more complex datasets while preserving interpretability under specific conditions. In particular, we show that the GFLSR framework naturally reduces to classical PLS under certain assumptions and retains identifiability; we refer to Generative-PLS. With additional constraints, the generative-PLS closely aligns with the majority of existing probabilistic PLS (PPLS) models. Moreover, unlike many PPLS formulations, Generative-PLS operates effectively without requiring strong distributional assumptions, offering robustness to non-Gaussian data.

Utilising a recurrent estimation algorithm, the proposed Generative-PLS model produces identical estimators to traditional PLS methods while simultaneously quantifying uncertainty. The estimation process involves recursively minimising loss functions, which recent studies \cite{hamilton2020likely} have linked to probabilistic interpretations, demonstrating alignment with specific distributions. The flexibility in loss function selection within our model facilitates competitive performance relative to existing probabilistic methods under given distributional assumptions.

Additionally,  we establish the theoretical convergence of both the estimated loading vectors and latent variables and empirically verify this through simulation. Specifically, we proved that the estimated loading vectors converge in $r$-th moment to the parameter in the data generation procedure at each iteration. Similarly, the difference between the fitted latent variables and their true counterparts converges to a specified probabilistic bound based on the error term. These results helped with the model inference; corrected estimators are established under model assumptions. To further address parameter uncertainty, we introduce a novel bootstrap approach. Extensive simulation studies and real-world data analyses substantiate the validity and effectiveness of our proposed framework.

The outline of the paper is as follows. Section \ref{sec: latent} presents the structure of the generative flexible latent structure regression model and demonstrates that many well-known methods, including partial least squares (PLS), canonical correlation analysis (CCA), and principal components analysis (PCA), are embedded within this framework. Section \ref{sec: pls} introduces the PLS model within the generative framework and explores the relationship between generative pls and existing Probabilistic PLS. Section \ref{subsec: theorem} presents the theoretical convergence of both the estimated loading vectors and latent variables and the identifiability of the proposed model. Section~\ref{sec: inference} discussed the model inference and the proposed bootstrap structure. Section \ref{sec:sim} presents the results of several simulation studies, and Section \ref{sec:real} demonstrates applications to real data. Finally, Section \ref{sec:conclusion} discusses the findings and outlines directions for future work.

\section{Generative Flexible Latent Structure Regression model}
\label{sec: latent}

In this section, we introduce our generative flexible latent structure regression model. We demonstrate the model definition and the model estimation procedure. Then, we discuss the relationship between the proposed model structure and current well-established linear latent structure algorithms.
\subsection{Mathematical Definition of the GFLSR model}
\label{subsec:generalmodel}

\subsubsection{Model Definition}
Let $p, q, H$ ($H<p$) and $K$ be given positive integers. For $\ell\in\{p,q\}$, denote  $\mathbb{S}^{\ell-1}$ the $\ell$-dimensional unit sphere in $\mathbb{R}^\ell$ centered at the origin. For each $h=1,\ldots,H$, let
\[
\left\{ t \in [0,1] \mapsto \left( \Psi_{h1}(t; \boldsymbol{\gamma}), \Psi_{h2}(t; \boldsymbol{\gamma}) \right)\in\mathbb{R}^2 \;\middle|\; \boldsymbol{\gamma} \in \Gamma \subset \mathbb{R}^m \right\}
\]
be a given family of parametric curves in \(\mathbb{R}^2\), where for each \(\boldsymbol{\gamma} \in \Gamma\), both components \(\Psi_{h1}(t; \boldsymbol{\gamma})\) and \(\Psi_{h2}(t; \boldsymbol{\gamma})\) are continuous and strictly increasing functions of \(t \in [0,1]\). These functions are thus invertible, and we assume additionally that \(\Psi_{h1}(0; \boldsymbol{\gamma})=\Psi_{h2}(0; \boldsymbol{\gamma})=0\). Then each such curve defines the graph of a continuous, strictly increasing, and invertible function \(f_{\boldsymbol{\gamma},h}: \mathbb{R} \rightarrow \mathbb{R}\) with $f_{\boldsymbol{\gamma},h}(0) = 0$, given by
\[
\forall x\in\mathbb{R},\qquad f_{\boldsymbol{\gamma},h}(x) = \Psi_{h2}\left( \Psi_{h1}^{-1}(x; \boldsymbol{\gamma}); \boldsymbol{\gamma} \right).
\]
We assume that $\mathbb{E}(\Psi_{h1}(U))=\mathbb{E}(\Psi_{h2}(U))=0$ and $\mathbb{E}[(\Psi_{h1}(U))^2]=\mathbb{E}[(\Psi_{h2}(U))^2]=1$ where $U$ is a random variable having a $\textrm{Unif}[0,1]$ distribution. Additionally, let  $\left\{\boldsymbol{f}_H(\cdot;\boldsymbol{\Theta}):\mathbb{R}^H\rightarrow\mathbb{R}^q\mid \boldsymbol{\Theta}\in\mathbb{R}^{K}\right\}$ be a given family of parametric vector-valued functions (some further restrictions on this family are provided below). For each $h=1,\ldots,H$, we denote by $\boldsymbol{f}_h$ the following restriction of $\boldsymbol{f}_H$ to $\mathbb{R}^h$:
$$\boldsymbol{f}_h(x_1,\ldots,x_h;\boldsymbol{\Theta}):=\boldsymbol{f}_H(x_1,\ldots,x_h,0,\ldots,0;\boldsymbol{\Theta}).$$  
When $h=0$, we write $\boldsymbol{f}_0$ to denote the vector $\boldsymbol{0}\in\mathbb{R}^q$.

Now, consider\footnote{Throughout this work, all random variables are defined on a common probability space $(\Omega,\mathcal{A}, P)$. For any $p$-dimensional random vector $\boldsymbol{X}$ and any function $f:\mathbb{R}^p\rightarrow\mathbb{R}^q$, we adopt the convention $f(\boldsymbol{X})$ in place of $f(\boldsymbol{X}(\omega))$ for some $\omega\in\Omega$, acknowledging this slight abuse of notation. Vector-valued quantities are represented in boldface throughout.} a pair of observable random vectors $(\boldsymbol{Y}_0, \boldsymbol{X}_0):=(\boldsymbol{Y}, \boldsymbol{X})\in\mathbb{R}^q\times\mathbb{R}^p$. It is said to follow the \textit{Generative Flexible Latent Structure Regression (GFLSR) model} with random \textit{$X$-latent vector} $\boldsymbol{\xi}=(\xi_1,\ldots,\xi_{H})^\top$ and random \textit{$Y$-latent vector} $ \boldsymbol{\omega}=(\omega_1,\ldots,\omega_H)^\top$ if there exist $s_{11}> s_{21}>\cdots> s_{H1}$ and $s_{h2}$ ($h=H,\ldots,h$), as well as (non-random) vectors $\boldsymbol{\Theta}_0, \boldsymbol{\Theta}_1,\ldots,\boldsymbol{\Theta}_H\in\mathbb{R}^{K}$ ($\boldsymbol{\Theta}_0$ denotes the empty set, and additional constraints on these parameters together with the $\boldsymbol{f}_h$ functions are provided below) and $\boldsymbol{\gamma}\in\Gamma$, and a (deterministic) matrix $W:p\times H$ whose columns are 
linearly independent vectors $\boldsymbol{w}_1,\ldots,\boldsymbol{w}_H$ in $\mathbb{R}^{p}$ (assuming, w.l.o.g., that the component of $\boldsymbol{w}_h$ with the largest magnitude is non-negative) such that the  relationship
\begin{equation}
\mathbb{E}[\boldsymbol{Y}_{0} \mid\xi_1,\ldots,\xi_{H}] =  \boldsymbol{f}_{H}(\xi_1,\ldots,\xi_{H};\boldsymbol{\Theta}_{H}),\label{eq: Generative_Flexible_Latent_Structure_Regression}
\end{equation}
holds via the following backward recursions: For $h=H,\ldots,1$,
\begin{eqnarray}
\left(\xi_h, \omega_h\right)^\top &=& 
\begin{cases}
s_{h1}(\Psi_{h1}(U_h;\boldsymbol{\gamma}) + \epsilon_{1,h}) \\
s_{h2}(\Psi_{h2}(U_h;\boldsymbol{\gamma}) + \epsilon_{2,h})
\end{cases}
~\text{with }U_h\overset{\text{ind}}{\sim}\text{Unif}[0,1]\text{ and }s_{h1},s_{h2}>0,
\label{eq: latent_relationship} \\
\boldsymbol{X}_{h-1} & = & \boldsymbol{w}_{h}\xi_h + \boldsymbol{X}_{h}, \label{eq: X_deflation}\\
\boldsymbol{Y}_{h-1} & = & \boldsymbol{f}_h(\xi_1,\ldots,\xi_h;\boldsymbol{\Theta}_{h})-\boldsymbol{f}_{h-1}(\xi_1,\ldots,\xi_{h-1};\boldsymbol{\Theta}_{h-1}) +\boldsymbol{Y}_h,
\label{eq: Y_deflation}
\end{eqnarray}
for some continuous zero-mean random error vectors $\boldsymbol{X}_{H}$, $\boldsymbol{Y}_H$ (with respective symmetric and positive semi-definite $p\times p$ covariance matrices $\boldsymbol{\Sigma}_X$ and $\boldsymbol{\Sigma}_Y$), and some continuous zero-mean random error variables $\epsilon_{1,h}$ and $\epsilon_{2,h}$ with respective finite variances $\sigma_{1,h}^2$ and $\sigma_{2,h}^2$, $h=H,\ldots,1$. All these random quantities are assumed to be mutually independent and no other assumptions are made about their marginal probability distributions. 

Now that we have introduced the model, we make a few comments to help the reader better understand various aspects of it.

Note that $\mathbb{V}(\xi_h)=s_{h1}^2(1+\sigma_{1,h}^2)$. Thus, with error variances $\sigma_{1,h}^2$ taken small enough, the strict decreasing of the $s_{h1}^2$'s will lead to a strict decreasing of the variances of the $\xi_h$'s, $h=1,\ldots,H$. They are thus ordered in terms of their "power of explainability". Also, imposing $1+\sigma_{1,h}^2=s_{h1}^{-2}$ allows us to retrieve the assumption made in CCA that  $\mathbb{V}(\xi_h)=1$. 

Because the $\xi_h$'s are generated from mutually independent $U_h$'s ($h=H,\ldots,1)$, they are themselves mutually independent by construction, and thus uncorrelated. But if we assume instead that, almost surely, $U_h=U$ (for $h=1,\ldots, H$) and if $\{\Psi_{h1},h=1,\ldots, H\}$ is chosen as a family of orthonormal functions in $L^2([0,1])$ (the assumption of strict monotonicity can be relaxed), $\{\Psi_{h2},h=1,\ldots, H\}$ is a function of $\{\Psi_{h1},h=1,\ldots, H\}$, then the components can be made uncorrelated, but dependent.

Also, since $\xi_h$ and $\omega_h$ are generated from the same $U_h$, they are maximally dependent by construction (if we neglect $\epsilon_{1,h}$ and $\epsilon_{2,h}$). These noises have been added to the model to allow for a slight departure (both in the horizontal and vertical directions, away from the parametric curve $u\mapsto(\Psi_{h1}(u),\Psi_{h2}(u))$) from this situation of perfect dependence. If the parametric curve is a straight line then (still neglecting the noises $\epsilon_{1,h}$ and $\epsilon_{2,h}$) $\xi_h$ and $\omega_h$ are maximally linearly dependent, with a correlation equal to $\pm1$, and a covariance equal to $\pm \sqrt{s_{h1}s_{h2}}$. These comments make it clear that one has to choose a measure of dependence $D$ tailored to detect the kind of assumed relationship between the two latent variables to be able to estimate them from an observed data set. Some examples include covariance, correlation, Spearman, and Hellinger.

Now, from equation~\eqref{eq: X_deflation}, it is easy to see that
\begin{equation}
    \boldsymbol{X}_h = \boldsymbol{X}_{h-1} - \xi_h\boldsymbol{w}_{h} =\boldsymbol{X}_{0} - \sum_{\ell=1}^h\xi_{\ell}\boldsymbol{w}_{\ell} \quad \text{ and } \quad \boldsymbol{X}_{0} = \sum_{\ell=1}^H\xi_{\ell}\boldsymbol{w}_{\ell}+\boldsymbol{X}_H, \label{eq: X_0}
\end{equation}
from which it appears that $\boldsymbol{X}_{h}$ is the error term in a simple multivariate linear regression model (without intercept) of $\boldsymbol{X}_{h-1}$ versus $\xi_h$. Thus $\boldsymbol{X}_{h}$ is the part of $\boldsymbol{X}_{h-1}$ that is not (linearly) explained by $\xi_h$. Similarly, we see that $\boldsymbol{X}_{h}$ is the part of $\boldsymbol{X}_0$ that is not (linearly) explained by the first $h$ latent variables $\xi_1,\ldots,\xi_h$.

From Equation~\eqref{eq: Y_deflation}, $\boldsymbol{Y_h}$ is defined as the error term in the multivariate non-linear regression model $\boldsymbol{Y}_{h-1} = \boldsymbol{f}_h(\xi_1,\ldots,\xi_h;\boldsymbol{\Theta}_{h}) - \boldsymbol{f}_{h-1}(\xi_1,\ldots,\xi_{h-1};\boldsymbol{\Theta}_{h-1}) + \boldsymbol{Y}_{h} =   
\boldsymbol{f}_{h+1}(\xi_1,\ldots,\xi_{h+1};\boldsymbol{\Theta}_{h+1}) - \boldsymbol{f}_{h-1}(\xi_1,\ldots,\xi_{h-1};\boldsymbol{\Theta}_{h-1}) + \boldsymbol{Y}_{h+1} = \boldsymbol{f}_{H}(\xi_1,\ldots,\xi_{H};\boldsymbol{\Theta}_{H}) - \boldsymbol{f}_{h-1}(\xi_1,\ldots,\xi_{h-1};\boldsymbol{\Theta}_{h-1}) + \boldsymbol{Y}_{H}$. From this, it is not difficult to see that $\boldsymbol{Y}_h$ is the part of $\boldsymbol{Y}_{0}$ that is not explained (through $\boldsymbol{f}_h$) by the first $h$ latent variables $\xi_1,\ldots,\xi_h$ because we have $\boldsymbol{Y}_{0} = \boldsymbol{f}_h(\xi_1,\ldots,\xi_h;\boldsymbol{\Theta}_{h}) + \boldsymbol{Y}_{h}$. This also implies that
\begin{equation}\boldsymbol{Y}_{0} = \boldsymbol{f}_H(\xi_1,\ldots,\xi_H;\boldsymbol{\Theta}_{H}) +\boldsymbol{Y_H}. \label{eq: Y_0}
\end{equation}
Note that the Equation~\eqref{eq: Y_0} can be used to generate observations from our model, but in the real data generation procedure, Equations~\eqref{eq: Y_deflation} are still required, becasure they are useful to characterize how $\boldsymbol{Y}_{h-1}$ is related to only the first $h$ $X$-latent variables. A discussion about all the $\{ \boldsymbol{\Theta_0},\ldots, \boldsymbol{\Theta}_H\}$ is in Remark 2.

\begin{figure}[H]
\includegraphics[width=10.5cm]{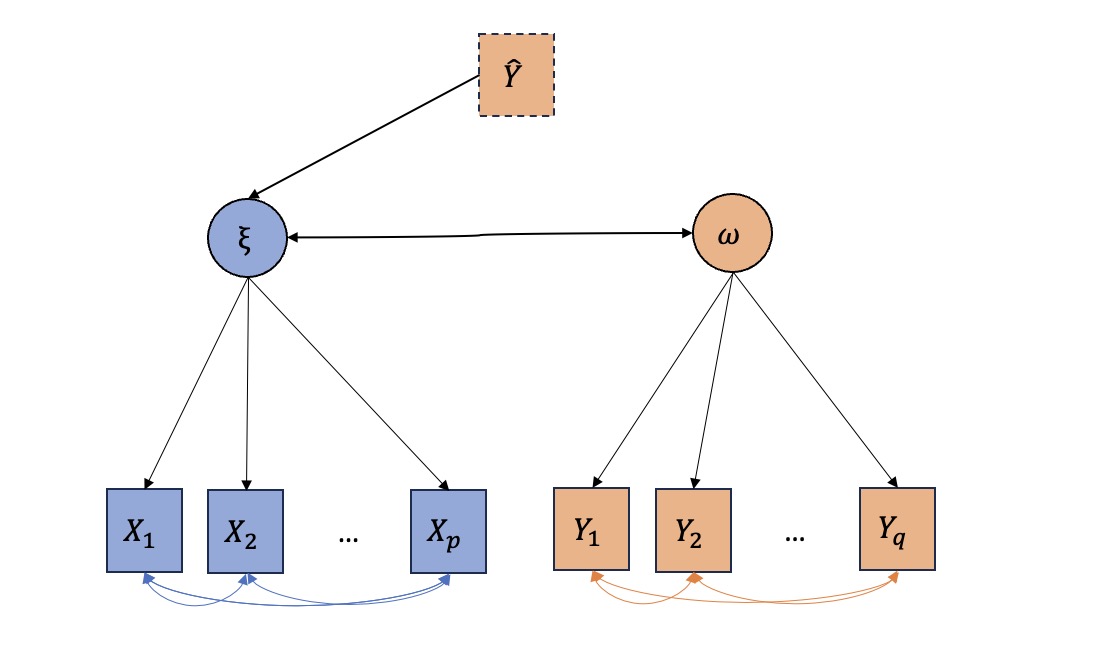}
\centering
\caption{Path Diagram for GFLSR model}
\label{fig: path}
\end{figure}

Figure~\ref{fig: path} shows a path diagram to illustrate the relationships among the variables. Observed variables are represented by solid squares, latent variables by circles, and predicted variables by dashed squares. A double-sided arrow indicates a non-zero dependence between two variables, while a single-sided arrow signifies that one variable can be explained as a function of the variable it is linked to.

\begin{remark}
    The relationship between $\xi_h$ and $\omega_h$ is generated from $\Psi_{h1}$ and $\Psi_{h2}$, which relationship ensure a maximal measure of dependence $D$, while small errors $\epsilon_{1,h}$ and $\epsilon_{2,h}$ can be added orthogonal to the correlation curve to have non-perfect correlation.  However, in real practice and Sec~\ref{sec: pls}, we can show that only $\epsilon_{1h}$ added to the model leads to the same result. Even the orthogonal error terms have a more straightforward interpretation; considering only $\epsilon_{1h}$ is simple and contributes to the identification of the model.
\end{remark}

\begin{remark}
The parameters $ \boldsymbol{\Theta_1},\ldots, \boldsymbol{\Theta}_H\in\mathbb{R}^K$ might have some specific structure related to the specific form of the $\boldsymbol{f}_h$ functions. For instance, if $\boldsymbol{f}_h$ is linear function, then $\boldsymbol{\Theta}_h = (\theta_1, \ldots, \theta_h,0,\ldots,0)^\top\in \mathbb{R}^K$. If $\boldsymbol{f}_h$ has a more complicated structure, then one might need to add other constraints to make sure the distance between $\boldsymbol{\Theta}_h$ and $\boldsymbol{\Theta}_{h+1}$ is small.
\end{remark}

\begin{remark}
It is worth noting that this does not simplify the task of generating observations from model~\eqref{eq: Generative_Flexible_Latent_Structure_Regression}--\eqref{eq: Y_deflation} due to the constraints imposed on the $\Psi_{h1}$, $\Psi_{h2}$ and $\boldsymbol{f}_H$ functions within Equations~\eqref{eq: latent_relationship} and~\eqref{eq: Y_deflation}. Additionally, the functions $\boldsymbol{\Psi}_h=(\Psi_{h1},\Psi_{h2}):[0,1]\rightarrow\mathbb{R}^2$ that define the dependence structure of the dependent variable should maximise the given dependence measure $D$.

In the model simulation, we distinguish between a single response and multiple responses.
\begin{enumerate}
    \item In the case of a single response, we can obtain the scaled value of $\boldsymbol{Y}_h$ directly from Equation~\eqref{eq: latent_relationship}. Therefore, the true relationship between $\boldsymbol{Y}$ and $\mathbf{X}$ should be embedded in the $\Psi_1, \Psi_2$ functions. 
    \item In the multiple-response situation, higher flexibility is allowed. The true relationship between $\mathbf{Y}$ and $\mathbf{X}$ is defined by both Equations~\eqref{eq: latent_relationship} and \eqref{eq: Y_deflation}. It is preferable to let Equation~\eqref{eq: latent_relationship} describe the relationship between $\boldsymbol{\xi}_h$, $h=1, \ldots, H$, and $\mathbf{X}$, while Equation~\eqref{eq: Y_deflation} describes the relationship between $\boldsymbol{\xi}_h$ and $\mathbf{Y}$. 
\end{enumerate} 

More details and a few simple examples are provided in Appendix \ref{app:simulation}
\end{remark}

\subsubsection{Model Estimation}

For the reader’s convenience, we introduce the following notation throughout the paper. We use a $\bullet$ symbol to refer to mathematical objects involved in the data-generating process detailed in the previous subsection. We use a $\,\widehat{\ }\,$ symbol to denote estimators (or predictors) of parameters (or random quantities) of the data-generating process computed from a random sample of size $n$, and we use a $\star$ symbol for their counterpart in the fitted model at the population level (i.e., when the sample size is infinite).

Applying Equation~\eqref{eq: X_deflation} successively,  we get $\boldsymbol{X}^\bullet_{h-1} = \boldsymbol{X}^\bullet_H + \sum_{k=h}^H\boldsymbol{w}^\bullet_k\xi^\bullet_k$. Associated to the $\boldsymbol{w}^\bullet_h$'s, there exists a (non necessarily unique) family $\{\boldsymbol{u}_1,\ldots,\boldsymbol{u}_H\}$ of vectors in $\mathbb{R}^p$ such that $\boldsymbol{u}_i^\top\boldsymbol{w}^\bullet_j=\delta_{ij}$ ($i,j=1,\ldots, H$) and $\|\boldsymbol{u}_h\|_2=1$ ($h=1,\ldots, H$), where $\delta_{ij}$ denotes the Kronecker symbol. Using such a $\boldsymbol{u}_h$, we get   $\boldsymbol{u}_h^\top\boldsymbol{X}^\bullet_{h-1}=\boldsymbol{u}_h^\top\boldsymbol{X}^\bullet_{H} + \sum_{k=h}^H(\boldsymbol{u}_h^\top\boldsymbol{w}^\bullet_k)\xi^\bullet_k=\boldsymbol{u}_h^\top\boldsymbol{X}^\bullet_{H}+\xi^\bullet_h$. 
The $h$th $X$-score can thus be represented as
\begin{equation}
\xi^\bullet_h = \boldsymbol{u}_h^\top\boldsymbol{X}^\bullet_{h-1} - \boldsymbol{u}_h^\top\boldsymbol{X}^\bullet_{H}, \label{eq: generated_xi}
\end{equation}
i.e., the sum of a linear component $\boldsymbol{u}_h^\top\boldsymbol{X}^\bullet_{h-1}$ and of a noise term $(-\boldsymbol{u}_h)^\top\boldsymbol{X}^\bullet_{H}$. 

With the assumptions made previously, the expectation of this noise term is zero while its variance is equal to $\boldsymbol{u}_h^\top\boldsymbol{\Sigma}_{X}^\bullet\boldsymbol{u}_h$. If we further assume that this latter term is small (see later for a discussion on this), it makes sense to search for a step-$h$ population-level fitted $X$-latent variable $\xi_h^{\star} = g_h(\boldsymbol{X}_{h-1}^{\star}; \boldsymbol{u}_{h}^{\star})$ (for some function $g_h:=g_h(\cdot;\boldsymbol{u}_h):\mathbb{R}^p\rightarrow\mathbb{R}$) under the form
\begin{eqnarray}
\xi_h^{\star} & = & g_h(\boldsymbol{X}_{h-1}^{\star}; \boldsymbol{u}_{h}^{\star}) = \boldsymbol{X}_{h-1}^{\star\top} \boldsymbol{u}_{h}^{\star}. \label{eq: xi_g_function}
\end{eqnarray}

Similarly, we can write $\boldsymbol{Y}_{h-1}^\bullet = \boldsymbol{Y}_{H}^\bullet + \boldsymbol{f}_{H}(\xi_1^\bullet,\ldots,\xi_{H}^\bullet;\boldsymbol{\Theta}_{H}^\bullet) - \boldsymbol{f}_{h-1}(\xi_1^\bullet,\ldots,\xi_{h-1}^\bullet;\boldsymbol{\Theta}_{h-1}^\bullet)$. 

Using multivariate Taylor expansions (in two steps, over the $\xi$ and then over the $\Theta$), we have:

\begin{eqnarray*}
\boldsymbol{f}_{H}(\xi_1^\bullet,\ldots,\xi_{H}^\bullet;\boldsymbol{\Theta}_{H}^\bullet) & = &  \boldsymbol{f}_{H}(\xi_1^\bullet,\ldots,\xi_{h-1}^\bullet,0,\ldots,0;\boldsymbol{\Theta}_{H}^\bullet) +  \sum_{j=h}^H\xi_j^\bullet\frac{\partial \boldsymbol{f}_H}{\partial x_j} (\xi_1^\bullet,\ldots,\xi_{h-1}^\bullet,0,\ldots,0;\boldsymbol{\Theta}_{H}^\bullet) + R_1 \\
& = & \boldsymbol{f}_{H}(\xi_1^\bullet,\ldots,\xi_{h-1}^\bullet,0,\ldots,0;\boldsymbol{\Theta}_{h-1}^\bullet) + (\boldsymbol{\Theta}_{H}^\bullet - \boldsymbol{\Theta}_{h-1}^\bullet)^\top \frac{\partial \boldsymbol{f}_H}{\partial \boldsymbol{\Theta}} (\xi_1^\bullet,\ldots,\xi_{h-1}^\bullet,0,\ldots,0;\boldsymbol{\Theta}_{h-1}^\bullet) \\
&&  + R_2 +  \sum_{j=h}^H\xi_j^\bullet\frac{\partial \boldsymbol{f}_H}{\partial x_j} (\xi_1^\bullet,\ldots,\xi_{h-1}^\bullet,0,\ldots,0;\boldsymbol{\Theta}_{H}^\bullet) + R_1 \\
& = & \boldsymbol{f}_{h-1}(\xi_1^\bullet,\ldots,\xi_{h-1}^\bullet;\boldsymbol{\Theta}_{h-1}^\bullet) + (\boldsymbol{\Theta}_{H}^\bullet - \boldsymbol{\Theta}_{h-1}^\bullet)^\top \frac{\partial \boldsymbol{f}_H}{\partial \boldsymbol{\Theta}} (\xi_1^\bullet,\ldots,\xi_{h-1}^\bullet,0,\ldots,0;\boldsymbol{\Theta}_{h-1}^\bullet) \\
&&  +   \sum_{j=h}^H\xi_j^\bullet\frac{\partial \boldsymbol{f}_H}{\partial x_j} (\xi_1^\bullet,\ldots,\xi_{h-1}^\bullet,0,\ldots,0;\boldsymbol{\Theta}_{H}^\bullet) + R \\
\end{eqnarray*}
where $R_1=O(\|(\xi_{h},\ldots,\xi_{H})\|^2)$ and $R_2=O(\|\boldsymbol{\Theta}_{H}^\bullet - \boldsymbol{\Theta}_{h-1}^\bullet\|^2)$ are two remainders, and $R=R_1+R_2$.

From this, we can write
$$
\boldsymbol{Y}_{h-1} = \boldsymbol{Y}_H +   \sum_{j=h}^H\xi_j^\bullet\frac{\partial \boldsymbol{f}_H}{\partial x_j} (\xi_1^\bullet,\ldots,\xi_{h-1}^\bullet,0,\ldots,0;\boldsymbol{\Theta}_{H}^\bullet) + (\boldsymbol{\Theta}_{H}^\bullet - \boldsymbol{\Theta}_{h-1}^\bullet)^\top \frac{\partial \boldsymbol{f}_H}{\partial \boldsymbol{\Theta}} (\xi_1^\bullet,\ldots,\xi_{h-1}^\bullet,0,\ldots,0;\boldsymbol{\Theta}_{h-1}^\bullet) + R.
$$

Neglecting the noises $\epsilon_{1,h}^\bullet$ and $\epsilon_{2,h}^\bullet$, we can write 
$\xi_h^\bullet=s_{h1}^\bullet f_{\boldsymbol{\gamma},h}^{\bullet-1}(\omega_h^\bullet/s_{h2}^\bullet)$. Another Taylor expansion leads to:
$$
f_{\boldsymbol{\gamma},h}^{\bullet-1}(\omega_h^\bullet/s_{h2}^\bullet) = f_{\boldsymbol{\gamma},h}^{\bullet-1}(0) + (\omega_h^\bullet/s_{h2}^\bullet) f_{\boldsymbol{\gamma},h}^{\bullet-1'}(0) = (\omega_h^\bullet/s_{h2}^\bullet) f_{\boldsymbol{\gamma},h}^{\bullet-1'}(0).
$$
As a result, we get
\begin{eqnarray*}
    \boldsymbol{Y}_{h-1} & = & \boldsymbol{Y}_H +   \sum_{j=h}^H \omega_h^\bullet (s_{h1}^\bullet/s_{h2}^\bullet) f_{\boldsymbol{\gamma},h}^{\bullet-1'}(0)\frac{\partial \boldsymbol{f}_H}{\partial x_j} (\xi_1^\bullet,\ldots,\xi_{h-1}^\bullet,0,\ldots,0;\boldsymbol{\Theta}_{H}^\bullet) + \\
    && (\boldsymbol{\Theta}_{H}^\bullet - \boldsymbol{\Theta}_{h-1}^\bullet)^\top \frac{\partial \boldsymbol{f}_H}{\partial \boldsymbol{\Theta}} (\xi_1^\bullet,\ldots,\xi_{h-1}^\bullet,0,\ldots,0;\boldsymbol{\Theta}_{h-1}^\bullet) + R.
    \end{eqnarray*}
Now, assuming that $\boldsymbol{\Theta}_h=(\theta_1,\ldots,\theta_h,0,\ldots,0)\in\mathbb{R}^K$, we get (see also Remark~2):
$$
(\boldsymbol{\Theta}_{H}^\bullet - \boldsymbol{\Theta}_{h-1}^\bullet)^\top \frac{\partial \boldsymbol{f}_H}{\partial \boldsymbol{\Theta}} (\xi_1^\bullet,\ldots,\xi_{h-1}^\bullet,0,\ldots,0;\boldsymbol{\Theta}_{h-1}^\bullet)
= \sum_{j=h}^H\theta_j\frac{\partial \boldsymbol{f}_H}{\partial \theta_j}(0)
$$
and we also assume that this term is small and can be neglected. 

So, if neglecting $R$ too, we can write:
$$
\boldsymbol{Y}_{h-1}^\bullet \approx \boldsymbol{Y}_{H}^\bullet +  \sum_{j=h}^H\omega_h^\bullet (s_{h1}^\bullet/s_{h2}^\bullet) f_{\boldsymbol{\gamma},h}^{\bullet-1'}(0)\frac{\partial \boldsymbol{f}_H}{\partial x_j} (\xi_1^\bullet,\ldots,\xi_{h-1}^\bullet,0,\ldots,0;\boldsymbol{\Theta}_{H}^\bullet)
$$
If the $\boldsymbol{f}_H$ is assumed to have a full-rank Jacobian matrix, then similarly, there exists a family $\{\boldsymbol{v}_1,\ldots,\boldsymbol{v}_H \} \in \mathbb{R}^q$, such that $\boldsymbol{z}_i^\top \boldsymbol{k}_j = \delta_{ij}$, with $\boldsymbol{k}_j = \{\frac{\partial \boldsymbol{f}_H}{\partial x_j} (\xi_1^\bullet,\ldots,\xi_{h-1}^\bullet,0,\ldots,0;\boldsymbol{\Theta}_{H}^\bullet)\}(s_{h1}^\bullet/s_{h2}^\bullet) f_{\boldsymbol{\gamma},h}^{\bullet-1'}(0)$.
Then we get
$$
\omega_h^\bullet \approx  \boldsymbol{v}_h^\top\boldsymbol{Y}^\bullet_{h-1} - \boldsymbol{v}_h^\top\boldsymbol{Y}^\bullet_{H}.
$$

And as above it makes sense to search for a step-$h$ population-level fitted $Y$-latent variable $\omega_h^{\star} = k_h(\boldsymbol{Y}_{h-1}^{\star}; \boldsymbol{v}_{h}^{\star})$ (for some function $k_h:=k_h(\cdot;\boldsymbol{v}_h):\mathbb{R}^q\rightarrow\mathbb{R}$) under the form
\begin{eqnarray}
\omega_h^{\star} & = & k_h(\boldsymbol{Y}_{h-1}^{\star}; \boldsymbol{v}_{h}^{\star}) = \boldsymbol{Y}_{h-1}^{\star\top} \boldsymbol{v}_{h}^{\star}. \label{eq: omega_k_function}
\end{eqnarray}

Note that the above procedure just shows that if the $\boldsymbol{f}_H$ is constructed well, then there exists a linear combination of $\boldsymbol{Y}^\bullet_{h-1}$ that approximates the $\omega_h^\bullet$ value. Thus, a linear combination of $\boldsymbol{Y}^\bullet_{h-1}$ can be used in the following optimisation problem. Even if the $\boldsymbol{f}_H$ is complicated or doesn't meet all the above constraints, the linear structure is still flavoured in the estimation procedure, since the $\xi_h^\star$ is searched such that it has most information from both $\boldsymbol{X}_{h-1}^\star$ and $\boldsymbol{Y}_{h-1}^\star$. The $\omega_h^\star$ is the representation of $\boldsymbol{Y}_{h-1}$, this linear structure can ensure the maximum information transfer. Moreover, from the model setting, only $\xi_h$ contributes to the prediction; hence, the identifiability and the error term between $\omega_h^\bullet$ and $\omega_h^\star$ is less important than the one from $\xi_h^\star$

The optimal fitted $\xi_h^\star$ and $\omega_h^\star$, linear combinations of the variables in $\boldsymbol{X}_{h-1}^\star$ and $\boldsymbol{Y}_{h-1}^\star$ respectively, are defined in such a way that their pairwise dependence (as measured by the dependence measure $D$) is maximised. Another benefit of considering linear combinations is that it is easy to see the contribution of each variable $X_{h-1,j}^\star$ and $Y_{h-1,k}^\star$ to this joint dependence, which contributes to the model interpretation. The vector $\boldsymbol{u}^\star_h$ is related to the vector $\boldsymbol{w}_h^\bullet \in \mathbb{R}^p$ used to generate the data. Similarly, we impose that $\boldsymbol{u}^\star_h \in \mathbb{S}^{p-1}$. Similarly, the parameters $\boldsymbol{v}^\star_h\in \mathbb{S}^{q-1}$ are related to the $\boldsymbol{\Theta}_h^\bullet \in \mathbb{R}^K$. 

In this subsection, we discuss ``fitting at the population level'' which we need to prove asymptotic results for the estimators of the unknown parameters in the data-generating model. Estimation (i.e., training) based on a given dataset of size $n$ is described in subsection~\ref{subsec:model_training}.

Now, given only $\boldsymbol{X}_0^\bullet$ and $\boldsymbol{Y}_0^\bullet$ following the GFLSR model, we set $(\boldsymbol{X}_0^\star, \boldsymbol{Y}_0^\star) = (\boldsymbol{X}_0^\bullet, \boldsymbol{Y}_0^\bullet)$. The goal is to recover all the model's parameters. For this task, let $L_X:\mathbb{R}^p\times\mathbb{R}^p\rightarrow\mathbb{R}^+$ be a loss function such as the squared loss function $L_X(\boldsymbol{X}_1,\boldsymbol{X}_2)=\|\boldsymbol{X}_1-\boldsymbol{X}_2\|_2^2$, for $\boldsymbol{X}_1,\boldsymbol{X}_2\in\mathbb{R}^p$ and let $L_Y:\mathbb{R}^q\times\mathbb{R}^q\rightarrow\mathbb{R}^+$ be another loss function. A convex loss function will help with convergence; otherwise more robust optimisation algorithm will be required in real practice. Define $D$ and $D_0$ be some dependence measures. (See the discussion above about $D$.)

We can solve recursively in a forward way (i.e., for $h = 1,\ldots, H$) the following oracle optimisation criteria. This allows one to retrieve the parameters $\boldsymbol{u}_h^\star$, $\boldsymbol{w}_h^\star$ and $\boldsymbol{\Theta}_h^\star$. 
\begin{itemize}
\item \textbf{Step 1}: solve    
\begin{equation}
(\boldsymbol{u}_h^\star,\boldsymbol{v}_h^\star)   = \underset{\substack{(\boldsymbol{u}_h,\boldsymbol{v}_h)\in\mathbb{R}^p\times\mathbb{R}^q;\|\boldsymbol{u}_h\|_2=\|\boldsymbol{v}_h\|_2=1\\D_0(\boldsymbol{u}_h^\top\boldsymbol{X}^\star_{h-1},\xi_{\ell}^\star)\text{ is minimal},~\ell=1,\ldots,h-1}}{\argmax}~D\left(\boldsymbol{u}_h^\top\boldsymbol{X}_{h-1}^\star, \boldsymbol{v}_h^\top\boldsymbol{Y}_{h-1}^\star\right),\label{eq: general_opti}
\end{equation}

and make sure while optimising the above that $\Psi_{h1}^{-1}(\xi_h^\star/s_{h1}^\star)$ is as close as possible to $\Psi_{h2}^{-1}(\omega_h^\star/s_{h2}^\star))$ (which is associated to Equation \eqref{eq: latent_relationship}) where we set $\xi_h^\star=(\boldsymbol{u}_h^\star)^\top\boldsymbol{X}_{h-1}^\star$ and $\omega_h^\star=(\boldsymbol{v}_h^\star)^\top\boldsymbol{Y}_{h-1}^\star$ (and $s_{h1}^\star$ and $s_{h2}^\star$ their respective standard deviations),
\item \textbf{Step 2}: solve
\begin{equation}
\boldsymbol{w}_{h}^\star  = \underset{\boldsymbol{w}_{h}\in\mathbb{R}^p}{\argmin}~\mathbb{E}_{(\boldsymbol{X}^\star_{h-1}, \xi^\star_h)}\left[L_X\left(\xi_h^\star \boldsymbol{w}_{h}, \boldsymbol{X}^\star_{h-1}\right)\right] \label{eq:opti between latent}
\end{equation}
(which is associated to Equation~\eqref{eq: X_deflation}) and set $\boldsymbol{X}_h^\star=\boldsymbol{X}_{h-1}^\star-\boldsymbol{w}_h^\star\xi_h^\star = \mathcal{P}_{\boldsymbol{\xi}_{h}^{\star\perp}}^{\perp} \boldsymbol{X}_{h-1}^\star \text{  if $L_x$ is square loss}$, 
\item \textbf{Step 3}: solve
\begin{equation}
\boldsymbol{\Theta}_{h}^\star  = \underset{\boldsymbol{\Theta}_h\in\mathbb{R}^K}{\argmin}~\mathbb{E}_{(\boldsymbol{Y}^\star_{h-1},\xi_1^\star,\ldots,\xi_h^\star)^{}}\left[L_Y\left(\boldsymbol{f}_H(\xi_1^\star,\ldots,\xi^\star_h,0,\ldots,0;\boldsymbol{\Theta}_h),\boldsymbol{Y}_{0}^\star\right)\right], \label{eq: generalloss}
\end{equation}
(which is associated to Equation~\eqref{eq: Y_deflation}) and set $\boldsymbol{Y}_h^\star=\boldsymbol{Y}_{h-1}^\star - \boldsymbol{f}_h(\xi^\star_1,\ldots,\xi^\star_h;\boldsymbol{\Theta}^\star_{h})+\boldsymbol{f}_{h-1}(\xi_1^\star,\ldots,\xi_{h-1}^\star;\boldsymbol{\Theta}_{h-1}^\star)$, then go back to Step 1 until $h=H$. 
\end{itemize}
The oracle latent variables $\xi_h^\star$ and $\omega_h^\star$ are also obtained, from which we get
$$
\mathbb{E}\left(\boldsymbol{Y}\mid \xi_1^\star,\ldots,\xi_H^\star\right) = \boldsymbol{f}_{H}(\xi_1^\star,\ldots,\xi_{H}^\star;\boldsymbol{\Theta}_{H}^\star)
$$
is obtained when $h=H$ from Equation \eqref{eq: Generative_Flexible_Latent_Structure_Regression}.

Before discussing how to estimate the unknown parameters of model \eqref{eq: Generative_Flexible_Latent_Structure_Regression}--\eqref{eq: Y_deflation} in practice, we will discuss the assumptions of the error term to help the reader gain a better understanding of several aspects of this model.

From the model setting, the latent variable $\xi_h$ contributes to the prediction of the response, while $\omega_h$ works more like an auxiliary variable. Thus, in the model estimation, $\xi_h$ is more important. From equation~\ref{eq: generated_xi} and equation~\ref{eq: xi_g_function}, even assume that $\boldsymbol{X}_{h-1}^\star$ is a good estimator of $\boldsymbol{X}_{h-1}^\bullet$, the distance between the estimated and the generated latent variable is still $\boldsymbol{u}_h^\top \boldsymbol{X}_H^\bullet$. This term has a mean zero, but a variance $\boldsymbol{u}_h^\top \boldsymbol{\Sigma}_X\boldsymbol{u}_h$. In real practice, we consider four cases for $\boldsymbol{\Sigma}_X:=\mathbb{V}(\boldsymbol{X}_H)$, assuming it is a symmetric and positive semi-definite $p\times p$ matrix:
\begin{enumerate}
    \item all the entries of $\boldsymbol{\Sigma}_X$ are equal to 0;
    \item $\boldsymbol{\Sigma}_X$ is such that $\boldsymbol{u}_h^\top\boldsymbol{\Sigma}_X\boldsymbol{u}_h=0$, $h=1,\ldots,H$;
    \item $\boldsymbol{\Sigma}_X=\sigma^2I_p$ for some $\sigma^2>0$;
    \item $\boldsymbol{\Sigma}_X$ is unconstrained.
\end{enumerate}

In case 1, $\boldsymbol{X}_H=\boldsymbol{0}$ in probability, which corresponds to the situation when there is no added noise, so the estimated latent variable will converge to the generated one if $\boldsymbol{X}_{h}^\star$ converges to $\boldsymbol{X}_{h}^\bullet$.

In case 2, since we have $\mathbb{V}(\boldsymbol{u}_h^\top\boldsymbol{X}_H)=\boldsymbol{u}_h^\top\boldsymbol{\Sigma}_X\boldsymbol{u}_h=0$ and $\mathbb{E}(\boldsymbol{u}_h^\top\boldsymbol{X}_H)=\boldsymbol{u}_h^\top\mathbb{E}(\boldsymbol{X}_H)=\boldsymbol{u}_h^\top\boldsymbol{0}=0$, the random variable $\boldsymbol{u}_h^\top\boldsymbol{X}_H$ is equal to 0 in probability. We have the same situation as in case 1.

Moreover, for the reader's interest, if we want to generate $\boldsymbol{\Sigma}_X$, we cannot have $\boldsymbol{\Sigma}_X>0$ (positive-definite), so $\boldsymbol{\Sigma}_X$ must be singular (i.e., non-invertible). Also, since $\boldsymbol{\Sigma}_X$ is (only) positive semi-definite, it has a unique positive (symmetric positive semidefinite) square root $\boldsymbol{\Sigma}_X^{1/2}$ such that $\boldsymbol{\Sigma}_X=\boldsymbol{\Sigma}_X^{1/2}\boldsymbol{\Sigma}_X^{1/2}$. Thus $0=\boldsymbol{u}_h^\top\boldsymbol{\Sigma}_X\boldsymbol{u}_h=\boldsymbol{u}_h^\top\boldsymbol{\Sigma}_X^{1/2}\boldsymbol{\Sigma}_X^{1/2}\boldsymbol{u}_h=\|\boldsymbol{\Sigma}_X^{1/2}\boldsymbol{u}_h\|^2$ which leads to $\boldsymbol{\Sigma}_X^{1/2}\boldsymbol{u}_h=0$ and in turn to $\boldsymbol{\Sigma}_X\boldsymbol{u}_h=0$ (by pre-multiplying both sides by $\boldsymbol{\Sigma}_X^{1/2}$). Thus $\boldsymbol{u}_h$ must live in the null space of $\boldsymbol{\Sigma}_X$ (the null space, denoted $\textrm{Ker}(\boldsymbol{\Sigma}_X)$, consists of the $d$ eigenvectors corresponding to the zero eigenvalue of $\boldsymbol{\Sigma}_X$). The converse is obviously true. So we must have $\boldsymbol{\Sigma}_X U=0$. 

In case 3, we have $\mathbb{V}(\boldsymbol{u}_h^\top\boldsymbol{X}_H)=\boldsymbol{u}_h^\top\boldsymbol{\Sigma}_{X}\boldsymbol{u}_h=\sigma_x^2\|\boldsymbol{u}_h\|^2=\sigma^2_x$ if we assume additionally that $\|\boldsymbol{u}_h\|^2=1$. Thus, if the size of $\sigma_x$ is controlled, the estimated and true latent variables will be close enough.

Case 4 is similar to case 3, except the variance of the error term is  $\boldsymbol{u}_h^\top\boldsymbol{\Sigma}_X\boldsymbol{u}_h$. 

\begin{remark}
    Note that the model is identifiable with known $\Psi$ functions, or how $\Psi_{h,1}$ and $\Psi_{h,2}$ are related is known. The models will then be partly identifiable with unique $\xi$ and $\omega$.

    Alternatively, imposing further constraints on the $\boldsymbol{w}_h$'s can lead to the uniqueness of the $\boldsymbol{u}_h$'s: for instance, if we impose the former to be orthonormal, then $\boldsymbol{u}_h=\boldsymbol{w}_h$ (which is the case in PLS). If we do not want to impose constraints on the $\boldsymbol{w}_h$'s, we can instead impose constraints on $\Sigma$, the variance of the noise term $\boldsymbol{\epsilon}_H$. We assume that, for $h=1,\ldots,H$, $\boldsymbol{u}_h^\top\Sigma\boldsymbol{u}_h=\|\Sigma^{1/2}\boldsymbol{u}_h\|_2^2\leq \lambda_h$ for some given positive real number $\lambda_h$. This condition is obviously satisfied in case 1, case 2 and case 3 above (just choose some $\sigma^2\leq\min(\lambda_h)$). In case 4, we need to further impose that $\Sigma>0$ and $0<\boldsymbol{u}_h^\top\Sigma\boldsymbol{u}_h=\lambda_h$ (with $\lambda_h=\min~\boldsymbol{u}_h^\top\Sigma\boldsymbol{u}_h$ among all unit-norm $\boldsymbol{u}_h$ satisfying $\boldsymbol{u}_h^\top\boldsymbol{w}_{h'}=\delta_{hh'}$)  which leads to a strictly convex problem and to uniqueness of the $\boldsymbol{u}_h$'s.
    
    In real practice, especially when $f_H$ function is complicated and focusing on prediction, the model is not necessarily identifiable with respect to the $\boldsymbol{\Theta}_h$ parameters and the $\boldsymbol{u}_h$'s, but as soon as the value of $\boldsymbol{f}_H(\xi_1,\ldots,\xi_h,0,\ldots,0;\boldsymbol{\Theta}_h)$ is the same, the models have the same prediction ability.
\end{remark}

As mentioned in Remark 3, if further information or further constraints are added in the data generation procedure, the model is identifiable. However, in real practice,  when applying the equation~\ref{eq: general_opti} to \eqref{eq: generalloss} on a real dataset, the parameter $\boldsymbol{u}_h$ might be unidentifiable, together with $\xi_h$. Thus, it makes sense to try and find the $\boldsymbol{u}_h$ that would maximise the signal variance $\text{Var}(\xi_h^\star)$, while at the same time, minimise the noise variance $\boldsymbol{u}_h^\top \boldsymbol{\Sigma}_X^\star \boldsymbol{u}_h$.

Among those $\boldsymbol{u}_h$ satisfying the constraints given in the above equation~\ref{eq: general_opti}, if any flexibility is left (because of non-unicity), so what we want to achieve is to maximise (in terms of $\boldsymbol{u}_h$) the following: 
\begin{equation}
  \boldsymbol{u}_h^\top \boldsymbol{\Sigma}_{h-1}^\star \boldsymbol{u}_h,  \label{eq: regularization_term}  
\end{equation}
which is naturally the variance of $\xi_h^\star$. Moreover, from the deflation we have:
$$
\boldsymbol{X}_0^\star = \sum_{i=1}^H \boldsymbol{w}_i^\star \xi_i^\star +\boldsymbol{X}_H^\star.
$$
Then, because the $\xi_h^\star$ are still uncorrelated, we have:
$$
\text{Var}(\boldsymbol{X}_0^\star) = \sum_{i=1}^H \boldsymbol{w}_i^{\star\top} \text{Var}(\xi_i^\star) \boldsymbol{w}_i^\star+\boldsymbol{\Sigma}_X^\star.
$$
Since the $\text{Var}(\boldsymbol{X}_0^\star) = \text{Var}(\boldsymbol{X}_0^\bullet)$ is fixed, maximizing $\text{Var}(\xi_h)$ at each step, is the same as minimizing the $\boldsymbol{\Sigma}_X^\star$, thus minimizing $\boldsymbol{u}_h^{\star\top} \boldsymbol{\Sigma}_X^\star \boldsymbol{u}_h^\star$.

With the regulaization term, the optimization problem~\eqref{eq: general_opti} becomes:
\begin{equation}
    (\boldsymbol{u}_h^\star,\boldsymbol{v}_h^\star)   = \underset{\substack{(\boldsymbol{u}_h,\boldsymbol{v}_h)\in\mathbb{R}^p\times\mathbb{R}^q;\|\boldsymbol{u}_h\|_2=\|\boldsymbol{v}_h\|_2=1\\D_0(\boldsymbol{u}_h^\top\boldsymbol{X}^\star_{h-1},\xi_{\ell}^\star)\text{ is minimal},~\ell=1,\ldots,h-1}}{\argmax}~D\left(\boldsymbol{u}_h^\top\boldsymbol{X}_{h-1}^\star, \boldsymbol{v}_h^\top\boldsymbol{Y}_{h-1}^\star\right)+\eta \boldsymbol{u}_h^\top \boldsymbol{\Sigma}_{h-1}^\star \boldsymbol{u}_h,
\end{equation}
for some $\eta>0$.

\subsubsection{Model Training}
\label{subsec:model_training}
Now, we will discuss the training phase. Suppose we observe a a random sample $\{(\boldsymbol{X}^{\bullet(i)}_0,{\boldsymbol{Y}}^{\bullet(i)}_0)\}_{i=1}^n = \left\{(\boldsymbol{Y}^\bullet_1,\boldsymbol{X}^\bullet_1),\ldots,(\boldsymbol{Y}^\bullet_n,\boldsymbol{X}^\bullet_n)\right\}$, and we assume that, for $i=1,\ldots,n$, $(\boldsymbol{Y}_i,\boldsymbol{X}_i)$ follows model \eqref{eq: Generative_Flexible_Latent_Structure_Regression}--\eqref{eq: Y_deflation} for some (true) unknown parameter vectors $\boldsymbol{\Theta}_h^{\bullet}$ and coefficient vectors $\boldsymbol{w}_h^{\bullet}$, $h=1,\ldots,H$. As before, the ${\bullet}$ sign indicates the true fixed and unknown value of the parameters. 

Set $(\underline{\hat{\boldsymbol{X}}}_{0}, \underline{\widehat{{\boldsymbol{Y}}}}_{0}) = (\underline{\boldsymbol{X}}_{0},\underline{{\boldsymbol{Y}}}_{0})$. The goal is to find estimators $\widehat{\boldsymbol{\Theta}}_{h}$ of  $\boldsymbol{\Theta}_h^{\bullet}$, $\hat{\boldsymbol{w}}_h$ of $\boldsymbol{w}_h^{\bullet}$, as well as auxiliary estimators $\hat{\boldsymbol{u}}_h$ and $\hat{\boldsymbol{v}}_h$ respectively, that minimize (in a sense made explicit below) the residuals corresponding to the errors $\boldsymbol{X}_H$, $\boldsymbol{Y}_H$, $(\epsilon_{1}, \epsilon_{2})_h^{\top}$. This will enable us to obtain a set of $H$ pairs of \textit{components} $\hat{\boldsymbol{\xi}}_h$ and $\hat{\boldsymbol{\omega}}_h$.

Moreover, based on a new (i.e., previously unseen test data)  $\boldsymbol{X}_{\textrm{new}}$ taking values in $\mathbb{R}^p$, we can then  predict the unobserved associated random vector $\boldsymbol{Y}_{\textrm{new}}$ using the conditional expectation
\begin{equation}
    \widehat{\boldsymbol{Y}} = \widehat{\mathbb{E}}\left[\boldsymbol{Y}_{\textrm{new}}\mid \boldsymbol{X}_{\textrm{new}}\right] = \widehat{\mathbb{E}}\left[\boldsymbol{Y}_{\textrm{new}}\mid \hat{\xi}_{1, \textrm{new}},\ldots,\hat{\xi}_{h, \textrm{new}}\right] = \boldsymbol{f}_{H}(\hat{\boldsymbol{u}}_1 x_{0,\textrm{new}},\ldots,\hat{\boldsymbol{u}}_h x_{h-1,\textrm{new}}; \widehat{\boldsymbol{\Theta}}_{H}). \label{eq:generalprediction}
\end{equation}

To achieve this, we consider empirical versions of the optimisation Equations \eqref{eq: general_opti}, \eqref{eq:opti between latent} and \eqref{eq: generalloss}, and (numerically) solve the following optimization problems, for $h=1,\ldots,H$:

\begin{align}
(\hat{\boldsymbol{u}}_h,\hat{\boldsymbol{v}}_h)   &=\underset{\substack{(\boldsymbol{u}_h,\boldsymbol{v}_h)\in\mathbb{R}^p\times\mathbb{R}^q;\|\boldsymbol{u}_h\|=\|\boldsymbol{v}_h\|=1\\ \widehat{D_{0}}(g_h(\underline{\hat{\boldsymbol{X}}}_{h-1};\boldsymbol{u}_h),\boldsymbol{\xi}_{\ell})\text{ is minimal},~\ell=1,\ldots,h-1}}{\argmax}~\widehat{D}\left(g_h(\underline{\hat{\boldsymbol{X}}}_{h-1};\boldsymbol{u}_h), k_h(\underline{\hat{\boldsymbol{Y}}}_{h-1};\boldsymbol{v}_h)\right)\label{eq: empirical_maxD}\\\nonumber
\hat{\boldsymbol{\xi}}_h &= g_h(\underline{\hat{\boldsymbol{X}}}_{h-1};\hat{\boldsymbol{u}}_h) = \underline{\hat{\boldsymbol{X}}}_{h-1}\hat{\boldsymbol{u}}_h\\\nonumber
\hat{\boldsymbol{\omega}}_h &= k_h(\underline{\hat{\boldsymbol{Y}}}_{h-1}
\hat{\boldsymbol{v}}_h) = \underline{\hat{\boldsymbol{Y}}}_{h-1}\hat{\boldsymbol{v}}_h.
\\
\hat{\boldsymbol{w}}_{h}  &= \underset{\boldsymbol{w}\in\mathbb{R}^p}{\argmin}~n^{-1}\sum_{i=1}^n\left[L_X\left(\hat{\xi}_{h,i}\boldsymbol{w}_h, \boldsymbol{X}_{h-1, i}\right)\right],\quad i=1,\ldots,n,\\\nonumber
\underline{\hat{\boldsymbol{X}}}_{h}&=\underline{\hat{\boldsymbol{X}}}_{h-1}-\hat{\boldsymbol{\xi}}_{h}\hat{\boldsymbol{w}}_{h}^\top,\quad\text{(deflation of the predictors)} \\
&=\mathcal{P}_{\hat{\boldsymbol{\xi}}_{h}^\perp}^\perp \underline{\hat{\boldsymbol{X}}}_{h-1} \text{  if $L_x$ is square loss}  \label{eq:empirical_Xdelfation}\\
\widehat{\boldsymbol{\Theta}}_{h}  &= \underset{\boldsymbol{\Theta}_h\in\mathbb{R}^K}{\argmin}~n^{-1}\sum_{i=1}^n\left[L_Y\left(\boldsymbol{f}_H(\hat{\boldsymbol{\xi}}_{1,i},\ldots,\hat{\boldsymbol{\xi}}_{h,i},0,\ldots,0;\boldsymbol{\Theta}_h),\boldsymbol{Y}_{0,i}\right)\right]\label{eq: empirical_loss}\\
\underline{\hat{\boldsymbol{Y}}}_{h} &= \underline{\hat{\boldsymbol{Y}}}_{0}-f_H(\hat{\boldsymbol{\xi}}_{1},\ldots,\hat{\boldsymbol{\xi}}_{h},0,\ldots,0;\widehat{\boldsymbol{\Theta}}_h),\quad\text{(deflation of the response variable)} \label{eq: empirical_Ydeflation}
\end{align}
where $\widehat{D}$ is the empirical version of $D$, $\widehat{D_0}$ is the empirical version of $D_0$, and $\mathcal{P}_{\hat{\boldsymbol{\xi}}_{h}^\perp}^\perp \underline{\hat{\boldsymbol{X}}}_{h-1}$ denotes the orthogonal projection of $\underline{\hat{\boldsymbol{X}}}_{h-1}$ onto the space orthogonal to $\hat{\boldsymbol{\xi}}_{h}$.

\subsection{Links to other methods}

The GFLSR model family encompasses numerous well-known methods used in the literature, but often not presented as models. 

\begin{figure}[H]
\includegraphics[width=17cm]{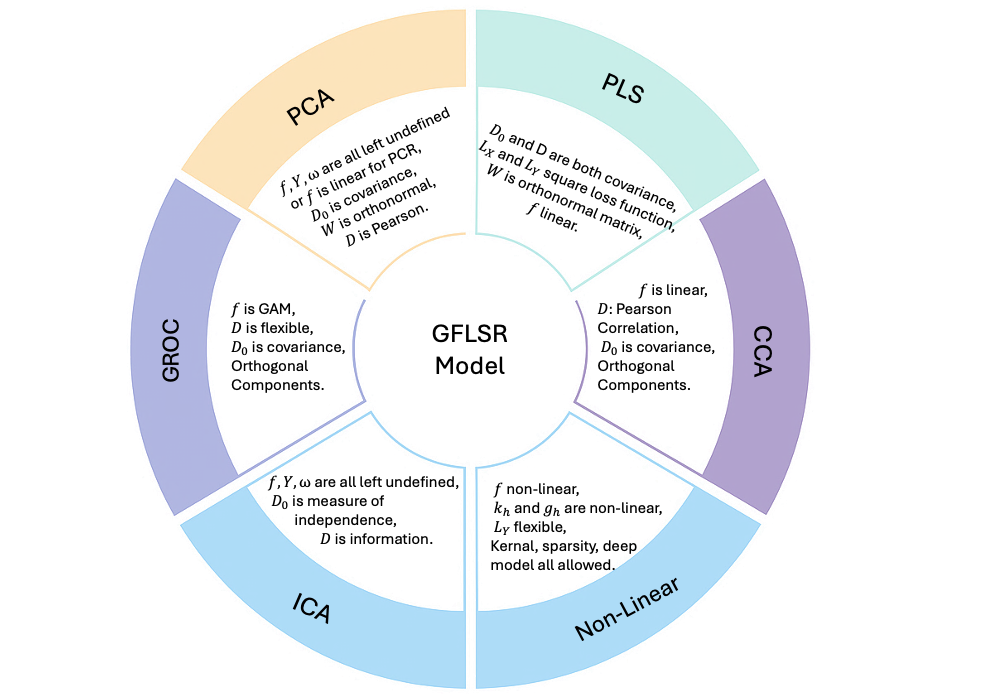}
\centering
\caption{The link between the GFLSR model and other methods}
\label{fig: link}
\end{figure}

As shown in Fig~\ref{fig: link}, if the $\omega$ in Equation~\eqref{eq: latent_relationship}, the $\boldsymbol{f}_H$ in Equations~\eqref{eq: Generative_Flexible_Latent_Structure_Regression} is linear and $\boldsymbol{Y}$ is left undefined, then the proposed GFLSR model is generalized to unsupervised learning. For example, in Equation \eqref{eq: general_opti}, if the dependence measure $D$ is the variance of the components, $\eta = 0$, and the matrix $\boldsymbol{W}$  in equation~\ref{eq: X_deflation} is orthonormal, the model reduces to traditional Principal Components Analysis (PCA) \cite{hotelling1933analysis}. Moreover, if the $\boldsymbol{f}_H$ in Equations \eqref{eq: Generative_Flexible_Latent_Structure_Regression} is linear, while still leaving the $Y$ latent variable $\omega$ undefined, the model is transferred to Principal Components Regression (PCR) \cite{jolliffe1982note}. Additionally, if in Equation \eqref{eq: general_opti}, $D_0$ denotes the measure of independence between latent variables, the $D$ denotes the information criterion, the method will change to Independent Component Analysis (ICA) \cite{comon1994independent}.

Since the GFLSR model is originally a supervised learning method, it's easier to convert to regression methods. If the dependence measure $D$ in Equation  \eqref{eq: general_opti} represents the covariance, while maintaining linear functions in Equations \eqref{eq: Generative_Flexible_Latent_Structure_Regression}. Again, the matrix $\boldsymbol{W}$  in equation~\ref{eq: X_deflation} is orthonormal and maintains the orthogonality between components; the model transitions to Partial Least Squares (PLS) \cite{cha1994partial}. Then, if the dependence measure $D$ is the Pearson correlation instead of the covariance and the matrix $\boldsymbol{W}$  is no longer orthonormal, the method corresponds to Canonical Correlation Analysis (CCA) \cite{hotelling1992relations}. Moreover, if the dependence measure $D$ is flexible and $\boldsymbol{f}_H$ in Equations \eqref{eq: Generative_Flexible_Latent_Structure_Regression} is a Generalised Additive Model (GAM), this model becomes Generalised Regression with the Orthogonal Components (Groc) \cite{bilodeau2015r}. Note that the $L_Y$ function in equation~\eqref{eq: general_opti} can incorporate a regularisation term to involve sparsity, or even change to classification in all the above-mentioned methods.

Additionally, remind the $g_h$ function $g_h(\boldsymbol{X};\boldsymbol{u}_h)$ and $k_h$ function $k_h(\boldsymbol{Y};\boldsymbol{v}_h)$, if these two functions are no longer linear, the GFLSR model can be extended to incorporate non-linear latent structure regression, for example: kernel PLS \cite{rosipal2001kernel, rosipal2002kernel}, kernel PCA \cite{scholkopf1997kernel, kim2005iterative, nguyen2010fault, lee2004nonlinear,wang2012kernel}, or even deep CCA \cite{yu2018category, zhang2021feature, benton2017deep, sun2020learning}. If the model structure is more flexible, with only $\boldsymbol{f}_H$, $g_h$ and $k_h$ left, then some deep latent structure methods can be incorporated, such as: Autoencoder, Generative Adversarial Network and so on. However, this flexibility comes at the cost of interpretability.

\section{Partial Least Square}
\label{sec: pls}
\subsection{Model Definition}
The partial least squares (PLS) method, also known as a projection to latent structures \cite{abdi2010partial}, is a powerful unsupervised method for dimension reduction. Keeping the same notation as in Section \ref{subsec:generalmodel}, PLS addresses the challenge of analyzing high-dimensional data by transforming $\boldsymbol{X} = \{\boldsymbol{X}_1, \ldots, \boldsymbol{X}_p \} \rightarrow \boldsymbol{\xi} = \{\xi_1, \ldots, \xi_H \}$ and $\boldsymbol{Y} = \{\boldsymbol{Y}_1, \ldots, \boldsymbol{Y}_q\}\rightarrow \boldsymbol{\omega} = \{\omega_1, \ldots, \omega_H \}$. The $\boldsymbol{\xi}$ and $\boldsymbol{\omega}$ latent variables are called $\boldsymbol{X}$ scores and $\boldsymbol{Y}$ scores respectively. The scores are linear combinations of the original variables that maximise covariance, and best describe the relationship between the predictors $\boldsymbol{X}$ and the response variables $\boldsymbol{Y}$. 

Starting from the proposed GFLSR model structure, to formally define a PLS model, we first need a linear $\boldsymbol{f}_H$ function in equation \eqref{eq: Generative_Flexible_Latent_Structure_Regression}. Then, from Proposition~\ref{prop: covariance} in appendix~\ref{app: theorem}, to maximise the covariance between the latent variables, a linear relationship between $\Psi_{h1}$ and $\Psi_{h2}$ is considered, so only a single $\Psi_h$ function is used in each step. Moreover, the proposed GFLSR model structure is naturally a supervised learning model, focusing on prediction. Thus, the resulting model will align with the PLS-R method. In order to further align with the symmetric PLS method (PLS-W2A and PLS-SVD), a symmetric version is also proposed. Further information about different types of PLS can be found here \cite{lafaye2019pls}.

With the proposed GFLSR model structure, the PLS method can be formally defined, and the details of the model are presented below as Definition 1. The proposed PLS version is referred to as Generative-PLS in the following.

\begin{definition}[The Generative PLS model]
In our generative FLSR model, set  $\Psi_{h1}=\Psi_{h2}:=\Psi_{h}$, $K=q$, $\boldsymbol{\Theta}_h=(\boldsymbol{\theta}_1, \ldots, \boldsymbol{\theta}_h,0,\ldots,0)$, $\boldsymbol{f}_{H}(\boldsymbol{X};\boldsymbol{\Theta}) = \boldsymbol{\Theta}^\top\boldsymbol{X}$, $b_h=s_{2h}/s_{1h}$, $U_h=U$ and $\epsilon_{1,h}=0$ and let $\epsilon_{h}=s_{2h}\epsilon_{2,h}$. Additionally, assume $\boldsymbol{W} = (\boldsymbol{w}_1, \ldots, \boldsymbol{w}_H)$ and $\boldsymbol{V} = (\boldsymbol{v}_1, \ldots, \boldsymbol{v}_H)$ are orthonormal, and set $\boldsymbol{\theta}_h = \boldsymbol{v}_hb_h$, $b_h\sigma_{\xi,h}^2$ decreases strictly. We then obtain the  \textbf{Generative PLS-R model}:
$$\boldsymbol{Y}_{0}  =  \boldsymbol{f}_{H}(\xi_1,\ldots,\xi_{H};\boldsymbol{\Theta}_{H}) + \widetilde{\boldsymbol{\epsilon}}_{H} = \boldsymbol{\theta}_1\xi_1+\cdots+\boldsymbol{\theta}_H\xi_H+\boldsymbol{Y}_H,$$
where, for $h=H,\ldots,1$:
\begin{eqnarray*}
\left(\xi_h, \omega_h\right)^\top &=& 
\begin{cases}
s_{1h}[\Psi_{h}(U)]\\
b_hs_{1h}[\Psi_{h}(U)] + \epsilon_{h}
\end{cases}
~\text{with }U\sim\text{Unif}[0,1], \\
\boldsymbol{X}_{h-1} & = & \boldsymbol{w}_{h}\xi_h + \boldsymbol{X}_{h}\\
\boldsymbol{Y}_{h-1} & = & \boldsymbol{\theta}_h\xi_h+\boldsymbol{Y}_h\\
\end{eqnarray*}
Moreover, setting for $h=0,\ldots,H$, $\widetilde{\boldsymbol{Y}}_h = \boldsymbol{Y}_h-\sum_{k=1}^h\boldsymbol{v}_k\epsilon_{k}$, we obtain the \textbf{Generative (symmetric) PLS-SVD  model}: 
$$\boldsymbol{Y}_{0} = \widetilde{\boldsymbol{Y}}_0  =  \boldsymbol{\theta}_1\xi_1+\cdots+\boldsymbol{\theta}_H\xi_H+\boldsymbol{Y}_H = \boldsymbol{v}_1\omega_1+\cdots+\boldsymbol{v}_H\omega_H+\widetilde{\boldsymbol{Y}}_H,$$
where, for $h=H,\ldots,1$:
\begin{eqnarray*}
\left(\xi_h, \omega_h\right)^\top &=& 
\begin{cases}
s_{1h}[\Psi_{h}(U)]\\
b_hs_{1h}[\Psi_{h}(U)] + \epsilon_{h}
\end{cases}
~\text{with }U\sim\text{Unif}[0,1], \\
\boldsymbol{X}_{h-1} & = & \boldsymbol{w}_{h}\xi_h + \boldsymbol{X}_{h}\\
\widetilde{\boldsymbol{Y}}_{h-1} &=& \boldsymbol{v}_h\omega_h
+\widetilde{\boldsymbol{Y}}_h
\end{eqnarray*}
In the symmetric case, extra interpretation can be made for $\omega$. Moreover, use $\boldsymbol{\epsilon}$ to denote the column vector of all $\{\epsilon_h\}_{h=1}^n$, to better align with the symmetric PLS method, instead of $\boldsymbol{Y}_H \perp \boldsymbol{\epsilon}$, we can assume $\widetilde{\boldsymbol{Y}}_H \perp \boldsymbol{\epsilon}$.
\end{definition}

One can define $\Psi_n(t)=H_n(\Phi^{-1}(t))$ where $\Phi^{-1}$ is the inverse CDF of the $N(0,1)$ and $H_n$ is a polynomial of degree $n$, so that $\{\Psi_n;n\in\mathbb{N}\}$ forms an orthonormal family of functions from $[0,1]$ to $\mathbb{R}$ (to ensure the $\xi_h$'s are uncorrelated, and similarly for the $\omega_h$'s). The $H_n$ functions are built so that  $\int_0^1\Psi_n(t)dt=0$ and $\int_0^1\Psi_n(t)\Psi_m(t)dt=\delta_{nm}$. Other orthonormal families (e.g., not using polynomials) or other distributions can be used. Moreover, if $\Psi_n(t)$ is not from an orthogonal family, instead of using the same $U \sim U(0,1)$ at each step, different $U_h$ can be used. This ensures the uncorrelated between the latent variables within different steps.

For the error term $\text{Var}(\boldsymbol{X}_H) = \boldsymbol{\Sigma}_X$ and $\text{Var}(\widetilde{\boldsymbol{Y}}_H) = \boldsymbol{\Sigma}_{\tilde{Y}}$, we make one of the following three  assumptions: 
\begin{itemize}
\item [(a)] $\boldsymbol{\Sigma}_X = \boldsymbol{\Sigma}_{\tilde{Y}} = \boldsymbol{0}$ or $\boldsymbol{\Sigma}_X=\boldsymbol{\Sigma}_X^2(I_p-\boldsymbol{W}(\boldsymbol{W})^+)(I_p-\boldsymbol{W}(\boldsymbol{W})^+)^\top=\sigma_X^2(I_p-\boldsymbol{W}(\boldsymbol{W})^\top)$ and $\boldsymbol{\Sigma}_{\tilde{Y}}=\sigma_y^2 (I_q-\boldsymbol{V}(\boldsymbol{V})^+)(I_q-\boldsymbol{V}(\boldsymbol{V})^+)^\top=\sigma_y^2(I_q-\boldsymbol{V}(\boldsymbol{V})^\top)$.
\item [(b)] $\boldsymbol{\Sigma}_X=\sigma^2_xI_p$ and $\boldsymbol{\Sigma}_X=\sigma^2_yI_q$.
\item [(c)]  $\boldsymbol{\Sigma}_X$ and $\boldsymbol{\Sigma}_{\tilde{Y}}$ are semi-positive definite matrices.
\end{itemize}
Note that the assumption is made based on $\boldsymbol{\Sigma}_{\tilde{Y}}$ is to align with the current Probabilistic models. If a generative PLS-R model is preferred, with the relationship $ \widetilde{\boldsymbol{Y}}_H=\boldsymbol{Y}_H-\sum_{k=1}^H\boldsymbol{v}_k\epsilon_{k}$, if we further impose $\text{Var}(\epsilon_h) = \sigma_{1}^2$ for all $h$, we can have 
$$
\boldsymbol{\Sigma}_{\tilde{Y}} = \boldsymbol{\Sigma}_Y + \sigma_{1}^2\boldsymbol{V}\boldsymbol{V}^\top
$$
\begin{remark}
    We note that, thanks to our specific choice of the $\Psi_n$ functions, the $\xi_h$'s (resp., the $\omega_h$'s) are uncorrelated. But they are \textbf{not} independent because the same uniform random variable $U$ is used for $h=1,\ldots, H$. If we decide to use different (independent) uniform random variables, $U_h$ say, then the $\xi_h$'s (resp., the $\omega_h$'s) will become independent as well. Also, because we used the same function $\Psi_h(\cdot)$ (and same $U$) to define both $\xi_h$ and $\omega_h$, we have that $\xi_h$ and $\omega_h$ are maximally dependent (and maximally correlated because $\omega_h=b_h\xi_h$, neglecting $\epsilon_{1,h}$). But if we use a different function, $\Psi_h$ and $\Psi_{h+1}$ say, (but same $U$) to define respectively $\xi_h$ and $\omega_h$, then $\xi_h$ and $\omega_h$ will still be maximally dependent (neglecting $\epsilon_{1,h}$) but uncorrelated. In this latter case, other advanced methods, such as Groc \cite{bilodeau2015r}, should work, but not PLS.
\end{remark}

\subsection{Model Properties and Parameter Estimation}
\label{subsec: pls_property}
From the model, clearly $\mathbb{E}(\xi_h)=\mathbb{E}(\xi_l)=\mathbb{E}(\omega_h)=\mathbb{E}(\omega_l)=0$ and thus we have (if $h\ne l$):
$$
\textrm{Cov}(\xi_h,\xi_l)=\mathbb{E}(\xi_h\xi_l)=\sigma_{\xi,h}\sigma_{\xi,l}\int_0^1(\Psi_h(t)-\mathbb{E}(\Psi_h(t)))(\Psi_l(t)-\mathbb{E}(\Psi_l(t)))dt=0.
$$
Symmetrically,
$$
\textrm{Cov}(\omega_h,\omega_l)=\mathbb{E}(\omega_h\omega_l)=\mathbb{E}\left((b_h\xi_h+\epsilon_{1,h})(b_l\xi_l+\epsilon_{1,l})\right)=0.
$$
We also have
$$
Cov(\xi_h,\omega_h)=Cov(\xi_h,b_h\xi_h+\epsilon_{1,h})=b_h\sigma_{\xi,h}^2; \quad Cov(\xi_h,\omega_j)=0 \quad (\text{if } j\ne h).
$$

For the variance of the latent variables, if we further denote $\boldsymbol{B} = \text{diag}(b_1,\ldots, b_H)$, we have 
$$
Var(\boldsymbol{\xi})=\textrm{diag}(s_{1,1}^2,\ldots,s_{1,H}^2); \quad Var(\boldsymbol{\omega})=\textrm{diag}(b_1^2s_{1,1}^2,\ldots,b_H^2s_{1,H}^2)+\sigma_1^2\boldsymbol{I}_H=\boldsymbol{B}^2Var(\boldsymbol{\xi})+\sigma_1^2\boldsymbol{I}_H.
$$

From equation~\ref{eq: generated_xi}, we have that that there exist $\boldsymbol{U} = \{\boldsymbol{u}_1, \ldots, \boldsymbol{u}_H\}$, such that $\xi_h = \boldsymbol{u}_h^\top\boldsymbol{X}_{h-1} - \boldsymbol{u}_h^\top\boldsymbol{X}_{H}$, and Lemma~\ref{lemma: UV} in Appendix~\ref{app: theorem}, we have the result that $\boldsymbol{U} = \boldsymbol{W}$ and $\boldsymbol{U}$ is unique, so in Generative-PLS, we have:
\begin{equation}
\xi_h = \boldsymbol{w}_h^\top\boldsymbol{X}_{h-1} - \boldsymbol{w}_h^\top\boldsymbol{X}_{H}.\label{eq: pls_generated_xi}
\end{equation}
Symmetrically, we have:
\begin{equation}
    \omega_h = {\boldsymbol{v}_h}^\top \widetilde{\boldsymbol{Y}}_{h-1}- {\boldsymbol{v}_h}^\top \widetilde{\boldsymbol{Y}}_{H} = {\boldsymbol{v}_h}^\top \boldsymbol{Y}_{h-1}- {\boldsymbol{v}_h}^\top \boldsymbol{Y}_{H}+\epsilon_h.\label{eq: pls_generated_omega}
\end{equation}

From the iterative model, we can further write down the matrix form of the Generative (Symmetric) PLS-SVD model:
\begin{eqnarray}
\boldsymbol{X}_0 =& \sum_{h=1}^H\boldsymbol{w}_h\xi_h+\boldsymbol{X}_H =\boldsymbol{W}\boldsymbol{\xi}+\boldsymbol{X}_H, \label{eq: matrix_pls_X}\\
\widetilde{\boldsymbol{Y}}_0=&\sum_{h=1}^H\boldsymbol{v}_h\omega_h+\widetilde{\boldsymbol{Y}}_H
=\boldsymbol{V}\boldsymbol{\omega}+\widetilde{\boldsymbol{Y}}_H  \label{eq: matrix_pls_Y}\\
\boldsymbol{\omega} =&  \boldsymbol{B}\boldsymbol{\xi}+\boldsymbol{\epsilon} \label{eq: matrix_pls_latent}
\end{eqnarray}
Note that, for Generative PLS-R, the equation~\eqref{eq: matrix_pls_Y} becomes 
\begin{equation}
    \boldsymbol{Y}_0 = \boldsymbol{\Theta}\boldsymbol{\xi}+\boldsymbol{Y}_H \tag{24-A} \label{eq: matrix_pls_Y_A}
\end{equation}
From the matrix form, it's clear that this iterative model has the same matrix form as the traditional PLS method \cite{Urbas_2024}. Moreover, we can show that most of the recent probabilistic PLS models can be incorporated into our structure. 
\begin{figure}[ht]
        \centering
        \includegraphics[width = 16cm]{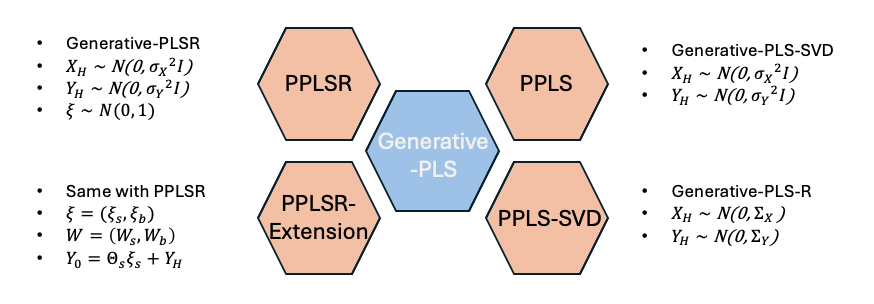}
        \caption{Relationship between Generative PLS and other PLS models}
        \label{fig: ppls}
\end{figure}

As shown in Figure~\ref{fig: ppls}, the initial probabilistic partial least squares regression model (PPLSR), proposed by  Li et al. \cite{Li2011, Li_2015}, is the Generative PLS-R under the assumption (a) and all $s_{1,h} = 1$. Moreover, the relationship between $\xi$ and $\omega$ is not considered, thus they further have $\epsilon_{h} = 0$ for all $h$.
Then, Zheng et al. \cite{Zheng2016, Zheng2018} extended the model, denoted as PPLSR-Extension. To incorporate with this extension model, we have the Generative PLS-R have all the assumptions same as PPLSR, with $\boldsymbol{\xi} = \{\boldsymbol{\xi_s,\boldsymbol{\xi_b}}\}$ and only $\boldsymbol{\xi}_s$ contributes to the $\boldsymbol{Y}_0$.

Moreover, the symmetric probabilistic model (PPLS) proposed by \cite{el_Bouhaddani_2018} is the Generative PLS-SVD under the assumption (b) and $\epsilon \sim N(0,\sigma_1^2)$. In 2022, \cite{Eti_vant_2022} proposed a more flexible PPLS-SVD model based on the PPLS, which has the same model structure as Generative PLS-R under assumption (c).

For the parameter estimation, the same estimation structure can be used.  Since from Lemma\ref{lemma: UV}, the constructed latent variable in PLS as equation~\eqref{eq: pls_generated_xi} is unique, the regularisation term in \eqref{eq: general_opti} is neglected (Also for alignment with the current PLS algorithms). Given only $\boldsymbol{X}_0^\bullet$ and $\boldsymbol{Y}_0^\bullet$ following the Generative PPLS-R or generative PPLS-SVD model, we set $(\boldsymbol{X}_0^\star, \boldsymbol{Y}_0^\star) = (\boldsymbol{X}_0^\bullet, \boldsymbol{Y}_0^\bullet)$. Set $D=D_0=Cov$. Iteratively for $h = 1,\ldots, H$, one can find the oracle minimizes 
\begin{equation}
(\boldsymbol{u}_h^\star,\boldsymbol{v}_h^\star)   = \underset{\substack{(\boldsymbol{u}_h,\boldsymbol{v}_h)\in\mathbb{R}^p\times\mathbb{R}^q;\|\boldsymbol{u}_h\|=\|\boldsymbol{v}_h\|=1\\
\boldsymbol{u}_{\ell}^\top\boldsymbol{u}^\star_h=\boldsymbol{v}_{\ell}^\top\boldsymbol{v}^\star_h=0,~\ell=1,\ldots,h-1\\
Cov(\boldsymbol{X}_{h-1}^{\star\top}\boldsymbol{u}_h,\xi_{\ell})\text{ is minimal},~\ell=1,\ldots,h-1}}{\argmax}~Cov\left(\boldsymbol{X}_{h-1}^{\star\top}\boldsymbol{u}_h, \boldsymbol{Y}_{h-1}^{\star\top}\boldsymbol{v}_h\right),\label{eq: pls_optim} \end{equation}
then set $\xi_h^\star=\boldsymbol{X}_{h-1}^{\star\top}\boldsymbol{u}_h^\star$ and $\omega_h^\star=\boldsymbol{X}_{h-1}^{\star\top}\boldsymbol{v}_h^\star$. We further proved in Lemma~\ref{lemma: star} in Section~\ref{subsec: theorem} that $\boldsymbol{w}^\bullet = \boldsymbol{u}^\star$, so the step equation \ref{eq:opti between latent} that solving $\boldsymbol{w}$ can be prevented. Then solve
\begin{equation}
b_{h}^\star = \underset{b_h\in\mathbb{R}}{\argmin}~\mathbb{E}_{(\boldsymbol{Y}_{0},\xi_1^\star,\ldots,\xi_h^\star)^{}}\left[L_Y\left(\boldsymbol{Y}_{0}, \sum_{j=1}^hb_j\boldsymbol{v}_j^\star\xi_j^\star\right)\right], \end{equation}

(note: $L_Y\left(\boldsymbol{Y}_{0}, \sum_{j=1}^hb_j\boldsymbol{v}_j^\star\xi_j^\star\right)=L_Y\left(\boldsymbol{Y}_{0}- \sum_{j=1}^{h-1}b_j\boldsymbol{v}_j^\star\xi_j^\star,b_h\boldsymbol{v}_h^\star\xi_h^\star\right)$)

or equivalently
\begin{equation}
b_{h}^\star = \underset{b_h\in\mathbb{R}}{\argmin}~\mathbb{E}_{(\boldsymbol{Y}^\star_{h-1},\xi_h^\star)}\left[L_Y\left(\boldsymbol{Y}^\star_{h-1}, b_h\boldsymbol{v}_h^\star\xi_h^\star\right)\right], \label{eq:pls_opti3}\end{equation}
and set
$$
\boldsymbol{\Theta}_{h}^\star = \{b_1^\star\boldsymbol{v}_1^\star,\ldots,b_h^\star\boldsymbol{v}_h^\star\}=\left(\cup_{j=1}^{h-1}\boldsymbol{\Theta}_{j}^\star\right)\cup\{b_h^\star\boldsymbol{v}_h^\star\}.
$$
Note that since 
$\boldsymbol{Y}_{h-1}=b_h^\star\boldsymbol{v}_h^\star\xi_h^\star+(\boldsymbol{v}_h^\star\epsilon_{1,h}+\boldsymbol{Y}_h)$, we get 
$(\boldsymbol{v}_h^\star)^\top\boldsymbol{Y}_{h-1}=b_h^\star(\boldsymbol{v}_h^\star)^\top\boldsymbol{v}_h^\star\xi_h^\star+(\boldsymbol{v}_h^\star)^\top(\boldsymbol{v}_h^\star\epsilon_{1,h}+\boldsymbol{Y}_h)=b_h^\star\xi_h^\star+(\epsilon_{1,h}+\epsilon_{y,h})$ which shows that the above optimisation problem (if $L_Y$ is the squared loss) can be solved by regressing $(\boldsymbol{v}_h^\star)^\top\boldsymbol{Y}_{h-1}$ on $\xi_h^\star$ (without intercept). 

For the Generative PLS-SVD model, we have $\widetilde{\boldsymbol{Y}}$ replace $\boldsymbol{Y}$ in all the above steps.

Then we set
\begin{eqnarray}
\boldsymbol{X}_{h} & = & -\boldsymbol{u}_{h}^\star\xi_h^\star + \boldsymbol{X}_{h-1}\\
\tilde{\boldsymbol{Y}}_{h} & = & -\boldsymbol{v}_h^\star\omega_h^\star
+\tilde{\boldsymbol{Y}}_{h-1} \text{ or } \boldsymbol{Y}_{h}  =  -\boldsymbol{\theta}_h^\star\xi_h^\star
+\boldsymbol{Y}_{h-1} \label{eq: pls_delfation}
\end{eqnarray}
before setting $h\leftarrow h+1$.

In real practice, an empirical version of the above equations~\eqref{eq: pls_optim} to \eqref{eq: pls_delfation} is used. The empirical version of parameter estimation can be referred to equation~\eqref{eq: empirical_maxD} to \eqref{eq: empirical_Ydeflation}

\subsection{Theoretical Results}
\label{subsec: theorem}
In this section, we present our main theorems and the lemmas that help the proof. In Theorem~\ref{theorem: main}, we show that with a random sample following the Generative (symmetric) PLS-SVD model, if the sample size goes to infinity, we have the following results. First, solving the empirical version of the optimisation problem~\eqref{eq: pls_optim}, the result will converge in the r-th moment to the parameters $\boldsymbol{w}_h^\bullet$ and $\boldsymbol{v}_h^\bullet$ used in the data generation procedure for any $v >0$. Then, the expected squared error between the estimated empirical latent variable $\hat{\boldsymbol{\xi}}$ and the generated ${\boldsymbol{\xi}}^\bullet$ converges to a deterministic limit based on the variance of generated error term $\text{Var}(\boldsymbol{X}_H^\bullet)$. Moreover, a Symmetric result can be obtained for $\hat{\boldsymbol{\omega}}$.

Then in Lemma~\ref{lemma: plsr result}, the result is extended to the Generative PLS-R model, except the result for $\hat{\boldsymbol{\omega}}$.

\begin{theorem}
\label{theorem: main}
    Let $\{(\boldsymbol{X}^{\bullet(i)}_0,\widetilde{\boldsymbol{Y}}^{\bullet(i)}_0)\}_{i=1}^n$ be a random sample (with respective population covariance matrices $\boldsymbol{\Sigma}_{\boldsymbol{X}_0}^\bullet$ and $\boldsymbol{\Sigma}_{\widetilde{\boldsymbol{Y}}_0}^\bullet$) following the Generative (symmetric) PLS-SVD model, and for $h=H,\ldots,1$, denote  $\boldsymbol{w}_h^\bullet:p\times 1$ and $\boldsymbol{v}_h^\bullet:q\times 1$ the parameters used in the data generation process,  $\boldsymbol{\epsilon}_h^\bullet:n\times 1$ the corresponding noise term, and $\boldsymbol{\xi}_h^\bullet:n\times 1$ and $\boldsymbol{\omega}_h^\bullet:n\times 1$ the two latent variable vectors. Moreover, denote $(\underline{\boldsymbol{X}}_0^\bullet,\underline{\widetilde{\boldsymbol{Y}}}_0^\bullet):n\times(p+q)$ the pair of random matrices formed by stacking as rows the elements of the random sample, by $(\underline{\boldsymbol{X}}_{h-1}^\bullet,\underline{\widetilde{\boldsymbol{Y}}}_{h-1}^\bullet):n\times(p+q)$ the pair of deflated matrices, and when $h=H$, by $(\underline{\boldsymbol{X}}_H^\bullet,\underline{\widetilde{\boldsymbol{Y}}}_H^\bullet):n\times(p+q)$ the error term matrices, with associated population covariances  $\boldsymbol{\Sigma_{X_H}}^\bullet$ and $\boldsymbol{\Sigma}_{\widetilde{Y}_{H}}^\bullet$, respectively.

    Set $(\underline{\hat{\boldsymbol{X}}}_{0}, \underline{\widehat{\widetilde{\boldsymbol{Y}}}}_{0}) = (\underline{\boldsymbol{X}}_{0},\underline{\widetilde{\boldsymbol{Y}}}_{0})$. Then iteratively for each $h=1,\ldots, H$, define the $n\times p$ random matrices $\underline{\hat{\boldsymbol{X}}}_{h} = \underline{\hat{\boldsymbol{X}}}_{h-1}-\hat{\boldsymbol{\xi}}_h\hat{\boldsymbol{u}}_h^\top$ with $\hat{\boldsymbol{\xi}}_h = \underline{\hat{\boldsymbol{X}}}_{h-1} \hat{\boldsymbol{u}}_h$, and symmetrically, define the $n\times q$ random matrices $\underline{\hat{\boldsymbol{Y}}}_{h} = \underline{\hat{\boldsymbol{Y}}}_{h-1}-\hat{\boldsymbol{\omega}}_h\hat{\boldsymbol{v}}_h^\top$ with $\hat{\boldsymbol{\omega}}_h = \underline{\hat{\boldsymbol{Y}}}_{h-1} \hat{\boldsymbol{v}}_h$, where $(\hat{\boldsymbol{u}}_h,\hat{\boldsymbol{v}}_h)$ is a solution to the empirical optimisation problem
    \begin{eqnarray*}
(\hat{\boldsymbol{u}}_h,\hat{\boldsymbol{v}}_h) \in \underset{(\boldsymbol{u}_h,\boldsymbol{v}_h)}{\argmax}~\widehat{\textrm{Cov}}\left({\hat{\boldsymbol{X}}^{\top}_{h-1}}\boldsymbol{u}_h, \hat{\widetilde{\boldsymbol{Y}}}_{h-1}^{\top}\boldsymbol{v}_h\right),
    \end{eqnarray*}
    under the following set of constraints:
    \begin{itemize}
    \item[(C1)] 
    $(\boldsymbol{u}_h,\boldsymbol{v}_h)\in\mathbb{R}^p\times\mathbb{R}^q;\|\boldsymbol{u}_h\|_2=\|\boldsymbol{v}_h\|_2=1$,
    \item[(C2)] $\boldsymbol{u}_{h}^\top\hat{\boldsymbol{w}}_{\ell}=\boldsymbol{v}_h^\top\hat{\boldsymbol{v}}_{\ell}=0,~\ell=1,\dots,h-1$,
    \item[(C3)] 
    $\textrm{Cov}(\hat{\boldsymbol{X}}_{h-1}^{\top}\boldsymbol{u}_h,\hat{\xi}_{\ell})\text{ is minimal},~\ell=1,\dots,h-1$,
    \end{itemize}
where we denoted $\widehat{\textrm{Cov}}\left({\hat{\boldsymbol{X}}^{\top}_{h-1}}\boldsymbol{u}_h, \hat{\widetilde{\boldsymbol{Y}}}_{h-1}^{\top}\boldsymbol{v}_h\right):=\boldsymbol{u}_h\widehat{\boldsymbol{\Sigma}}_{X,\widetilde{Y}}^{(h-1)}\boldsymbol{v}_h^\top$ whith $\widehat{\boldsymbol{\Sigma}}_{X,\widetilde{Y}}^{(h-1)}=\frac{1}{n} \sum_{i = 1}^{n} \hat{\boldsymbol{X}}_{h-1}^{(i)} {\hat{\widetilde{\boldsymbol{Y}}}_{h-1}^{(i)\top}}:p\times q$ the empirical covariance matrix at step $h$.

    Then at each step $h$, as the sample size $n \to \infty$, we have:
    \begin{itemize}
    \item[(1)]for all $r>0$, 
        $$
            \mathbb{E}\left[\|\hat{\boldsymbol{u}}_h - \boldsymbol{w}_h^\bullet\|_2^r\right]\longrightarrow0,
        $$
    \item[(2)] the per-sample expected squared difference between the estimated latent variable and the population latent variable satisfies: 
        $$
            \frac{1}{n}\mathbb{E}\left\|\hat{\boldsymbol{\xi}}_h - \boldsymbol{\xi}_h^\bullet\right\|_2^2 \longrightarrow {\boldsymbol{w}_h^\bullet}^\top \boldsymbol{\Sigma}_{X}^\bullet \boldsymbol{w}_h^\bullet.
        $$
    \item[(3)] Similar results can be obtained in the symmetric case:
    $$
\forall r>0,\mathbb{E}\left[\|\hat{\boldsymbol{v}}_h - \boldsymbol{v}_h^\bullet\|_2^r\right]\longrightarrow0 \quad\text{and} \quad \frac{1}{n}\mathbb{E}\left\|\hat{\boldsymbol{\omega}}_h - \boldsymbol{\omega}_h^\bullet\right\|_2^2 \longrightarrow {\boldsymbol{v}_h^\bullet}^\top \widetilde{\boldsymbol{\Sigma}}_{Y}^\bullet \boldsymbol{v}_h^\bullet.
    $$
    \end{itemize}
\end{theorem}

\begin{proof} 
\label{proof: main}
The proof proceeds in three steps that we prove recursively, for $h=1,\ldots,H$. First, we show that
\begin{equation}
\underset{(\boldsymbol{u}_h,\boldsymbol{v}_h)\textrm{ s.t. }(C1)}{\sup}\left|\widehat{\textrm{Cov}}\left({\hat{\boldsymbol{X}}^{\top}_{h-1}}\boldsymbol{u}_h, \hat{\widetilde{\boldsymbol{Y}}}_{h-1}^{\top}\boldsymbol{v}_h\right) - \textrm{Cov}\left({\boldsymbol{X}^{\star\top}_{h-1}}\boldsymbol{u}_h, \widetilde{\boldsymbol{Y}}_{h-1}^{\star\top}\boldsymbol{v}_h\right)\right|\stackrel{P}{\longrightarrow}0,\label{eq1MainTheorem}
\end{equation}
then we show that 
\begin{equation}
(\hat{\boldsymbol{u}}_h, \hat{\boldsymbol{v}}_h) \stackrel{r}{\longrightarrow} (\boldsymbol{u}_h^\star, \boldsymbol{v}_h^\star) := (\boldsymbol{w}_h^\bullet, \boldsymbol{v}_h^\bullet),
\label{eq2MainTheorem}
\end{equation}
and finally, we prove that 
\begin{equation}
\frac{1}{n}\mathbb{E}\left\|\hat{\boldsymbol{\xi}}_h - \xi_h^\bullet\right\|_2^2 \longrightarrow {\boldsymbol{w}_h^\bullet}^\top \boldsymbol{\Sigma}_{X}^\bullet \boldsymbol{w}_h^\bullet,\qquad \text{as } n\rightarrow\infty.\label{eq3MainTheorem}
\end{equation}

Denote  $\boldsymbol{\Sigma}_{X,\tilde{Y}}^{\star(h-1)}$ the covariance matrix between $\boldsymbol{X}^\star_{h-1}$ and $\widetilde{\boldsymbol{Y}}^\star_{h-1}$,  the population level counterparts of $\underline{\hat{\boldsymbol{X}}}_{h-1}$ and $\underline{\widehat{\widetilde{\boldsymbol{Y}}}}_{h-1}$, respectively. 
We get that
$$
    \widehat{\boldsymbol{\Sigma}}_{X,\tilde{Y}}^{(h-1)} \stackrel{P}{\longrightarrow} \boldsymbol{\Sigma}_{X,\tilde{Y}}^{\star(h-1)} \quad \text{ (elementwise)},
$$
by using the Law of Large Numbers when $h=1$, and by using the following relationships when $h>1$:
$$ \widehat{\boldsymbol{\Sigma}}_{X,\tilde{Y}}^{(h-1)}=(I_p-\hat{\boldsymbol{u}}_{h-1}\hat{\boldsymbol{u}}_{h-1}^\top)\widehat{\boldsymbol{\Sigma}}_{X,\widetilde{Y}}^{(h-2)}(I_q-\hat{\boldsymbol{v}}_{h-1}\hat{\boldsymbol{v}}_{h-1}^\top)\quad\text{ and }\quad\boldsymbol{\Sigma}_{X,\tilde{Y}}^{\star(h-1)}=(I_p-\boldsymbol{u}_{h-1}^\star\boldsymbol{u}_{h-1}^{\star\top})\boldsymbol{\Sigma}_{X,\tilde{Y}}^{\star(h-2)}(I_q-\boldsymbol{v}_{h-1}^\star\boldsymbol{v}_{h-1}^{\star\top}).
$$
Next, by the continuous mapping theorem \cite{mann1943stochastic}, for any fixed $(\boldsymbol{u}_h, \boldsymbol{v}_h) \in \mathbb{S}^{p-1}\times\mathbb{S}^{q-1}$ (obviously a compact metric set endowed with the Euclidean product (chordal) metric), we have
$$
\boldsymbol{u}_h^\top\widehat{\boldsymbol{\Sigma}}_{X,\tilde{Y}}^{(h-1)}\boldsymbol{v}_h -\boldsymbol{u}_h^\top\boldsymbol{\Sigma}_{X,\tilde{Y}}^{\star(h-1)}\boldsymbol{v}_h\stackrel{P}{\longrightarrow}0.
$$
The function in the singleton set $\{(\boldsymbol{u}_h,\boldsymbol{v}_h)\mapsto\boldsymbol{u}_h^\top\boldsymbol{\Sigma}_{X,\tilde{Y}}^{\star(h-1)}\boldsymbol{v}_h\}$ is Lipschitz continuous with Lipschitz constant $K=\sqrt{2}\|{\boldsymbol{\Sigma}}_{X,\widetilde{Y}}^{\star(h-1)}\|_F<\infty$ since
\begin{eqnarray*}
\left|\boldsymbol{u}_h^\top\boldsymbol{\Sigma}_{X,\tilde{Y}}^{\star(h-1)}\boldsymbol{v}_h-\boldsymbol{u}_h'^{\top}\boldsymbol{\Sigma}_{X,\tilde{Y}}^{\star(h-1)}\boldsymbol{v}_h'\right| =& \left|(\boldsymbol{u}_h-\boldsymbol{u}_h')^\top\boldsymbol{\Sigma}_{X,\tilde{Y}}^{\star(h-1)}\boldsymbol{v}_h+ \boldsymbol{u}_h'^\top\boldsymbol{\Sigma}_{X,\tilde{Y}}^{\star(h-1)}(\boldsymbol{v}_h-\boldsymbol{v}_h')\right| \\ \leq& \| \boldsymbol{\Sigma}_{X,\tilde{Y}}^{\star(h-1)}\|_F (\|\boldsymbol{u}_h-\boldsymbol{u}_h' \|_2+\|\boldsymbol{v}_h-\boldsymbol{v}_h' \|_2)\\ \leq& \sqrt{2}\| \boldsymbol{\Sigma}_{X,\tilde{Y}}^{\star(h-1)}\|_F \sqrt{\|\boldsymbol{u}_h-\boldsymbol{u}_h' \|_2^2+\|\boldsymbol{v}_h-\boldsymbol{v}_h' \|_2^2},
\end{eqnarray*}
thus the set is equicontinuous.

Similarly, the random functions in the set $\{(\boldsymbol{u}_h,\boldsymbol{v}_h)\in \mathbb{S}^{p-1}\times\mathbb{S}^{q-1}\mapsto\boldsymbol{u}_h^\top\widehat{\boldsymbol{\Sigma}}_{X,\tilde{Y}}^{(h-1)}\boldsymbol{v}_h;~n\in\mathbb{N}\}$ satisfy
\begin{eqnarray*}
\left|\boldsymbol{u}_h^\top\widehat{\boldsymbol{\Sigma}}_{X,\tilde{Y}}^{(h-1)}\boldsymbol{v}_h-\boldsymbol{u}_h'^{\top}\widehat{\boldsymbol{\Sigma}}_{X,\tilde{Y}}^{(h-1)}\boldsymbol{v}_h'\right|  \leq& \sqrt{2}\| \widehat{\boldsymbol{\Sigma}}_{X,\tilde{Y}}^{(h-1)}\|_F \sqrt{\|\boldsymbol{u}_h-\boldsymbol{u}_h' \|_2^2+\|\boldsymbol{v}_h-\boldsymbol{v}_h' \|_2^2},
\end{eqnarray*}
with 
$$\| \widehat{\boldsymbol{\Sigma}}_{X,\tilde{Y}}^{(h-1)}\|_F  \leq \|\boldsymbol{\Sigma}_{X,\tilde{Y}}^{\star(h-1)}\|_F+\| \widehat{\boldsymbol{\Sigma}}_{X,\tilde{Y}}^{(h-1)}-\boldsymbol{\Sigma}_{X,\tilde{Y}}^{\star(h-1)}\|_F = \|\boldsymbol{\Sigma}_{X,\tilde{Y}}^{\star(h-1)}\|_F+ o_p(1),$$
so that $$\| \widehat{\boldsymbol{\Sigma}}_{X,\tilde{Y}}^{(h-1)}\|_F = O_p(1).$$
Applying \cite[Corollary~2.2]{newey1991uniform}, we  obtain~\eqref{eq1MainTheorem}.

By Lemma~\ref{lemma: star}, the population-level optimisation problem 
$$
\underset{(\boldsymbol{u}_h,\boldsymbol{v}_h)}{\argmax}~\textrm{Cov}\left({\boldsymbol{X}^{\star\top}_{h-1}}\boldsymbol{u}_h, \widetilde{\boldsymbol{Y}}_{h-1}^{\star\top}\boldsymbol{v}_h\right)
$$
has a unique solution and from~\cite[Theorem~5.7, Chapter~5]{van2000asymptotic}, we get~\eqref{eq2MainTheorem}. And because $\|\hat{\boldsymbol{u}}_h\|_2=1$, we get $\mathbb{E}(\|\hat{\boldsymbol{u}}_h\|_2^r)=1$ for any $r>0$, and thus we have $\mathbb{E}(\|\hat{\boldsymbol{u}}_h-\boldsymbol{u}^\star\|_2^r)\longrightarrow0$ as $n\rightarrow\infty$ (see, e.g., Exercise 3 (a) page 11 in the Ferguson book \cite{ferguson1996}).

Now, denote
\begin{equation*}
    \boldsymbol{\xi}_h^\star = \underline{\boldsymbol{X}}^\star_{h-1}\boldsymbol{u}_h^\star:n\times1
\end{equation*}
The population counterpart of the empirical latent variable
\begin{equation*}
    \hat{\boldsymbol{\xi}}_h = \hat{\underline{\boldsymbol{X}}}_{h-1}\hat{\boldsymbol{u}}_h:n\times1.
\end{equation*}

Note that, due to the deflation formula, we have:
$$
\underline{\hat{\boldsymbol{X}}}_{h-1}  = \underline{\hat{\boldsymbol{X}}}_{h-2} - \underline{\hat{\boldsymbol{X}}}_{h-2} \hat{\boldsymbol{u}}_{h-1}\hat{\boldsymbol{u}}_{h-1}^\top = (\boldsymbol{I}-\hat{\boldsymbol{u}}_{h-1}\hat{\boldsymbol{u}}_{h-1}^\top)\underline{\hat{\boldsymbol{X}}}_{h-2}  = \mathcal{P}_{\hat{\boldsymbol{u}}_{h-1}}\underline{\hat{\boldsymbol{X}}}_{h-2}
$$
and because the Frobenius norm of an orthogonal projection matrix is equal to 1, we get 
$$
\left\| \underline{\hat{\boldsymbol{X}}}_{h-1}\right\|_F = \left\| \mathcal{P}_{\hat{\boldsymbol{u}}_{h-1}}\underline{\hat{\boldsymbol{X}}}_{h-2}\right\|_F \leq \left\| \underline{\hat{\boldsymbol{X}}}_{h-2}\right\|_F \leq \cdots \leq \left\| \underline{\hat{\boldsymbol{X}}}_{0}\right\|_F.
$$
We have $\underline{\hat{\boldsymbol{X}}}_{0} = \underline{\boldsymbol{X}}_{0}^\bullet$, and  $$\underline{\boldsymbol{X}}_{0}^\bullet = \sum_{i = 1}^H \boldsymbol{\xi}_h^\bullet\boldsymbol{w}_i^{\bullet\top} +\underline{\boldsymbol{X}}_{H}^\bullet.$$
Note that the $\boldsymbol{\xi}_h^\bullet$'s are independent with finite variance, $\|\boldsymbol{w}_i^\bullet \| = 1$, and $\underline{\boldsymbol{X}}_{H}^\bullet$ has finite variance matrix too, so we have $$
\frac{1}{\sqrt{n}}\left\| \underline{\hat{\boldsymbol{X}}}_{0}\right\|_F  = \frac{1}{\sqrt{n}}\left\| \underline{\boldsymbol{X}}_{0}^\bullet\right\|_F = \sqrt{\textrm{tr}(\widehat{\boldsymbol{\Sigma}}_{X_0}^\bullet)} < \infty.$$
Now, since $\left\|\boldsymbol{u}_h^\star \right\|_2=1$ 
and
$$
\hat{\underline{\boldsymbol{X}}}_h-\underline{\boldsymbol{X}}_h^\star=-(\hat{\boldsymbol{u}}_h\hat{\boldsymbol{u}}_h^\top-\boldsymbol{u}_h^\star\boldsymbol{u}_h^{\star\top})\hat{\underline{\boldsymbol{X}}}_{h-1},
$$
we obtain
\begin{align*}
\frac{1}{\sqrt{n}} \|\hat{\boldsymbol{\xi}}_h - \boldsymbol{\xi}_h^\star\|_2 
&= \frac{1}{\sqrt{n}} \left\| \underline{\hat{\boldsymbol{X}}}_{h-1}(\hat{\boldsymbol{u}}_h - \boldsymbol{u}_h^\star + \boldsymbol{u}_h^\star) - \underline{\boldsymbol{X}}_{h-1}^\star \boldsymbol{u}_h^\star \right\|_2 \\
&= \frac{1}{\sqrt{n}} \left\| \underline{\hat{\boldsymbol{X}}}_{h-1}(\hat{\boldsymbol{u}}_h - \boldsymbol{u}_h^\star) + (\underline{\hat{\boldsymbol{X}}}_{h-1} - \underline{\boldsymbol{X}}_{h-1}^\star) \boldsymbol{u}_h^\star \right\|_2 \\
&\leq \frac{1}{\sqrt{n}} \left\| \underline{\hat{\boldsymbol{X}}}_{h-1}(\hat{\boldsymbol{u}}_h - \boldsymbol{u}_h^\star) \right\|_2 + \frac{1}{\sqrt{n}} \left\| (\underline{\hat{\boldsymbol{X}}}_{h-1} - \underline{\boldsymbol{X}}_{h-1}^\star) \boldsymbol{u}_h^\star \right\|_2 \\
&\leq \frac{1}{\sqrt{n}} \left\|\underline{\hat{\boldsymbol{X}}}_{h-1} \right\|_F \left\| \hat{\boldsymbol{u}}_h - \boldsymbol{u}_h^\star \right\|_2 + \frac{1}{\sqrt{n}}\left\| \underline{\hat{\boldsymbol{X}}}_{h-1} - \underline{\boldsymbol{X}}_{h-1}^\star \right\|_F \left\|\boldsymbol{u}_h^\star \right\|_2 \\
&\leq \frac{1}{\sqrt{n}} \left\|\underline{\hat{\boldsymbol{X}}}_{0} \right\|_F \left\| \hat{\boldsymbol{u}}_h - \boldsymbol{u}_h^\star \right\|_2 + \frac{1}{\sqrt{n}}\left\| \underline{\hat{\boldsymbol{X}}}_{h-1} - \underline{\boldsymbol{X}}_{h-1}^\star \right\|_F \\
&= \sqrt{\textrm{tr}(\widehat{\boldsymbol{\Sigma}}_{X_0}^\bullet)}\left\| \hat{\boldsymbol{u}}_h - \boldsymbol{u}_h^\star \right\|_2 + \frac{1}{\sqrt{n}}\left\| \underline{\hat{\boldsymbol{X}}}_{h-1} - \underline{\boldsymbol{X}}_{h-1}^\star \right\|_F \\
&\leq \sqrt{\textrm{tr}(\widehat{\boldsymbol{\Sigma}}_{X_0}^\bullet)}\left\| \hat{\boldsymbol{u}}_h - \boldsymbol{u}_h^\star \right\|_2 + \frac{1}{\sqrt{n}}\left\|\hat{\boldsymbol{u}}_{h-1}\hat{\boldsymbol{u}}_{h-1}^\top-\boldsymbol{u}_{h-1}^\star\boldsymbol{u}_{h-1}^{\star\top} \right\|_2\left\|\underline{\hat{\boldsymbol{X}}}_{0} \right\|_F \\
&= \sqrt{\textrm{tr}(\widehat{\boldsymbol{\Sigma}}_{X_0}^\bullet)}\left\| \hat{\boldsymbol{u}}_h - \boldsymbol{u}_h^\star \right\|_2 + \sqrt{\textrm{tr}(\widehat{\boldsymbol{\Sigma}}_{X_0}^\bullet)}\left\|\hat{\boldsymbol{u}}_{h-1}\hat{\boldsymbol{u}}_{h-1}^\top-\boldsymbol{u}_{h-1}^\star\boldsymbol{u}_{h-1}^{\star\top} \right\|_2 \\
&\leq \sqrt{\textrm{tr}(\widehat{\boldsymbol{\Sigma}}_{X_0}^\bullet)}\left\| \hat{\boldsymbol{u}}_h - \boldsymbol{u}_h^\star \right\|_2 + \sqrt{\textrm{tr}(\widehat{\boldsymbol{\Sigma}}_{X_0}^\bullet)}\left\| \hat{\boldsymbol{u}}_h - \boldsymbol{u}_h^\star \right\|_2(\left\|\hat{\boldsymbol{u}}_{h-1}\right\|_2+\left\|\boldsymbol{u}_{h-1}^\star\right\|_2) \\
&= \sqrt{\textrm{tr}(\widehat{\boldsymbol{\Sigma}}_{X_0}^\bullet)}\left\| \hat{\boldsymbol{u}}_h - \boldsymbol{u}_h^\star \right\|_2\left(1 + \left\|\hat{\boldsymbol{u}}_{h-1}\right\|_2+\left\|\boldsymbol{u}_{h-1}^\star\right\|_2 \right)\\
&=3\sqrt{\textrm{tr}(\widehat{\boldsymbol{\Sigma}}_{X_0}^\bullet)}\left\| \hat{\boldsymbol{u}}_h - \boldsymbol{u}_h^\star \right\|_2\qquad(\text{which converges in probability to 0}).
\end{align*}

Then, given that
\begin{align*}
\frac{1}{n} \|\hat{\boldsymbol{\xi}}_h - \boldsymbol{\xi}_h^\star\|_2^2 &\leq 9\textrm{tr}(\widehat{\boldsymbol{\Sigma}}_{X_0}^\bullet))\left\| \hat{\boldsymbol{u}}_h - \boldsymbol{u}_h^\star \right\|_2^2\leq36\textrm{tr}(\widehat{\boldsymbol{\Sigma}}_X^\bullet)\leq36(1+\textrm{tr}(\boldsymbol{\Sigma}_X^\bullet))\qquad(\text{for $n$ large enough})
\end{align*}
and because we assumed $\mathbb{E}(X_{0,j})<\infty$, we can conclude from \cite[Exercise~3(a)]{ferguson1996} that $\mathbb{E}(\frac{1}{n} \|\hat{\boldsymbol{\xi}}_h - \boldsymbol{\xi}_h^\star\|_2^2)\longrightarrow0$, as $n\rightarrow\infty$.\\

Now, consider the expected squared difference:
\begin{equation*}
    \mathbb{E}\left\| \hat{\boldsymbol{\xi}}_h - \boldsymbol{\xi}_h^\bullet \right\|_2^2
= \mathbb{E}\left[ \left\| \hat{\boldsymbol{\xi}}_h - \boldsymbol{\xi}_h^\star + \boldsymbol{\xi}_h^\star - \boldsymbol{\xi}_h^\bullet \right\|_2^2 \right]
= \mathbb{E}\left[ \left\| \hat{\boldsymbol{\xi}}_h - \boldsymbol{\xi}_h^\star \right\|_2^2
+ 2\left\langle \hat{\boldsymbol{\xi}}_h - \boldsymbol{\xi}_h^\star,\; \boldsymbol{\xi}_h^\star - \boldsymbol{\xi}_h^\bullet \right\rangle
+ \left\| \boldsymbol{\xi}_h^\star - \boldsymbol{\xi}_h^\bullet \right\|_2^2 \right].
\end{equation*}
Additionally, from Lemma~\ref{lemma: star}, we have $\boldsymbol{\xi}_h^\star - \boldsymbol{\xi }_h^\bullet =  \underline{\boldsymbol{X}}_H^\bullet\boldsymbol{w}_h$ (matrix form of the result), so that $n^{-1}\mathbb{E}[\| \boldsymbol{\xi}_h^\star - \boldsymbol{\xi }_h^\bullet\|_2^2]=\boldsymbol{w}_h^\bullet\boldsymbol{\Sigma}_X^\bullet \boldsymbol{w}_h^{\bullet\top}$. Using the Cauchy-Schwarz inequality, we get
$$
n^{-1}\mathbb{E} \left[ \left\langle \hat{\boldsymbol{\xi}}_h - \boldsymbol{\xi}_h^\star,\; \boldsymbol{\xi}_h^\star - \boldsymbol{\xi}_h^\bullet \right\rangle \right]
\le n^{-1}\sqrt{ \mathbb{E} \left\| \hat{\boldsymbol{\xi}}_h - \boldsymbol{\xi}_h^\star \right\|_2^2 }
\cdot \sqrt{ \mathbb{E} \left\| \boldsymbol{\xi}_h^\star - \boldsymbol{\xi}_h^\bullet \right\|_2^2 }
 =  \sqrt{n^{-1} \mathbb{E} \left\| \hat{\boldsymbol{\xi}}_h - \boldsymbol{\xi}_h^\star \right\|_2^2 }\times\sqrt{\boldsymbol{w}_h^{\bullet\top}\boldsymbol{\Sigma}_X^\bullet \boldsymbol{w}_h^\bullet}\longrightarrow0.
$$
So,
\begin{equation*}
   n^{-1}\mathbb{E} \left[ \left\| \hat{\boldsymbol{\xi}}_h - \boldsymbol{\xi}_h^\bullet \right\|_2^2 \right]
\longrightarrow{\boldsymbol{w}_h^\bullet}^\top \boldsymbol{\Sigma}_X^\bullet \boldsymbol{w}_h^\bullet.
\end{equation*}

\end{proof}

\begin{lemma}
    \label{lemma: plsr result}
    Under the same condition with Theorem~\ref{theorem: main}, let$\{(\boldsymbol{X}^{\bullet(i)}_0,\boldsymbol{Y}^{\bullet(i)}_0)\}_{i=1}^n$ be a random sample following the Generative PLS-R model, consider the same optimisation problem with $\hat{\boldsymbol{Y}}_{h-1}$ replace $\hat{\widetilde{\boldsymbol{Y}}}_{h-1}$ (Generative PLS-R instead of PLS-SVD), we still have: (1) $\mathbb{E}\left[\|\hat{\boldsymbol{u}}_h - \boldsymbol{w}_h^\bullet\|_2^r\right]\longrightarrow0$ and $\mathbb{E}\left[\|\hat{\boldsymbol{v}}_h - \boldsymbol{v}_h^\bullet\|_2^r\right]\longrightarrow0$ for all $r >0$. (2) $ \frac{1}{n}\mathbb{E}\left\|\hat{\boldsymbol{\xi}}_h - \boldsymbol{\xi}_h^\bullet\right\|_2^2 \longrightarrow {\boldsymbol{w}_h^\bullet}^\top \boldsymbol{\Sigma}_{X}^\bullet \boldsymbol{w}_h^\bullet$
\end{lemma}
\begin{proof}
    It's clear from the proof in Theorem~\ref{theorem: main} and Lemma~\ref{lemma: star}
\end{proof}
The Lemma~\ref{lemma: star} is the result for population-level estimated parameters, then Lemma~\ref{lemma: bullet} and Lemma~\ref{lemma: cov_solution} are the middle results helping with the proof. The proofs of these lemmas are in Appendix~\ref{app: theorem}
\begin{lemma}
\label{lemma: star}
With the notation introduced in Theorem 2. Additionally, let $\boldsymbol{X}_{0}^\star = \boldsymbol{X}_{0}^\bullet$ and $\boldsymbol{X}_{h}^\star = \boldsymbol{X}_{h-1}^\star-\boldsymbol{u}_h^\star \xi_h^\star$, where $\xi_h^\star = {\boldsymbol{u}_h^\star}^\top \boldsymbol{X}_{h-1}^\star$ and 
$\boldsymbol{u}_h^\star$ and $\boldsymbol{v}_h^\star$ be the solutions that maximize the theoretical optimization problem:
$$(\boldsymbol{u}_h^\star,\boldsymbol{v}_h^\star) \in \underset{(\boldsymbol{u}_h,\boldsymbol{v}_h) \in C}{\argmax}~\textrm{Cov}\left({\boldsymbol{X}^{\star\top}_{h-1}}\boldsymbol{u}_h, \widetilde{\boldsymbol{Y}}_{h-1}^{\star\top}\boldsymbol{v}_h\right),
$$
where C contains the same constraint as in Theorem 2 (except that C3 becomes $\text{C3}^\star$ $\textrm{Cov}({\boldsymbol{X}}_{h-1}^{\star\top}\boldsymbol{u}_h,{\xi}_{\ell}^\star)\text{ is minimal},~\ell=1,\dots,h-1$).

We have  (1) $\boldsymbol{u}_h^\star= \boldsymbol{w}_h^\bullet$ and $\boldsymbol{v}_h^\star= \boldsymbol{v}_h^\bullet$. (2) $\xi_h^\star - \xi_h^\bullet ={\boldsymbol{u}_h^\bullet}^\top \boldsymbol{X}^\bullet_{H}$ and $\boldsymbol{X}^\bullet_{h} - \boldsymbol{X}^\star_{h} = \sum_{i=1}^{h}\boldsymbol{u}_i^\bullet {\boldsymbol{u}_i^\bullet}^\top \boldsymbol{X}^\bullet_{H}$. 
(3) Symmetric results can be obtained for $\omega$ under the assumption of symmetric model.

Moreover, for Generative PLS-R (same notation in Lemma~\ref{lemma: plsr result}), we also have (1) $\boldsymbol{u}_h^\star= \boldsymbol{w}_h^\bullet$ and $\boldsymbol{v}_h^\star= \boldsymbol{v}_h^\bullet$. (2) $\xi_h^\star - \xi_h^\bullet ={\boldsymbol{u}_h^\bullet}^\top \boldsymbol{X}^\bullet_{H}$ and $\boldsymbol{X}^\bullet_{h} - \boldsymbol{X}^\star_{h} = \sum_{i=1}^{h}\boldsymbol{u}_i^\bullet {\boldsymbol{u}_i^\bullet}^\top \boldsymbol{X}^\bullet_{H}$.
\end{lemma}

\begin{lemma}
\label{lemma: bullet}
Under the same setting in Theorem 2, for each $h=1,\ldots, H$, consider respectively $(\boldsymbol{w}_h^\star,\boldsymbol{\theta}_h^\star)$ and $(\widetilde{\boldsymbol{w}}_h^\star,\boldsymbol{v}_h^\star)$ any solution of the two following optimisation problems:
\begin{eqnarray*}
(\boldsymbol{u}_h^{opti},\boldsymbol{v}_h^{opti}) \in \underset{(\boldsymbol{w}_h,\boldsymbol{v}_h) \in C1}{\arg\max}~\textrm{Cov}\left({\boldsymbol{X}^{\bullet\top}_{h-1}}\boldsymbol{u}_h, {\boldsymbol{Y}}_{h-1}^{\bullet\top}\boldsymbol{v}_h\right)\quad\text{ and 
 }\quad(\widetilde{\boldsymbol{u}}_h^{opti},\widetilde{\boldsymbol{v}}_h^{opti}) \in \underset{(\boldsymbol{w}_h,\boldsymbol{v}_h)\in C1}{\arg\max}~\textrm{Cov}\left({\boldsymbol{X}^{\bullet\top}_{h-1}}\boldsymbol{u}_h, \widetilde{\boldsymbol{Y}}_{h-1}^{\bullet\top}\boldsymbol{v}_h\right),
\end{eqnarray*}
where C1 contains the same constraint as C1 in Theorem 2.
We have $(\boldsymbol{u}_h^{opti},\boldsymbol{v}_h^{opti})=(\widetilde{\boldsymbol{u}}_h^{opti},\widetilde{\boldsymbol{v}}_h^{opti})=(\boldsymbol{w}_h^\bullet, \boldsymbol{v}_h^\bullet)$. (A unique common solution to both constrained optimisation problems exists. It is equal to the pair of vectors that were used to generate the data.)
\end{lemma}

\begin{lemma}
\label{lemma: cov_solution}
    Let $\boldsymbol{X}_0^\bullet \in \mathbb{R}^p$ be data generated by the Generative PLS-R model or the Generative symmetric PLS-SVD model with parameters $\boldsymbol{w}_1^\bullet,\ldots, \boldsymbol{w}_H^\bullet$ and latent variables $\xi_1^\bullet,\ldots, \xi_H^\bullet$. Let $\mathcal{W} := \text{span}\{\boldsymbol{w}_1^\bullet,\ldots, \boldsymbol{w}_H^\bullet\}$. 

    Then for any optimisation problem of the form:
    $$
    \underset{\boldsymbol{w}}{\arg\max}\text{ Cov}(\boldsymbol{w}^\top (\boldsymbol{X}^\bullet_{h-1}-\boldsymbol{X}_H^\bullet),\cdot),\quad h=1,\ldots,H,
    $$
    (1) If a solution exists, then there exists a solution $\boldsymbol{w}^{opti} \in \mathcal{W}$.
    
    (2) If, additionally, we impose the constraints $\boldsymbol{w} \perp \{\boldsymbol{w}_j^{opti} \}_{j<h}$,  then if there are solutions, there exists an optimal solution $\boldsymbol{w}^{opti} \in \text{span}\{\boldsymbol{w_j^\bullet} \}_{j=h}^H \subset \mathcal{W}$.
\end{lemma}

Moreover, similar to \cite{el_Bouhaddani_2018}, we can prove that the proposed model is identifiable.
From equation~\eqref{eq: matrix_pls_X} and \eqref{eq: matrix_pls_Y} and use $\boldsymbol{\Sigma}_{\xi}$ to represent $\text{Var}(\boldsymbol{\xi})$, we have $$Var(\boldsymbol{X}_0)=\boldsymbol{W}\boldsymbol{\Sigma}_{\xi}\boldsymbol{W}^\top+\boldsymbol{\Sigma}_X,$$ 
and
$$Var(\boldsymbol{Y}_0)=\boldsymbol{V}(\boldsymbol{B}^2\boldsymbol{\Sigma}_{\xi}+\sigma_1^2\boldsymbol{I}_H)\boldsymbol{V}^\top+\boldsymbol{\Sigma}_{\tilde{Y}} =\boldsymbol{V}\boldsymbol{B}^2\boldsymbol{\Sigma}_{\xi}\boldsymbol{V}^\top + \boldsymbol{\Sigma}_Y.$$

\begin{theorem}
If $\mathbb{P}$ is used to denote all the parameters in the model, in the Generative (symmetric) PLS-SVD model $\mathbb{P} = \{\boldsymbol{w}_h, \boldsymbol{v}_h, b_h, \boldsymbol{\Sigma}_X, \boldsymbol{\Sigma}_{\tilde{Y}}, \boldsymbol{\Sigma}_{\xi}, \sigma_1^2\}_{h = 1,\ldots, H}$ or $\boldsymbol{\Sigma}_Y$ replace $\boldsymbol{\Sigma}_{\tilde{Y}}$ if it's Generative PLS-R model. 

With $H\leq min(p,q)$, $s_{1,h}^2 b_h$ is strictly decreasing with h, the proposed Generative-PLS model (both PLS-SVD and PLS-R) is identifiable, which means if $(\boldsymbol{X}^1,\boldsymbol{Y}^1)$ and $(\boldsymbol{X}^2,\boldsymbol{Y}^2)$ are generated by the proposed model with parameters set $\mathbb{P}_1$ and $\mathbb{P}_2$. If $Cov((\boldsymbol{X}^1,\boldsymbol{Y}^1)) = cov((\boldsymbol{X}^2,\boldsymbol{Y}^2))$, the parameter sets $\mathbb{P}_1 = \mathbb{P}_2$
\end{theorem}

\begin{proof}
With a pair of data points $(\boldsymbol{X},\boldsymbol{Y})$ simulated from the Generative PLS-SVD model, the variance-covariance matrix of the simulated data point is
$$\text{Cov}((\boldsymbol{X},\boldsymbol{Y})) = \left( \begin{array}{cc} \boldsymbol{W}\boldsymbol{\Sigma}_\xi \boldsymbol{W}^\top +\boldsymbol{\Sigma}_X & \boldsymbol{W}\boldsymbol{\Sigma}_\xi \boldsymbol{B}\boldsymbol{V}^\top \\ \boldsymbol{V}\boldsymbol{B}\boldsymbol{\Sigma}_\xi \boldsymbol{W}^T & \boldsymbol{V} (\boldsymbol{B}^\top \boldsymbol{\Sigma}_\xi \boldsymbol{B}+\sigma_1^2\boldsymbol{I}_H) \boldsymbol{V}^\top +\boldsymbol{\Sigma}_{\tilde{Y}} \end{array} \right).$$
If $\text{Cov}((\boldsymbol{X}^1,\boldsymbol{Y}^1)) = \text{Cov}((\boldsymbol{X}^2,\boldsymbol{Y}^2))$, firstly $\boldsymbol{W}^1\boldsymbol{\Sigma}^1_\xi \boldsymbol{B}^1\boldsymbol{V}^1 = \boldsymbol{W}^2\boldsymbol{\Sigma}^2_\xi \boldsymbol{B}^2\boldsymbol{V}^2$.

With the orthogonal matrix of $\boldsymbol{W}$, $\boldsymbol{V}^T$ and diagonal matrix $\boldsymbol{\Sigma}_\xi \boldsymbol{B}$ with strictly decreasing elements, this naturally gives an SVD decomposition. Using Lemma 1 from \cite{el_Bouhaddani_2018}, $\boldsymbol{W}^1 = \boldsymbol{W}^2 \boldsymbol{\Delta}$, $\boldsymbol{V}^1 = \boldsymbol{V}^2 \boldsymbol{\Delta}$ and $\boldsymbol{\Sigma}^1_\xi \boldsymbol{B}^1 = \boldsymbol{\Sigma}^2_\xi \boldsymbol{B}^2$. Moreover, under the model setting, the largest value in each $\boldsymbol{w}_h$ and $\boldsymbol{v}_h$ is positive, so we have:
$$
\boldsymbol{W}^1 = \boldsymbol{W}^2; \quad \boldsymbol{V}^1 = \boldsymbol{V}^2.
$$
Additionally, with the constraint $H \leq \min(p,q)$, for the orthonormal matrix $\boldsymbol{W}^T$, there exists a non-zero vector $\boldsymbol{c}$, s.t. $\boldsymbol{W}^T\boldsymbol{c} = 0$. Then we have, 
$$
\boldsymbol{W}\boldsymbol{\Sigma}^1_\xi \boldsymbol{W}^\top +\boldsymbol{\Sigma}^1_X = \boldsymbol{W}\boldsymbol{\Sigma}^2_\xi \boldsymbol{W}^\top +\boldsymbol{\Sigma}^2_X
$$
$$
\boldsymbol{W}\boldsymbol{\Sigma}^1_\xi \boldsymbol{W}^\top \boldsymbol{c} +\boldsymbol{\Sigma}^1_X \boldsymbol{c} = \boldsymbol{W}\boldsymbol{\Sigma}^2_\xi \boldsymbol{W}^\top \boldsymbol{c} +\boldsymbol{\Sigma}^2_X \boldsymbol{c}
$$
So, 
$$\boldsymbol{\Sigma}^1_X \boldsymbol{c} =\boldsymbol{\Sigma}^2_X \boldsymbol{c},$$
since $\boldsymbol{c}$ is a non-zero vector, we can get 
$$\boldsymbol{\Sigma}^1_X =\boldsymbol{\Sigma}^2_X.$$
Moreover, we have
$\boldsymbol{W}\boldsymbol{\Sigma}^1_\xi \boldsymbol{W}^\top +\boldsymbol{\Sigma}^1_X = \boldsymbol{W}\boldsymbol{\Sigma}^2_\xi \boldsymbol{W}^\top +\boldsymbol{\Sigma}^2_X$ gets reduced to $\boldsymbol{W}\boldsymbol{\Sigma}^1_\xi \boldsymbol{W}^\top = \boldsymbol{W}\boldsymbol{\Sigma}^2_\xi \boldsymbol{W}^\top$, $\boldsymbol{W}$ is an orthonormal matrix, it's easy to see 
$$\boldsymbol{\Sigma}^1_\xi= \boldsymbol{\Sigma}^2_\xi.$$ 
Then, because $\boldsymbol{\Sigma}_\xi \boldsymbol{B}^1 = \boldsymbol{\Sigma}_\xi \boldsymbol{B}^2$, we must have $$\boldsymbol{B}^1 = \boldsymbol{B}^2.$$
Similarly, we can get $$\boldsymbol{\Sigma}^1_{\tilde{Y}} =\boldsymbol{\Sigma}^2_{\tilde{Y}} \text{ ,and } (\sigma_1^2\boldsymbol{I}_H)^1 = (\sigma_1^2\boldsymbol{I}_H)^2.$$
Moreover, if the Generative PLS-R model is using, we have  
$$\boldsymbol{\Sigma}^1_Y = \boldsymbol{\Sigma}_{\tilde{Y}}+\sigma_1^2\boldsymbol{V}\boldsymbol{V}^\top = \boldsymbol{\Sigma}^2_Y.$$
\end{proof}

\subsection{Further Inference}
\label{sec: inference}
\subsubsection{Model Inference}
\label{subsubsec: model_inference}

In the Generative (Symmmetric) PLS-SVD model,  $\mathbb{P} = \{\boldsymbol{u}_h, \boldsymbol{v}_h, b_h, \boldsymbol{\Sigma}_X, \boldsymbol{\Sigma}_{\tilde{Y}}, s_{1,h}, \sigma_1\}_{h = 1,\ldots, H}$ represent the unknown parameters. If the Generative PLS-R model is used, the unknown parameters are reduced to $\mathbb{P} = \{\boldsymbol{u}_h, \boldsymbol{v}_h, \boldsymbol{\Sigma}_X, \boldsymbol{\Sigma}_{Y}, s_{1,h}, \}_{h = 1,\ldots, H}$. Since we have proved that in the two models, $\boldsymbol{u}_h, \boldsymbol{v}_h, \boldsymbol{\Sigma}_X, \text{ and } s_{1,h}$ have the same performance and $\boldsymbol{\Sigma}^1_Y = \boldsymbol{\Sigma}_{\tilde{Y}}+\sigma_1^2\boldsymbol{V}\boldsymbol{V}^\top = \boldsymbol{\Sigma}^2_Y$, we will discuss the inference based on Generative (Symmmetric) PLS-SVD model in this section.

With a given data point $(\boldsymbol{X}^\bullet_0,\boldsymbol{Y}^\bullet_0)$, the optimisations \eqref{eq: pls_optim} to \eqref{eq:pls_opti3} are applied to solve the unknown parameters. The estimators of latent variables $({\boldsymbol{\xi}_h}^\star, {\boldsymbol{\omega}_h}^\star)_{h = 1, \ldots, H}$ together with part of the estimated parameters $\{{\boldsymbol{u}_h}^\star, {\boldsymbol{v}_h}^\star, {b_h}^\star\}_{h = 1,\ldots, H}$ will be available. The rest of the parameters $\{\boldsymbol{\Sigma}_X, \boldsymbol{\Sigma}_{\tilde{Y}}, \sigma_{\xi_h}, \sigma_1\}$ are not available from the optimization steps directly, however, the $\{{\boldsymbol{X}_h}^\star, {\boldsymbol{Y}_h}^\star\}_{h = 1, \ldots, H}$, the fitted latent variables $({\boldsymbol{\xi}_h}^\star, {\boldsymbol{\omega}_h}^\star)_{h = 1, \ldots, H}$, together with the error term $\{{\boldsymbol{\epsilon}_{1,h}}^\star\}_{h = 1, \ldots, H}$  are available, which are closely related to the rest of these parameters.

Although from the data generation procedure, these parameters are the variance of the $\{{\boldsymbol{X}}_H, {\boldsymbol{Y}}_H, {\boldsymbol{\xi}}_h, {\boldsymbol{\epsilon}}\}_{h = 1, \ldots, H}$, the estimation is not that simple with the fitted variables. From Lemma~\ref{lemma: star}, it's clear that although ${\boldsymbol{u}}_h^\star$ converges to the true value ${\boldsymbol{u}}_h^\bullet$, ${\boldsymbol{X}}^\star_h$ does not converge to the true value ${\boldsymbol{X}}_h^\bullet$ and ${\boldsymbol{\xi}}^\star_h$ does not converge to the true value ${\boldsymbol{\xi}}_h^\bullet$. The estimator needs to be constructed carefully.

From Lemma~\ref{lemma: star}, we have:
$$
{\boldsymbol{X}}^\star_{H}  =  \boldsymbol{X}_{H}^\bullet-\sum_{i=1}^{H}\boldsymbol{w}^\bullet_i {\boldsymbol{w}^\bullet_i}^\top \boldsymbol{X}_{H}^\bullet = \boldsymbol{X}_{H}^\bullet-\sum_{i=1}^{H}\boldsymbol{u}^\star_i {\boldsymbol{u}^\star_i}^\top \boldsymbol{X}_{H}^\bullet.
$$

In this case, $(\boldsymbol{I-\sum_{i=1}^{H}\boldsymbol{u}^\star_i {\boldsymbol{u}^\star_i}^\top})^{-1}{\boldsymbol{X}}^\star_{H}$ will be a better estimator of the $\boldsymbol{X}_H^\bullet$ if $\boldsymbol{I-\sum_{i=1}^{H}\boldsymbol{u}^\star_i {\boldsymbol{u}^\star_i}^\top}$ is invertible.

However {\bf in case (b)}, under the assumption $\boldsymbol{\Sigma}_X = \sigma_x^2\boldsymbol{I}_p$, we have:
$$
{\sigma_x^\star}^2I=(\boldsymbol{I-\sum_{i=1}^{H}{\boldsymbol{u}}^\star_i {{\boldsymbol{u}_i^\star}}^\top})(\boldsymbol{I-\sum_{i=1}^{H}{\boldsymbol{u}}^\star_i {{\boldsymbol{u}_i}^\star}^\top})^{\top}{\sigma_x^\bullet}^2.
$$ 
So, we can obtain the corrected estimator:

$${{\sigma_x^\star}^2}^{corrected} = \frac{1}{p}\sum_{i = 1}^p \text{diag}(\text{Var}({\boldsymbol{X}_H}^\star))/\frac{1}{p}\sum_{i = 1}^p \text{diag}(\text{Var}((\boldsymbol{I-\sum_{i=1}^{H}{\boldsymbol{u}}^\star_i {{\boldsymbol{u}_i^\star}}^\top})(\boldsymbol{I-\sum_{i=1}^{H}{\boldsymbol{u}}^\star_i {{\boldsymbol{u}_i}^\star}^\top})^{\top})).$$

Similarly,

$${{\sigma_{\tilde{y}}^\star}^2}^{corrected} = \frac{1}{q}\sum_{i = 1}^q \text{diag}(\text{Var}({\tilde{\boldsymbol{Y}}_H}))/\frac{1}{q}\sum_{i = 1}^q \text{diag}(\text{Var}((\boldsymbol{I-\sum_{i=1}^{H}{\boldsymbol{v}}^\star_i {{\boldsymbol{v}_i^\star}}^\top})(\boldsymbol{I-\sum_{i=1}^{H}{\boldsymbol{v}}^\star_i {{\boldsymbol{v}^\star_i}}^\top})^{\top})).$$

then because $\boldsymbol{W}^{\bullet\top} Var(\boldsymbol{X}_0^\bullet)\boldsymbol{W}^\bullet=\boldsymbol{\Sigma}_{\xi}^\bullet+\boldsymbol{W}^{\bullet\top}\sigma_x^{2\bullet}\boldsymbol{W}^\bullet$, we have the estimated $\boldsymbol{\Sigma}_{\xi}^\star$ with diagnol elements:
$$
(s_{1,h}^2)^\star = (\boldsymbol{u}_h^\star)^\top Var(\boldsymbol{X}_0^\bullet)\boldsymbol{u}_h^\star-{(\sigma_x^2)^\star}^{corrected}.
$$
Similarly, we have the estimated $\boldsymbol{\Sigma}_{\omega}^\star$ with diagnol elements:
$$
(\sigma_{\omega,h}^2)^\star = (\boldsymbol{v}_h^\star)^\top Var(\boldsymbol{Y}_0^\bullet)\boldsymbol{v}_h^\star-{(\sigma_{\tilde{y}}^2)^\star}^{corrected}.
$$
Because $\text{Var}(\omega_h^\bullet) = b_h^{\bullet2}\text{Var}(\xi_h^\bullet)+\sigma_1^{\bullet2}$, we have:
$$
{\sigma_1^2}^\star = \text{mean}(\text{diag}((\boldsymbol{V}^\star)^\top Var(\boldsymbol{Y}_0^\bullet)\boldsymbol{V}^\star))-{(\sigma_y^2)^\star}^{corrected}-{b_h^\star}^2{\sigma_{\xi,h}^2}^\star.
$$
However, since $b_h^\star$ doesn't converge to $b_h^\bullet$, a corrected version is also required. From above, it's clear that obtaining the corrected version of $\boldsymbol{X}_H$ is hard, so it's not easy to construct an unbiased ${b}_h^{corrected}$. However, instead of constructing an unbiased estimator ${b}_h^\star$, as in Thereom 1.2.1 in Chapter 1 in \cite{fuller2009measurement}, we can get a consistency estimator when the sample size $n \rightarrow \infty$, which is:
$$\hat{b}_h^{corrected} = \frac{\hat{\boldsymbol{\xi}}_h^\top \hat{\boldsymbol{\omega}}_h}{\hat{\boldsymbol{\xi}}_h^\top \hat{\boldsymbol{\xi}}_h - {{\sigma_x^\star}^2}^{corrected} }.$$
In practice, with a sample size $n$, we use the empirical version for all the above equations.

In {\bf case (c)}, there are more parameters to estimate, but we can still correct the estimators with the Moore–Penrose Pseudo-inverse approximation of the non-invertible matrix. 

We can get:
$$\boldsymbol{\Sigma}_X^{\star corrected} = (\boldsymbol{I-\sum_{i=1}^{H}{\boldsymbol{u}}^\star_i {{\boldsymbol{u}_i^\star}}^\top})^+\text{Var}(\boldsymbol{X}_H^\star)$$
and 
$$\boldsymbol{\Sigma}_{\tilde{Y}}^{\star corrected} = (\boldsymbol{I-\sum_{i=1}^{H}{\boldsymbol{v}}^\star_i {{\boldsymbol{v}_i^\star}}^\top})^+\text{Var}(\tilde{\boldsymbol{Y}}_H)$$
Then,
$$(\Sigma_{\xi}^2)^\star = {\boldsymbol{U}^\star}^\top Var(\boldsymbol{X}_0)\boldsymbol{U}^\star-{\boldsymbol{U}^\star}^\top \boldsymbol{\Sigma}_X^{corrected}\boldsymbol{U}^\star$$
Similarly,
$${b}_h^{\star corrected} = \frac{\hat\xi_h^\top \hat\omega_h}{\hat\xi_h^\top \hat\xi - {\boldsymbol{u}_h^\star}^\top{\boldsymbol{\Sigma}_X^{corrected}}\boldsymbol{u}_h^\star}$$
and then
$$
{\sigma_1^2}^\star = \text{mean}(\text{diag}((\boldsymbol{V}^\star)^\top Var(\boldsymbol{Y}_0)\boldsymbol{V}^\star))-(\boldsymbol{V}^\star)^\top{(\sigma_y^2)^\star}^{corrected}\boldsymbol{V}^\star-{{b}_h^{corrected}}^2{\sigma_{\xi,h}^2}^\star 
$$

\subsubsection{Further Inference}
\label{subsubsec: bootstrap}
The presented Generative-PLS structure proposed in Section~\ref{sec: pls} has already enabled the estimation of error terms and the variance terms in the model. However, with a relatively small sample size, the result might be unstable and unreliable. Additionally, these results will only offer an inference structure for the latent variable and the original data. When a new data point is considered, the structure can still only offer a point estimation. 

Many recent works have discussed a probabilistic version of PLS that can obtain the confidence interval of the variables. However, within all these models and the structure we proposed, only the distribution of the variables is considered, not the parameters in the model. Moreover, we have shown that these PPLS models are involved in our proposed structure. The same drawback is that the prediction interval is not available.

The bootstrap method has been used a lot with PLS for model inference \cite{streukens2016bootstrapping,aguirre2018statistical, bras2008bootstrap, Odgers2023}. However, among the two types of bootstrap, most bootstrap studies on PLS utilise pair bootstrap, due to the lack of a model structure for PLS. The residual bootstrap on PLS is studied in \cite{denham1997prediction} and \cite{faber2002uncertainty}, however, they both only focus on the single response regression situation, instead of a symmetric model. The residual bootstrap on the algorithm, instead of a proper statistical model, lacks reliability due to the residuals being heteroscedasticity and not independent and identically distributed from each other. Thus, this method is less used in current PLS studies instead of the pairs bootstrap.

\begin{figure}[ht]
        \centering
        \includegraphics[scale = 0.43]{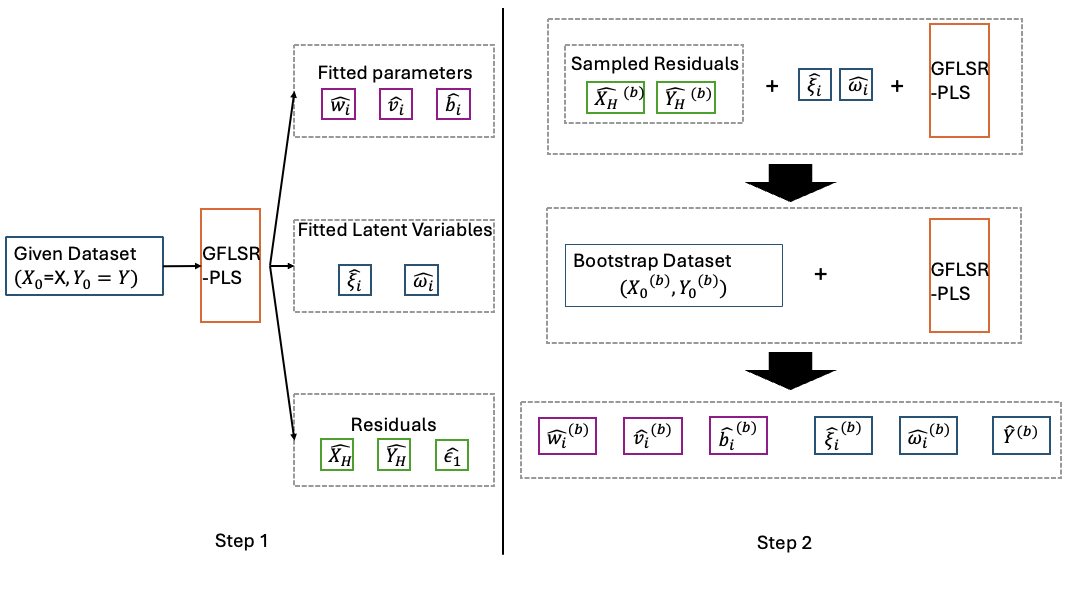}
        \caption{Bootstrap Procedure}
        \label{fig:boot}
\end{figure}

With the novel model structure, a novel bootstrap procedure is proposed. With the estimated residual terms, a new bootstrapped dataset can be built just with the estimated latent variables and the estimated parameters. The details of the data reconstruction are introduced in Fig~\ref{fig:boot}.

As shown in Fig \ref{fig:boot}, In step 1, with a given dataset $(\boldsymbol{X},\boldsymbol{Y}) \in \mathbb{R}^{n\times p} \times \mathbb{R}^{n\times q}$. If we assume that the given dataset follows the Generative-PLS structure, after applying the model and the optimization progress with $h = (1,\ldots, H)$ with a given or selected H, the fitted latent variables $(\hat{\boldsymbol{\xi}_h},\hat{\boldsymbol{\omega}_h})$, fitted parameters $(\hat{\boldsymbol{w}}_h, \hat{\boldsymbol{v}}_h,\hat{b}_h)$ together with fitted residuals $(\hat{\boldsymbol{X}}_H,\hat{\boldsymbol{Y}}_H,\hat{\boldsymbol{\epsilon}}_{1,h})$can be obtained. 

The Generative-PLS model structure proposed in section \ref{sec: pls} has three residuals terms $({\boldsymbol{X}}_H,{\boldsymbol{Y}}_H,{\boldsymbol{\epsilon}}_{1,h})$. However, the residual terms $\boldsymbol{X}_H$ and $\boldsymbol{Y}_H$ have already contained all the total errors from the dataset that the model can't explain. The residual term ${\boldsymbol{\epsilon}}_{1,h}$ can be interpret as the residual after fitting $\hat{\omega}$ on $\hat{\xi}$ without intercept term. With $\hat{\omega}$, $\hat{\xi}$ together with the error term $\hat{\boldsymbol{X}}_H,\hat{\boldsymbol{Y}}_H$, all the information from the original dataset can be recovered. Moreover, from Theorem~\ref{theorem: main}, it's clear that the distance between $\boldsymbol{\xi}^\bullet$ and $\hat{\boldsymbol{\xi}}$ contains $\boldsymbol{X}_H^\bullet$. Similarly, the distance between $\boldsymbol{\omega}^\bullet_h$ and $\hat{\boldsymbol{\omega}}_h$ contains the term $\boldsymbol{Y}_H^\bullet$; then the fitted error term $\epsilon_1$ is no longer independent of the other two error terms. Due to the above two reasons, the error term ${\boldsymbol{\epsilon}}_{1,h}$ is not used in the bootstrap procedure.

Thus, in step 2 of the proposed residual bootstrap, instead of three fitted versions of residuals $(\hat{\boldsymbol{X}}_H,\hat{\boldsymbol{Y}}_H,\hat{\boldsymbol{\epsilon}}_{1,h})$, $(\hat{\boldsymbol{X}}_H,\hat{\boldsymbol{Y}}_H)$ are used. These two residuals are resampled in groups $(\hat{\boldsymbol{X}}_H^{(b)},\hat{\boldsymbol{Y}}_H^{(b)})$, then the bootstrap samples can be obtained just with fitted $\hat{\boldsymbol{\xi}}_h, \hat{\boldsymbol{\omega}}_h$. The two residuals are assumed to be independent of each other; however, to keep the potential inner connection of the dataset, the residuals are sampled together in real practice. 

With a number of $B$ bootstrapped datasets, applying the optimisation again, $B$ number of fitted parameters will be available. With a new dataset, $B$ number of predictions based on the new dataset will also be available. Then, the empirical confidence interval and prediction interval for the new dataset are obtained. The example is illustrated in the simulation study 4.

\section{Simulation Study}
\label{sec:sim}
To evaluate the proposed model, parameter estimation, theoretical results, and the proposed bootstrap method, four simulation studies were conducted. The aims are (1) to evaluate the proposed Generative-PLS model, and whether the generated dataset follows a PLS structure. (2) Evaluate the performance of the estimators from optimisation steps in. (3) Verify the theoretical result we proposed in Theorem~\ref{theorem: main}. (4) Evaluate the confidence interval and prediction interval from bootstrapped Generative-PLS.

\subsection*{Simulation 1}
The first simulation study is designed to evaluate the proposed generative model. The Partial Least Squares is always known as an algorithm rather than a model, and the training algorithm is already a well-known, stable algorithm. In simulation 1, to verify that the proposed Generative-PLS model aligns well with the PLS data structure, we generate data from the proposed Generative-PLS model and apply the current PLS algorithm to estimate the parameters. 

In simulation 1, we set $p = 10$, $q = 10$ and $H = 2$ and the parameters in the model are generated randomly under all the model constraints, then data is simulated based on the data generation procedure with the random parameters. We consider a combination of small and large sample size $n = (50, 200, 1000, 5000)$, three types of noise: (1) $\boldsymbol{\Sigma}_X^\bullet = (\sigma_x^2 \boldsymbol{I})^\bullet$ with $\sigma_x^{2\bullet} = 0.01$, (2) $\boldsymbol{\Sigma}_X^\bullet = (\sigma_x^2 \boldsymbol{I})^\bullet$ with $\sigma_x^{2^\bullet} = 2$ and (3) $\boldsymbol{\Sigma}_X^\bullet$ is a random positive semidefinite matrix. Within type (3), the $\boldsymbol{\Sigma}_X^\bullet$ is generated from the Inverse Wishart Distribution, with scale matrix $\sigma_x^2\boldsymbol{I}$ and degree of freedom $p+1$. The noise $\boldsymbol{\Sigma}_{\tilde{Y}}^\bullet$ is set to be equal to $\boldsymbol{\Sigma}_X^\bullet$ or follow the same distribution. Moreover, both normally distributed data and non-normal data are considered.

As discussed in Section~\ref{subsec: pls_property}, under the setting of a diagonal matrix and normal data situation, the data generation procedure in our proposed model is similar to the PPLS model proposed in \cite{el_Bouhaddani_2018}. Moreover, if a positive semidefinite error matrix is used, our proposed model is similar to PPLS-SVD. So, the PPLS model and the PPLS-SVD models are involved to compare in comparing these specific situations.

In all the situations, the three loading weights matrix is recorded and then the distance between the simulated $\boldsymbol{w}_h^\bullet$ and the $\hat{\boldsymbol{u}}_h^{PLS}$ is recorded.
$$d(\hat{\boldsymbol{u}}_h^{PLS},\boldsymbol{u}_h^\bullet) = \frac{1}{p} \sqrt{\sum_{i = 1}^{p} (\hat{\boldsymbol{u}}_{h,i} - \boldsymbol{w}_{h,i}^\bullet)^2}$$

Since the PLS algorithm in the plsr package is not identifiable, we actually record:
$$d(\hat{\boldsymbol{u}}_h^{PLS},\boldsymbol{w}_h^\bullet) = \text{min}(\frac{1}{p} \sqrt{\sum_{i = 1}^{p} (\hat{\boldsymbol{u}}_{h,i} - \boldsymbol{w}_{h,i}^\bullet)^2},\frac{1}{p} \sqrt{\sum_{i = 1}^{p} (-\hat{\boldsymbol{u}}_{h,i} - \boldsymbol{w}_{h,i}^\bullet)^2)}$$
The experiments are repeated 50 times, and the mean value is recorded for a more reliable result.

\begin{table}[hbt!]
    \centering
    \caption{Result for normal situation}
    \label{tab:simulation1_normal}
    \begin{tabular}{|l|c|c|c|c|}\hline
         & 50 & 200 & 1000 & 5000 \\\hline
         \multicolumn{5}{|c|}{Generative-PLS+$\sigma_x^2 = 0.01$} \\\hline
        $d(\hat{\boldsymbol{u}}_1^{PLS},\boldsymbol{u}_1^\bullet)$ & 0.0005 & 0.0002 & 0.0001 & $4.8422 \times 10^{-5}$ \\\hline
        $d(\hat{\boldsymbol{u}}_2^{PLS},\boldsymbol{u}_2^\bullet)$ & 0.0005 & 0.0003 & 0.0001 & $5.4177 \times 10^{-5}$ \\\hline
        \multicolumn{5}{|c|}{PPLS+$\sigma_x^2 = 0.01$} \\\hline
        $d(\hat{\boldsymbol{u}}_1^{PLS},\boldsymbol{u}_1^\bullet)$ & 0.0848 & 0.0914 & 0.0975 & 0.0744 \\\hline
        $d(\hat{\boldsymbol{u}}_2^{PLS},\boldsymbol{u}_2^\bullet)$ & 0.1010 & 0.1012 & 0.1033 & 0.0906 \\\hline
        \multicolumn{5}{|c|}{Generative-PLS+$\sigma_x^2 = 2$} \\\hline
        $d(\hat{\boldsymbol{u}}_1^{PLS},\boldsymbol{u}_1^\bullet)$ & 0.0075 & 0.0039 & 0.0024 & 0.0006 \\\hline
        $d(\hat{\boldsymbol{u}}_2^{PLS},\boldsymbol{u}_2^\bullet)$ & 0.0095 & 0.0045 & 0.0027 & 0.0012 \\\hline
        \multicolumn{5}{|c|}{$PPLS+\sigma_x^2 = 2$} \\\hline
        $d(\hat{\boldsymbol{u}}_1^{PLS},\boldsymbol{u}_1^\bullet)$ & 0.0898 & 0.0926 & 0.1032 & 0.0863 \\\hline
        $d(\hat{\boldsymbol{u}}_2^{PLS},\boldsymbol{u}_2^\bullet)$ & 0.1019 & 0.0985 & 0.0878 & 0.0927 \\\hline
        \multicolumn{5}{|c|}{Generative-PLS+$\boldsymbol{\Sigma}_X \text{ positive semidefinite}$} \\\hline
        $d(\hat{\boldsymbol{u}}_1^{PLS},\boldsymbol{u}_1^\bullet)$ & 0.0041 & 0.0053 & 0.0008 & 0.0008 \\\hline
        $d(\hat{\boldsymbol{u}}_2^{PLS},\boldsymbol{u}_2^\bullet)$ & 0.0089 & 0.0062 & 0.0020 & 0.0094 \\\hline
        \multicolumn{5}{|c|}{PPLS-SVD+$\boldsymbol{\Sigma}_X \text{ positive semidefinite}$} \\\hline
        $d(\hat{\boldsymbol{u}}_1^{PLS},\boldsymbol{u}_1^\bullet)$ & 0.0978 & 0.0807 & 0.0865 & 0.0796 \\\hline
        $d(\hat{\boldsymbol{u}}_2^{PLS},\boldsymbol{u}_2^\bullet)$ & 0.1087 & 0.0878 & 0.1151 & 0.0910 \\\hline
    \end{tabular}
\end{table}

From the table \ref{tab:simulation1_normal}, with an increase in sample size, the data generated from the PPLS or PPLS-SVD model aligns better with PLS data structure. However, thanks to the recursive structure, within all the situations, data simulated from the Generative-PLS model always performs better in recovering the PLS structure, even with a small sample size 50. 

Then, within the non-normal case, we consider the $\xi_h^\bullet$ follow an exponential distribution with rate 1, then the $\xi_h^\bullet$ are scaled to have mean 0. For simplicity, the error term is still simulated from a normal distribution with mean 0 and variance $0.1*\boldsymbol{I}$. Then the result is recorded in Table~\ref{tab:simulation1_exponential}.

\begin{table}[hbt!]
    \centering
    \caption{Result for exponential data}\label{tab:simulation1_exponential}
    \begin{tabular}{|l|c|c|c|c|}\hline
         & 50 & 200 & 1000 & 5000 \\\hline
        $d(\hat{\boldsymbol{u}}_1^{PLS},\boldsymbol{u}_1^\bullet)$ & 0.0014 & 0.0009 & 0.0004 & 0.0002 \\\hline
        $d(\hat{\boldsymbol{u}}_2^{PLS},\boldsymbol{u}_2^\bullet)$ & 0.0016 & 0.0011 & 0.0004 & 0.0004 \\\hline
    \end{tabular}
\end{table}

From the table \ref{tab:simulation1_exponential}, we can conclude that the proposed Generative-PLS model can recover the PLS data structure well without a constraint on distribution or sample size. The proposed Generative-PLS model indeed aligns with the PLS algorithms.

\subsection*{Simulation 2}
To verify the convergence theory we introduced in section \ref{sec: latent}, a simple simulation example is introduced. 

In the simulation study, we set $p = q = 20$, and the number of components is $H = 3$. The parameter matrices $\boldsymbol{W}^\bullet$ and $\boldsymbol{V}^\bullet$ are random orthonormal matrices. $\boldsymbol{B}^\bullet$ and $\boldsymbol{\Sigma}_{\xi}^\bullet$ are also generated randomly, but with a strongly decreasing order by $h$. Without loss of generality, but simpler, the $b_h^\bullet$ and $s_{1,h}^\bullet$ are set to be integers in this simulation study. With a random seed, the simulated $\boldsymbol{\Sigma}_{\xi}^\bullet = \text{diag}(5,3,2)$ and $\boldsymbol{B}^\bullet = \text{diag}(9,6,4)$. 

In order to verify that  $\frac{1}{n}\mathbb{E}\left\|\hat{\boldsymbol{\xi}}_h - \boldsymbol{\xi}_h^\bullet\right\|_2^2 \longrightarrow {\boldsymbol{w}_h^\bullet}^\top \boldsymbol{\Sigma}_{X}^\bullet \boldsymbol{w}_h^\bullet$, the $\boldsymbol{\Sigma}_X^\bullet$ is set to be diagonal, where $\boldsymbol{\Sigma}_X^\bullet = \sigma_x^{2\bullet} \boldsymbol{I}$. The $\sigma_x^{2\bullet}$ is chosen to be $(1,15)$. The first case is when $\sigma_x^2$ is smaller than the smallest $s_{1,h}^2$, which is an error term. The second case is when $\sigma_x^{2\bullet}$ is larger than all the $s_{1,h}^2$. In this case, the residual is more significant than all the components; however, it is uncorrelated with the response variable.

The sample size is chosen as $(50, 100, 1000, 5000, 10000)$. The empirical distance between $\hat{\boldsymbol{u}}_h$ and $\boldsymbol{w}_h^\bullet$ and empirical distance between $\hat{\boldsymbol{\xi}}_h$ and $\boldsymbol{\xi}_h^\bullet$ are recorded, which are
$$d(\hat{\boldsymbol{u}}_h,\boldsymbol{w}_h^\bullet) = \frac{1}{p} \sum_{i = 1}^{p} (\hat{u}_{h,i} - w_{h,i}^\bullet)^2$$
$$d(\hat{\boldsymbol{\xi}}_h,\boldsymbol{\xi}_h^\bullet) = \frac{1}{n} \sum_{i = 1}^{n} (\hat{\xi}_{h,i} - \xi_{h,i}^\bullet)^2.$$

For each sample size, the experiments are repeated 50 times. The mean value of the 50 repetitions is recorded together with the empirical $95\%$ interval.

\begin{figure}[ht]
        \centering
        \includegraphics[scale = 0.18]{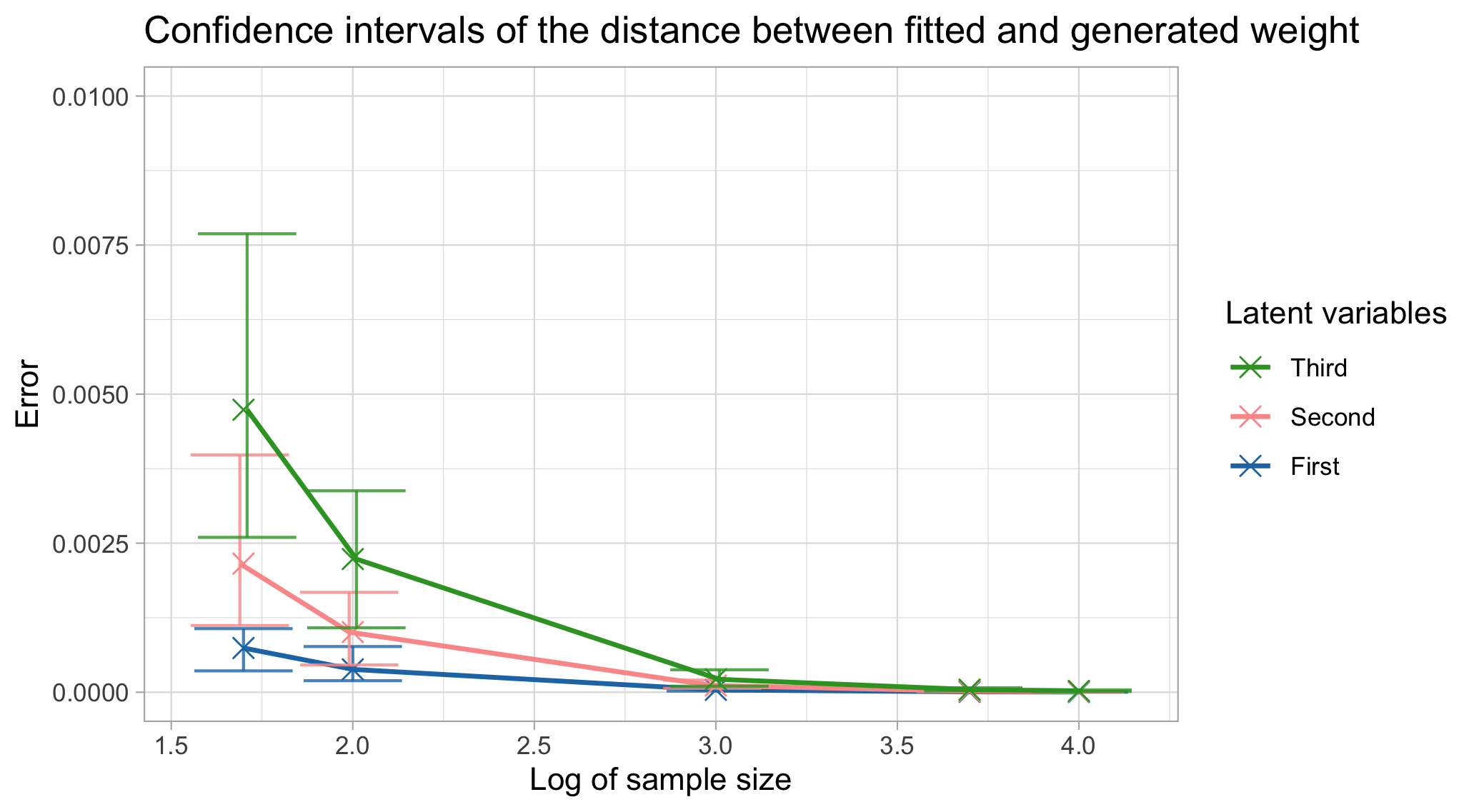}
        \caption{Plot of Confidence Intervals for the distance between $\hat{\boldsymbol{u}}_h$ and $\boldsymbol{u}_h$ with $\sigma_x^2 = 1$}
        \label{fig:converge1_1}
\end{figure}

\begin{figure}[ht]
        \centering
        \includegraphics[scale = 0.18]{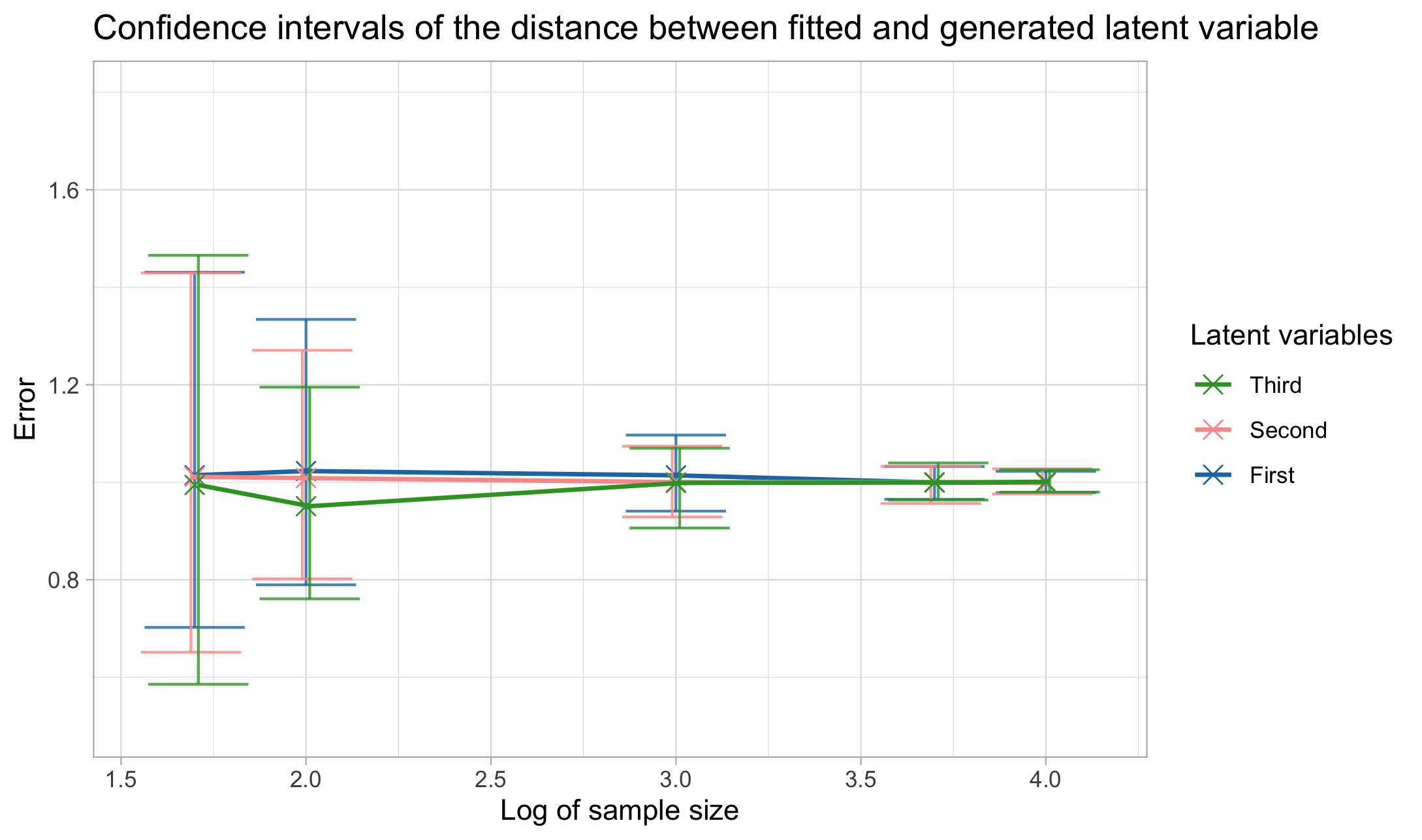}
        \caption{Plot of Confidence Intervals for the distance between $\hat{\boldsymbol{u}}_h$ and $\boldsymbol{u}_h$ with $\sigma_x^2 = 1$}
        \label{fig:converge1_2}
\end{figure}

\begin{figure}[ht]
        \centering
        \includegraphics[scale = 0.18]{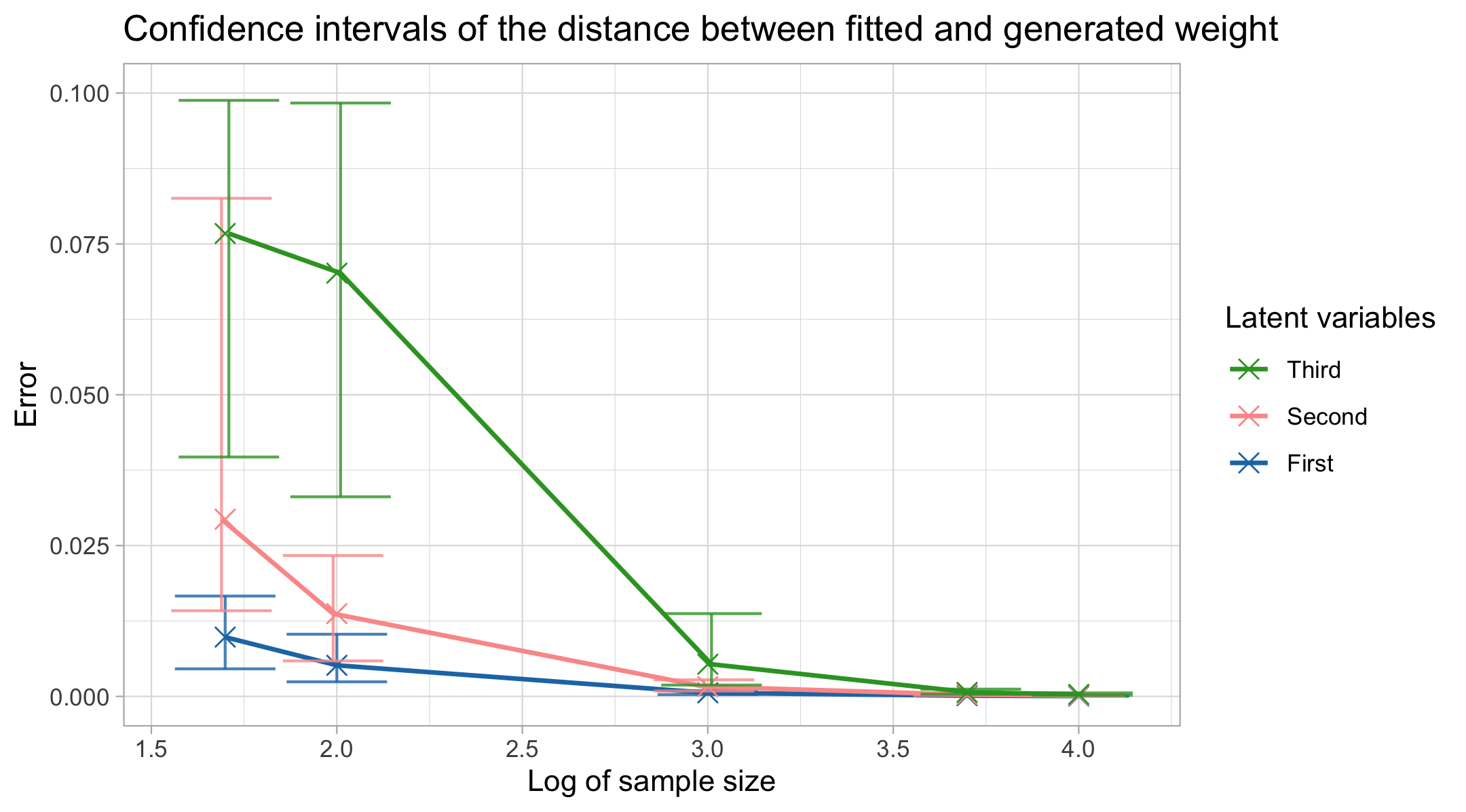}
        \caption{Plot of Confidence Intervals for the distance between $\hat{\boldsymbol{u}}_h$ and $\boldsymbol{u}_h$ with $\sigma_x^2 = 15$}
        \label{fig:converge1}
\end{figure}

\begin{figure}[ht]
        \centering
        \includegraphics[scale = 0.18]{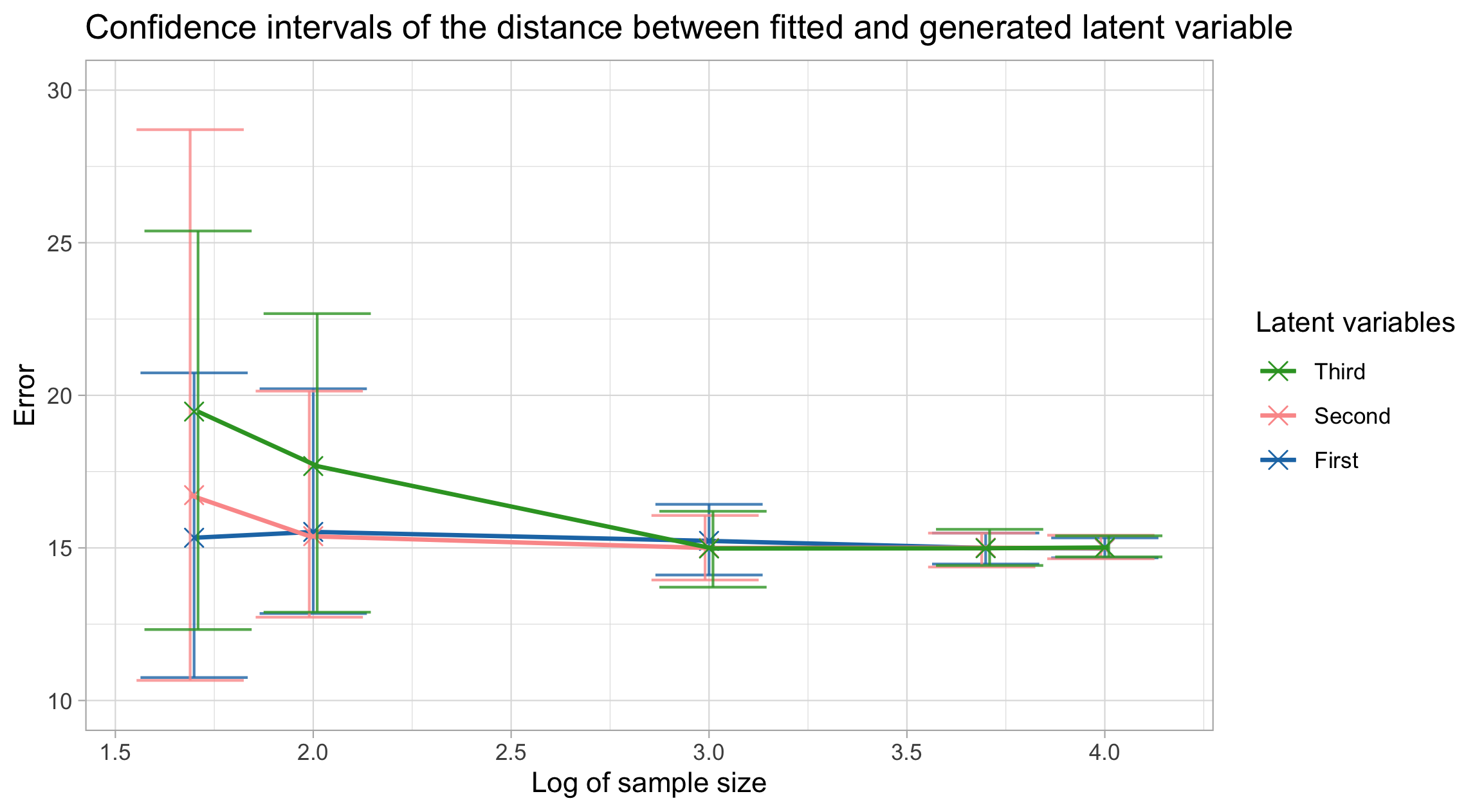}
        \caption{Plot of Confidence Intervals for distance between $\hat{\boldsymbol{\xi}}_h$ and $\boldsymbol{\xi}_h$ $\sigma_x^2 = 15$}
        \label{fig:converge2}
\end{figure}

Figure~\ref{fig:converge1_1} and \ref{fig:converge1_2} are the plots of the distance between $\hat{\boldsymbol{u}}_h$ and $\boldsymbol{w}_h^\bullet$ and the distance between $\hat{\boldsymbol{\xi}}_h$ and $\boldsymbol{\xi}_h^\bullet$ when $\sigma_x^{2,\bullet} = 1$. The figure \ref{fig:converge1} and \ref{fig:converge2} are the Plots when $\sigma_x^{2\bullet} = 15$. The green colour represents when $h = 3$, while the pink is related to $h = 2$, and the blue is $h =1$. The "x" dots are the mean values from the 50 times repetition, and the error bar represents the 95\% empirical intervals. For both figures, the y-axis represents the error, and the x-axis is the 10-log sample size.

From Figure~\ref{fig:converge1_1} and  \ref{fig:converge1}, it's clear that the distance between $\hat{\boldsymbol{u}}_h$ and $\boldsymbol{w}_h^\bullet$ is large when the sample size is smaller, also with a larger $h$, the error is larger. However, with a sample size larger than 1000, all the errors converge to around 0; meanwhile, the 95\% intervals also all vanish to 0. Additionally, with a smaller $\sigma_x^2$, the estimation of $\boldsymbol{u}_h$ is better when the sample size is small, note that the y range of \ref{fig:converge1} is 10 times larger than it in the figure \ref{fig:converge1_1}, though they both converge to 0 with a larger sample size.

From Figure~\ref{fig:converge1_2} and \ref{fig:converge2}, the order of components doesn't affect the error terms. In the meantime, with a larger sample size, the error term converges to $\sigma_x^{2\bullet}$ no matter how large the $\sigma_x^{2\bullet}$ is, and the interval is getting narrower with an increment in the sample size.

Thus, this simulation study validates the proposed Theorem~\ref{theorem: main}.

\subsection*{Simulation 3}
In simulation study 3, a similar setting is used as in \cite{el_Bouhaddani_2018} and \cite{Eti_vant_2022}. This simulation study is first introduced in \cite{el_Bouhaddani_2018}, then in \cite{Eti_vant_2022}, this study is extended to a more general case. From \cite{Eti_vant_2022}, they show that the PPLS method proposed in \cite{el_Bouhaddani_2018} will fail when the error term is not a diagonal matrix. Thus, in our simulation study, we consider both situations.

The first situation is under the assumption (b) that the error matrix is diagonal. The same setting as in \cite{el_Bouhaddani_2018} is applied, where $p = q = 20$, and $H = 3$. We also consider combinations of small and large proportions of noise $\alpha = \{0.1, 0.5 \}$. For the sample size, we consider small and large sample sizes $n = \{50, 1000\}$, we slightly enlarge the sample size than the setting in \cite{el_Bouhaddani_2018} due to the consistency estimator introduced in Section~\ref{subsubsec: model_inference}. The true loading weights are simulated from a normal density function
$$w_{j,k}^\bullet = \varphi\left(\left(\frac{1}{2} + \frac{1}{10}j\right)k, \frac{1}{10}k\right), \quad
v_{j,k}^\bullet = \varphi\left(\left(\frac{3}{5} + \frac{1}{10}j\right)k, \frac{1}{10}k\right).$$
Then, the simulated values are made to be orthogonal to each other. The parameter $B$ is set as $\boldsymbol{B}^\bullet = (1.5,1.11,0.82)$ and the $\boldsymbol{\Sigma}_{\xi}^\bullet = (1,0.9,0.82)$. The $\sigma_x^{2\bullet}$, $\sigma_y^{2\bullet}$ and $\sigma_1^{2\bullet}$ are set based on the error rate $\alpha$. Then, this simulation study is extended to a more general case, with a positive semi-definite error term.

In summary, all the parameters are set to be the same as the simulation study in \cite{el_Bouhaddani_2018} and \cite{Eti_vant_2022}, while our proposed model is used to generate the dataset. From simulation study 1, it's clear that our proposed model can lead to a dataset closer to the PLS data structure. All the experiments are repeated 100 times for a stable result.

In the diagonal error term case, we recorded all the estimated parameters, and we focus on the estimated weight for the first component $\hat{\boldsymbol{u}}_1$, fitted fitting parameter between two latent variables $\hat{\boldsymbol{B}} = \text{diag}(\hat{b}_1,\hat{b}_2,\hat{b}_3)$, the estimated variance of the latent variable $\hat{\boldsymbol{\Sigma}}_{\xi} = (\hat{s}_{1,1},\hat{s}_{1,2},\hat{s}_{1,3})$ and estimated error terms $\hat{\sigma}_x$, $\hat{\sigma}_y$ and $\hat{\sigma}_1$. 

\begin{figure}[ht]
        \centering
        \includegraphics[scale = 0.16]{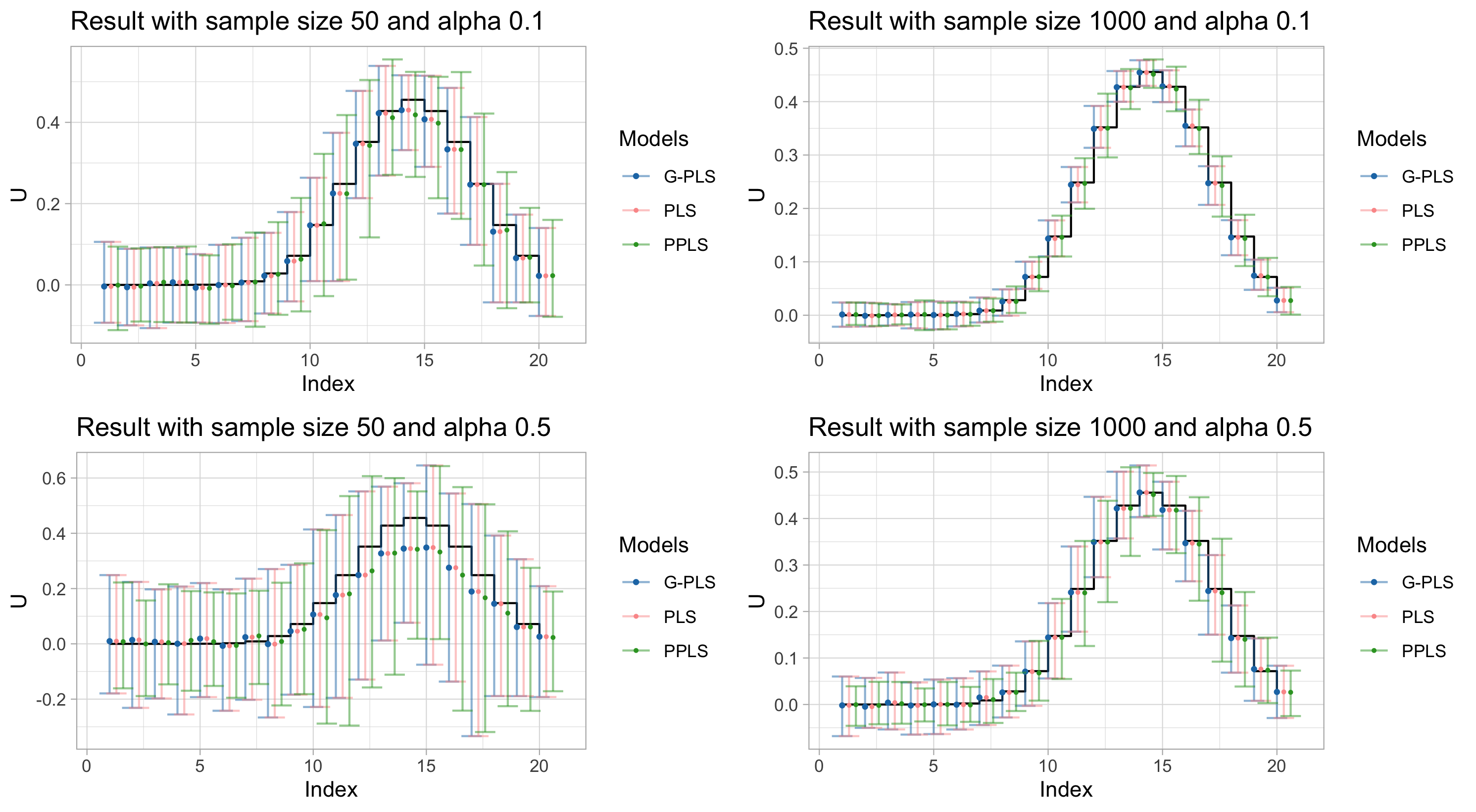}
        \caption{Plot of $\hat{\boldsymbol{u}}_1$ with 95 \% interval with diagonal error term}
        \label{fig:simu2_U}
\end{figure}

\begin{figure}[ht]
        \centering
        \includegraphics[scale = 0.17]{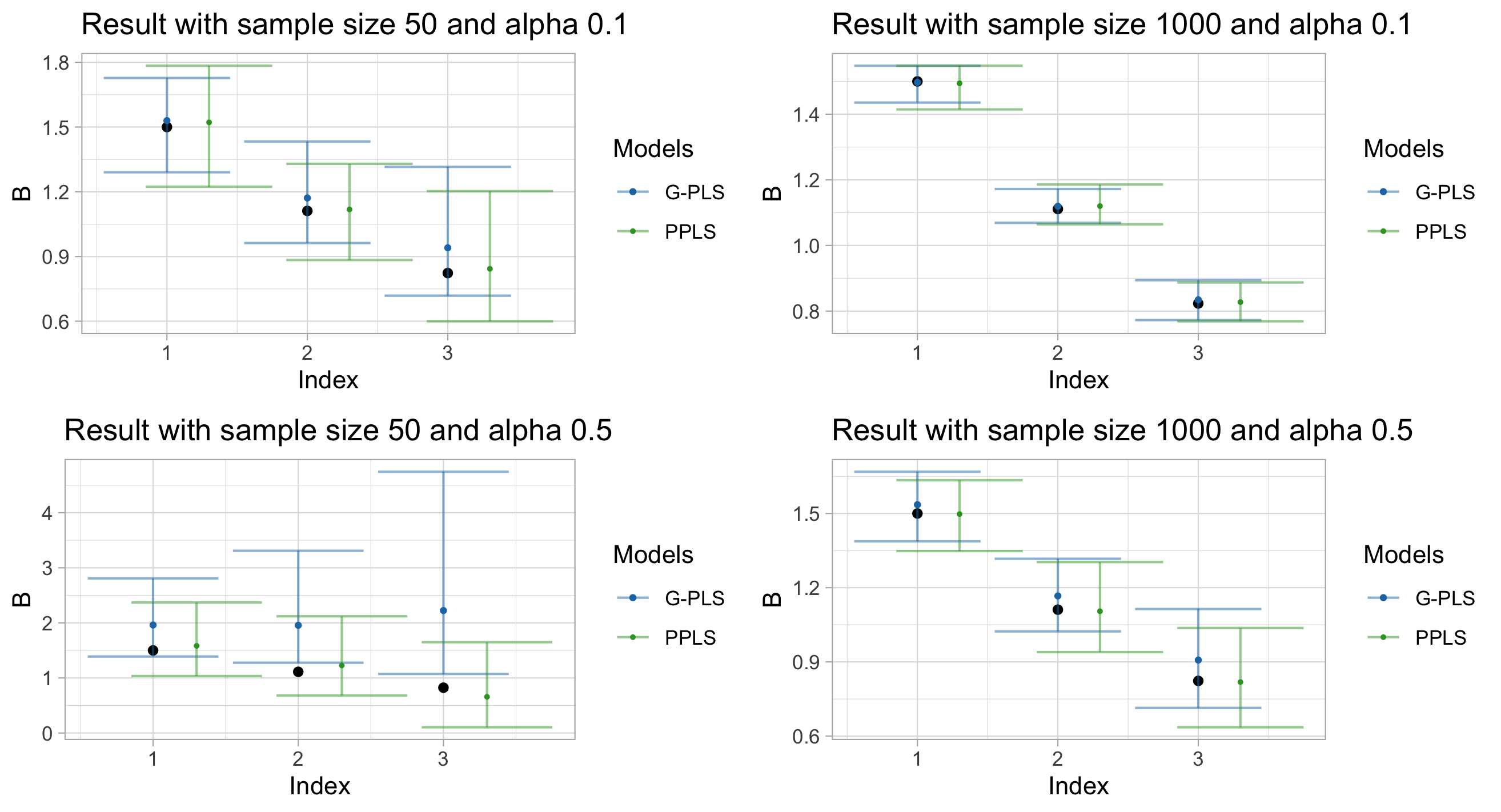}
        \caption{Plot of B with 95 \% interval with diagonal error term}
        \label{fig:simu2_B}
\end{figure}

\begin{figure}[ht]
        \centering
        \includegraphics[scale = 0.16]{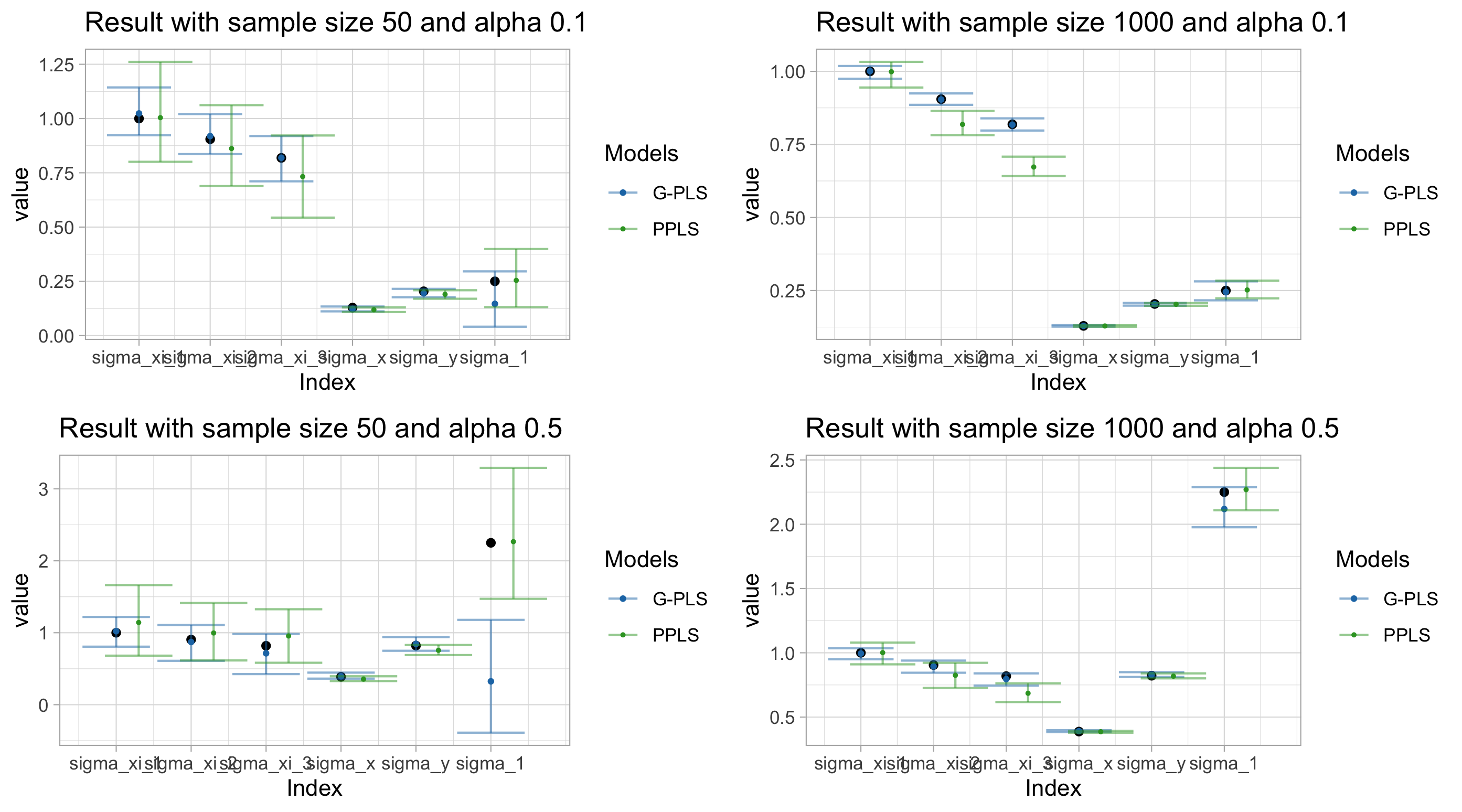}
        \caption{Plot of $\sigma^2$s with 95 \% interval with diagonal error term}
        \label{fig:simu2_sigma}
\end{figure}

Figure~\ref{fig:simu2_U}, \ref{fig:simu2_B}, and \ref{fig:simu2_sigma} are the results for the diagonal error term case (assumption (a)). In Figure~\ref{fig:simu2_U}, the results for the estimated weights for the first component are recorded. The black line is the true value, then the blue colour is the result of Generative-PLS, and the pink one is the result of the traditional PLS algorithm from the "PLS" package in R, and the green indicates the result from PPLS. For all the results from these three methods, the dots indicate the mean value from 100 replicates, and the error bar is the 95\% interval based on these replicates. Note that the traditional PLS from the "PLS" package couldn't provide an estimation for all the $\sigma^2$s and $\boldsymbol{B}$, so only the proposed method and PPLS are considered in \ref{fig:simu2_B} and \ref{fig:simu2_sigma}.

In \ref{fig:simu2_U}, we recorded the $\hat{\boldsymbol{u}}_1$ under different error terms together with different sample sizes. From the plot, we can see that the three methods work similarly in estimating $U$ in all the situations. The proposed method is slightly more stable when the error is small. Also, in all four scenarios, we can see that the performance of the PLS algorithm is closer to the proposed method than PPLS. All the methods can converge to the true value when the sample size is large, but when the sample size is small and the error rate is high, all the methods are affected.

Figure \ref{fig:simu2_B} records the estimated $\hat{\boldsymbol{B}}$ and the 95\% interval from PPLS and Generative-PLS and \ref{fig:simu2_sigma} records all the estimated errors. Within both plots, the black dots denote the true value used in data simulation; again, the green and blue dots denote the mean value of the estimators from PPLS and proposed Generative-PLS, respectively. Again, the error bar is the 95\% interval from the 100 replicates.

From the plot, it's clear that PPLS can perform better in estimating the $B$ when the sample size is smaller. However, with a larger sample size or when the error rate is small, the proposed method performs almost as well as the PPLS, with a slightly narrower 95\% interval. This is because the proposed $\hat B$ is a consistent estimator; it converges to the true value with a large sample size. 

From figure \ref{fig:simu2_sigma}, it's clear that the proposed method is better at estimating the variance of the latent variables. In all scenarios, the proposed method results in a closer mean estimated value to the true value and a narrower 95\% interval, which means more stable. For $\sigma_x^2$ and $\sigma_y^2$, again, the proposed method leads to a slightly better estimation, especially when the sample size is small.

However, for $\sigma_1^2$, the proposed method performs well only when the noise is low or the sample size is large, while the PPLS method performs well in all scenarios. Overall, the proposed method can always lead to a good and stable estimation when the sample size is large. Moreover, the proposed method can achieve competitive inference results without using the EM algorithm.

Then the simulation study is extended to a more general case. Instead of a diagonal error matrix (assumption (b)), a more general positive semi-definite error matrix is used (assumption (c)). In \cite{Eti_vant_2022}, they have shown that the PPLS method will perform close to PPCA when a positive semi-definite error matrix is used. In their simulation study, the error term is selected to make sure that the eigenvectors of $\text{Var}(\boldsymbol{X})$ and $\text{Var}(\boldsymbol{Y})$ are different to the singular value of the $\text{Cov}(\boldsymbol{X}, \boldsymbol{Y})$.

In this simulation study here, to further compare the performance of our proposed method and PPLS in the more general case, we simulated the error matrix randomly with Inverse Wishart Distribution, with scale matrix $\sigma_x^2\boldsymbol{I}$ and degree of freedom $p+1$, where $\sigma_x^2$ is the same as the $\sigma_x^2$ used when noise rate = 0.5. The sample size is chosen to be 1000 to ensure all methods have more stable results. Again, the experiment is repeated 100 times to ensure a stable result.

\begin{figure}[ht]
        \centering
        \includegraphics[scale = 0.15]{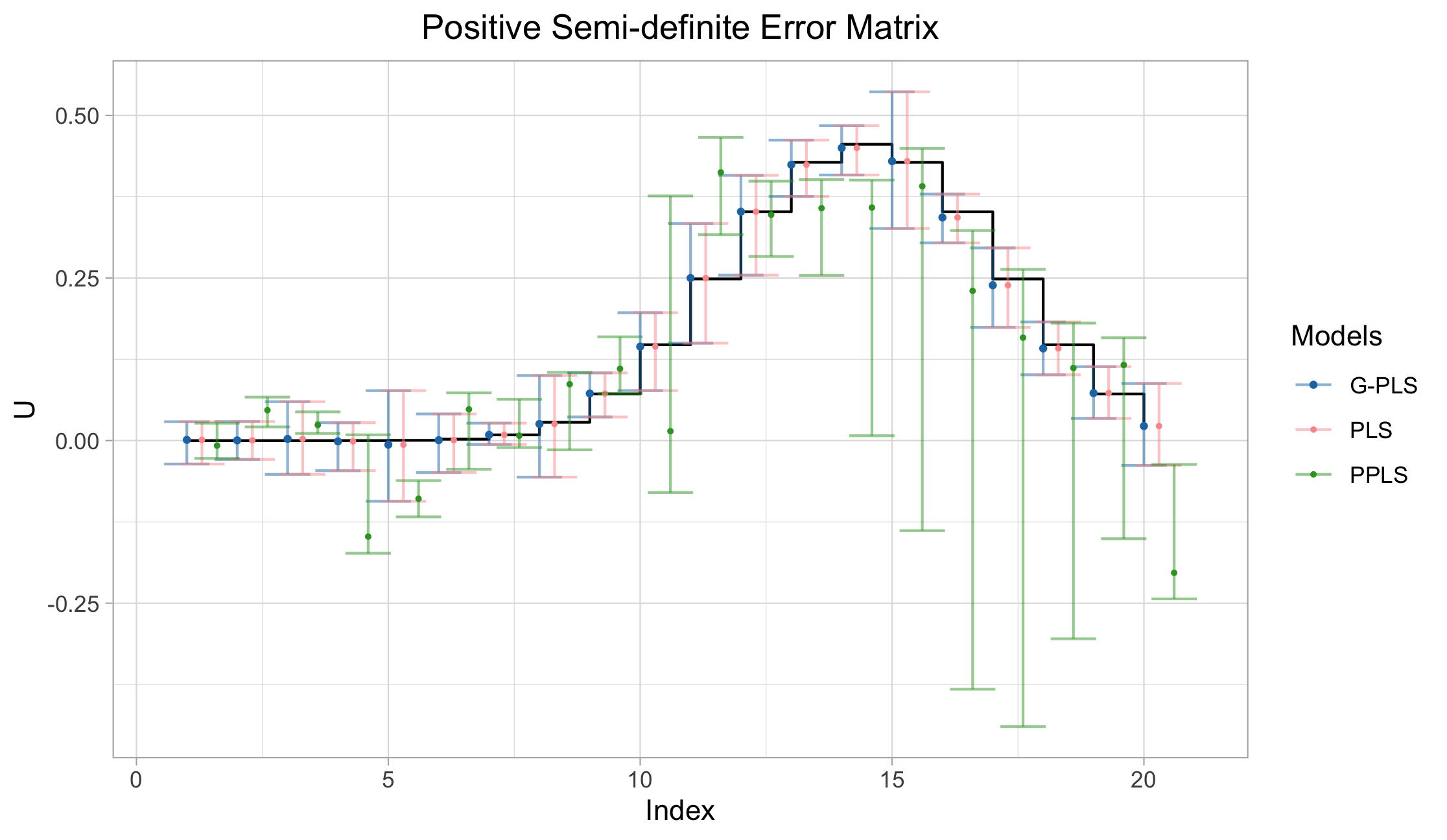}
        \caption{Plot of U with 95 \% interval with random positive semi-definite error term}
        \label{fig:simu2_positive_semi}
\end{figure}
 
In \ref{fig:simu2_positive_semi}, again the $\hat{\boldsymbol{u}}_1$ from the three methods are recorded. It's clear that when the error term is a general positive semi-definite matrix instead of a diagonal matrix, the PPLS method will be unstable and lead to biased estimators. However, the proposed Generative-PLS method still performs similarly to traditional PLS and converges to the true value.

\begin{figure}[ht]
        \centering
        \includegraphics[scale = 0.15]{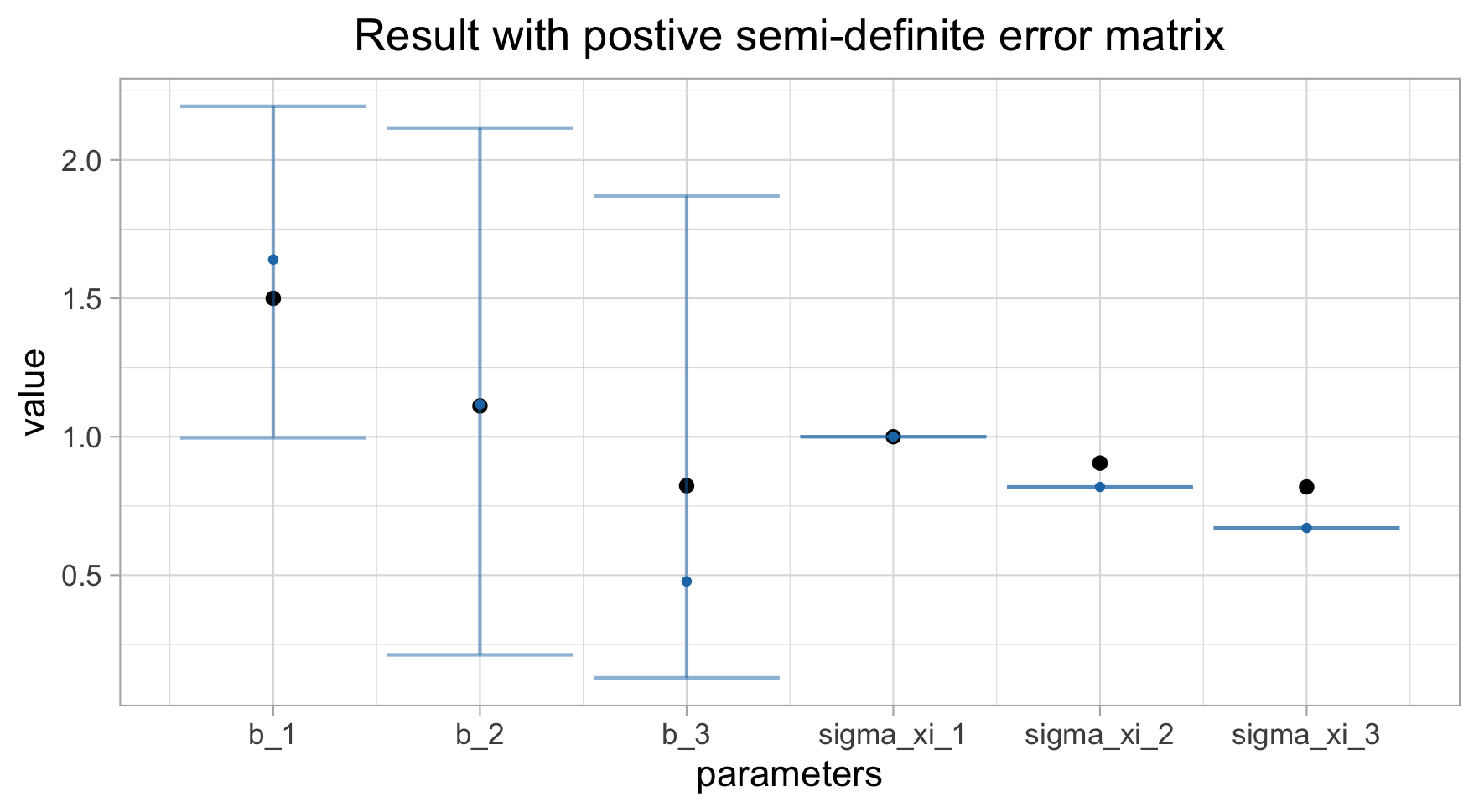}
        \caption{Plot of other estimators with 95 \% interval with random positive semi-definite error term}
        \label{fig:simu2_positive_others}
\end{figure}

The estimators for $\hat{\boldsymbol{B}}$ and $\hat{s}_{1,h}^2$ are plotted in \ref{fig:simu2_positive_others}. The estimated result is close to the true value; the bias comes from the non-invertible matrix while correcting the estimators when the $\boldsymbol{\Sigma}_X$ is not diagonal, and further improvement can be obtained with a better approximation of the inverse non-invertible matrix.

\subsection*{Simulation 4} 

In simulation study 3, although a 95\% interval for the estimated parameter is provided, that's based on known information about the data distribution and the data generation procedure. When the data distribution is unknown, it will be hard to simulate a number of datasets and get the interval. However, the bootstrap method proposed in \ref{sec: inference} ensures that we can get the confidence interval and prediction interval from the bootstrapped dataset. So, this simulation study is conducted to verify the bootstrap method proposed in \ref{sec: inference}.

Within this simulation study, for simplicity, the parameter setting is the same as in the simulation study 3, except we choose an error term with noise rate $\alpha = 0.1$. Then, only one dataset with sample size $n = 1000$ is generated, and another 10 data points are generated as the test set. The bootstrap method will lead to the confidence interval of the parameters and the prediction interval of the new data. Moreover, the Bayesian PLS from \cite{Vidaurre_2013} is involved in comparing predictions.

\begin{figure}[ht]
        \centering
        \includegraphics[scale = 0.24]{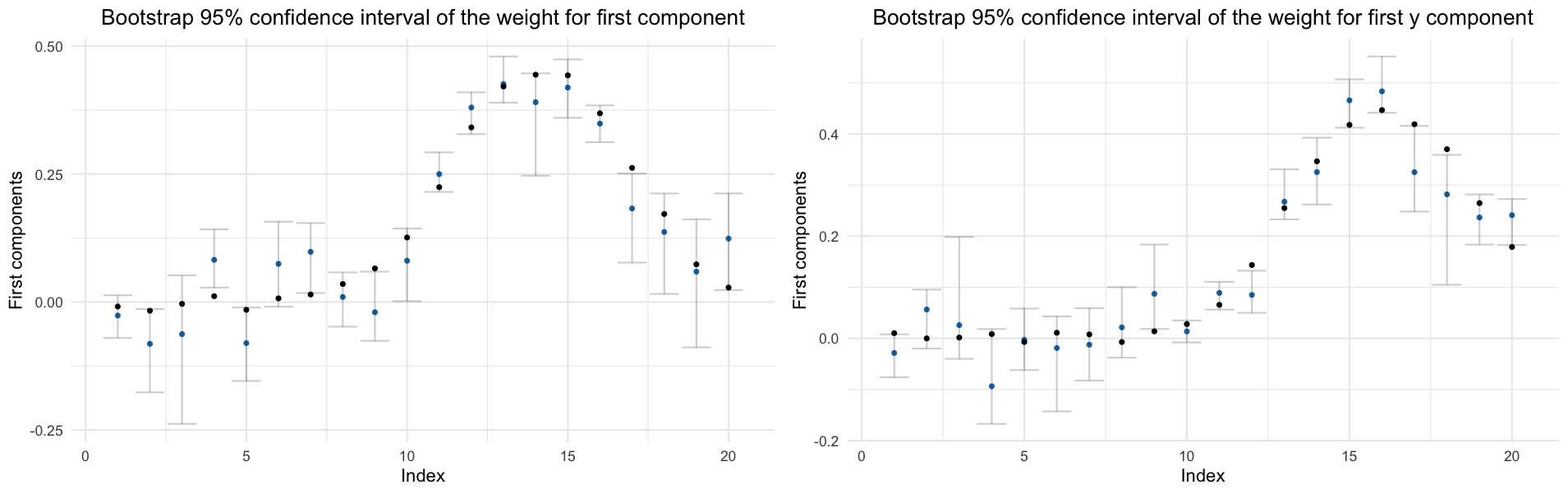}
        \caption{Plot of 95 \% confidence interval for $\hat{\boldsymbol{u}}_1$ and $\hat{\boldsymbol{v}}_1$}
        \label{fig:simu4_CI}
\end{figure}

The Figure~\ref{fig:simu4_CI} is the plot of 95\% confidence interval of the first estimated weight $\hat{\boldsymbol{u}}_1$ and $\hat{\boldsymbol{v}}_1$. The black dots indicate the true values, then the blue points are the middle points in the bootstrapped results. The bar indicates the 95\% interval. From the plot, it's clear that most of the true values are within the interval, while the algorithm is more sensitive to the values around 0, leading to wider intervals.

\begin{figure}[ht]
        \centering
        \includegraphics[scale = 0.15]{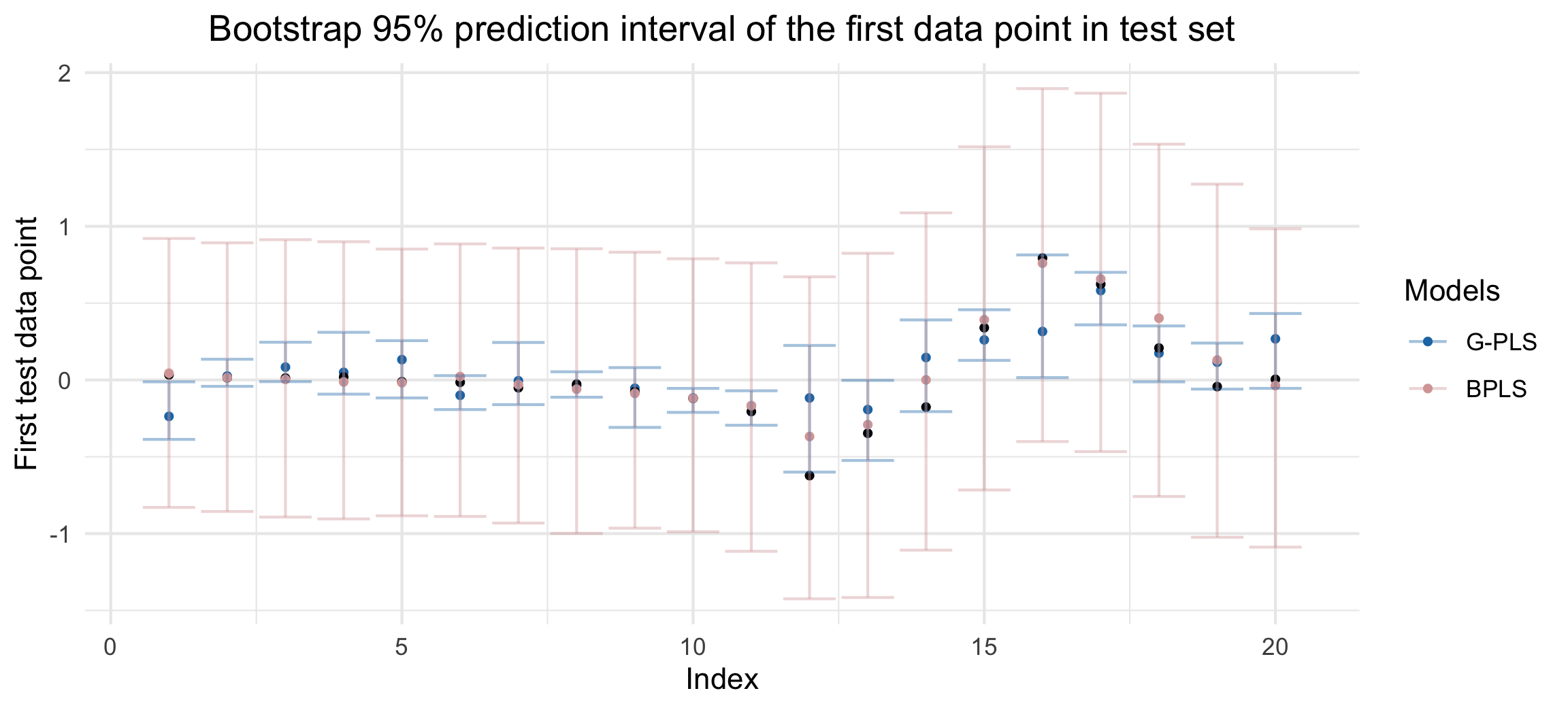}
        \caption{Plot of 95 \% prediction interval for the first data point in the test set}
        \label{fig:simu4_PI}
\end{figure} 

The Figure~\ref{fig:simu4_PI} is the plot of  95\% prediction interval of the first data point in the test set. Again, the black dots indicate the true value, the blue points and the error bar are the mean value and 95\% prediction interval from the proposed bootstrap algorithm. The pink points and bars are the results of the Bayesian PLS. From the result, our proposed bootstrap algorithm results in a much narrower prediction interval. However, the BPLS model results in a slightly better prediction result if the mean value of the interval is considered.

\section{Real Data}
\label{sec:real}
In the real data section, the Near-Infrared (NIR) spectra of the Corn Samples for Standardisation Benchmarking dataset is used. This dataset is an open dataset from Eigenvector Research \cite{eigenvector_corn_nir}, and it's a commonly used dataset in latent structure modelling. This dataset contains spectral data from three different spectrometers measuring 80 corn samples. The wavelength range is 1100-2498 nm, and the data is recorded at each 2 nm interval. Thus, the whole dataset contains three sub-datasets, each has a sample size of 80, and the number of explanatory variables is 700.

In this section, the dataset with m5 spectrometers is used, which results in an explanatory matrix with size $80\times 700$. Then, all the response properties are used as the response. The response matrix has a size $80 \times 4$, in which the columns contain the measured value for moisture, oil, protein, and starch content for each sample. Hence, in this dataset, $n = 80$, $p = 700$, $q = 4$.

Then the proposed model is applied to the dataset, and the number of components is chosen to be $H = 4$. Same as the result in the simulation chapters, the proposed model can provide estimators for the unknown parameters, together with estimators for the variance terms. If the assumption (a) in Section~\ref{sec: pls} is used, the estimated error term is listed in Table~\ref {tab: corn_result}. Moreover, if a predictive model is used, the proposed model results in fitting mean square errors $0.0541, 0.0124, 0.0662, 0.3411$ for the four responses, respectively.

\begin{table}[hbt!]
    \centering
    \caption{The result under assumption (a)}
    \begin{tabular}{|c|c|c|c|c|c|c|}\hline
         $\hat{\sigma}_x^2$& $\hat{\sigma}_y^2$ & $\hat{\sigma}_y^2$ & $\hat{s}_{1,1}^2$ & $\hat{s}_{1,2}^2$ &$\hat{s}_{1,3}^2$ &$\hat{s}_{1,4}^2$\\\hline
         $2.2045 \times 10^{-6}$ & 0.0272 & 0.2034 & 0.9437 & 0.0549 & 0.0724 & 0.0157 \\\hline
    \end{tabular}
    \label{tab: corn_result}
\end{table}

Moreover, from the model interpretation, the components $\xi_h$ can represent the original dataset, especially the first several components. Since the original dataset contains spectrometers' data, from the estimated latent variables, the importance of the spectrometer range can be analysed.

\begin{figure}[ht]
        \centering
        \includegraphics[scale = 0.16]{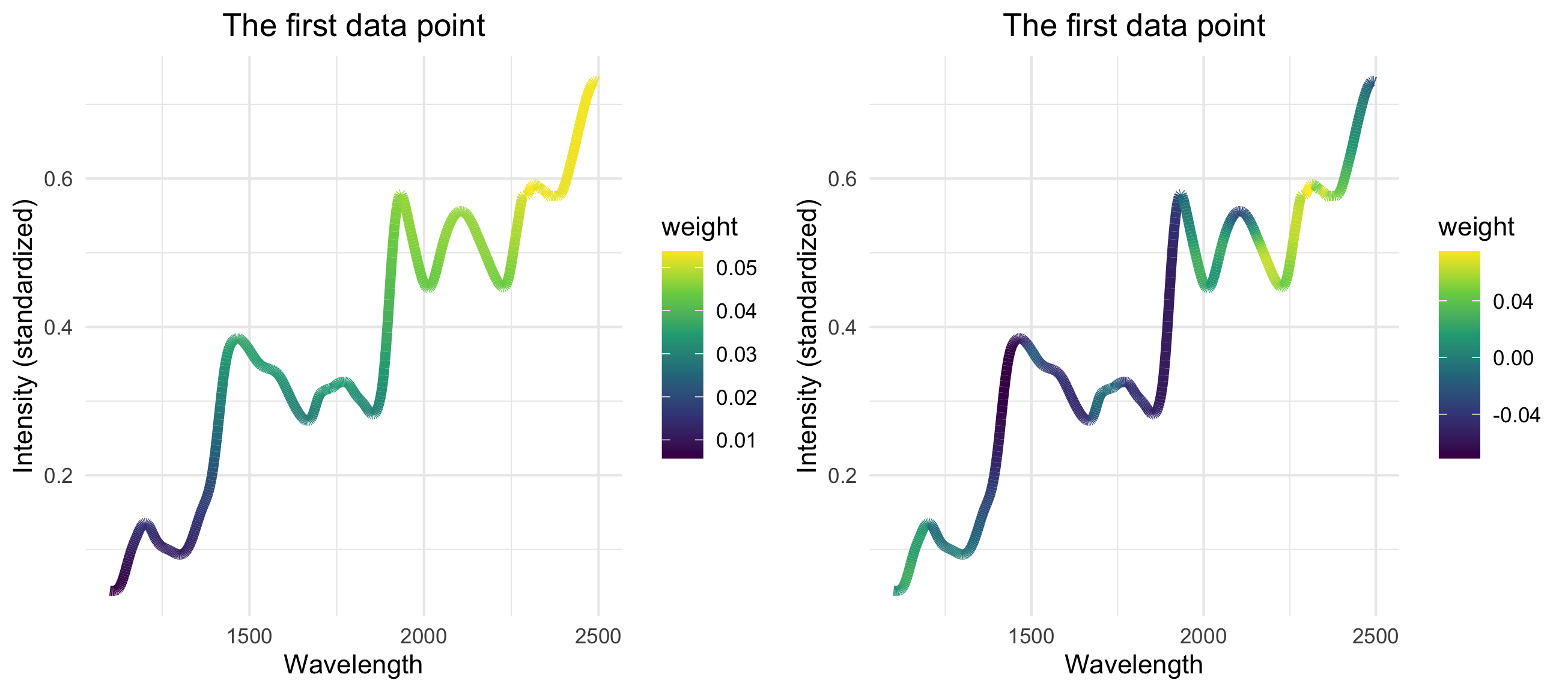}
        \caption{The first data point with colour gradient from fitted latent variables}
        \label{fig: corn_result}
\end{figure}

Figure~\ref{fig: corn_result} plots the first sample from the dataset. The X axis is the wavelength and the Y axis is the standardized intensity. The plotted line is the spectral wave from the first sample. Then, the weights $\hat{\boldsymbol{u}}_1$ from $\hat{\boldsymbol{\xi}}_1$ are used as the colour gradient, which indicates the importance of each data point to the fitted latent variables. The colour in the left figure is from the weight $\hat{\boldsymbol{u}}_1$, and the right one is from $\hat{\boldsymbol{u}}_2$.

From left figure in Figure~\ref{fig: corn_result}, we can see the wave have wavelength between 2250-2500 has larger $\hat{\boldsymbol{u}}_{1, i}$ values, which means the tail part is linearly affect the first component $\hat{\boldsymbol{\xi}}_1$ more, then linearly impact the response. These peaks in the spectra correspond to the C-H, N-H and O-H bonds, which means the position and quantity of these chemical bonds affect the contents of moisture, oil, protein, and starch content in the corn samples. Then, for the second component, again, the waves with large wavelengths have more significant importance. Moreover, the peaks around 1400 nm and 1800 nm negatively impact the response, which corresponds to the C-H and the C=O bonds.

Then, the proposed bootstrapped method in Section~\ref{subsubsec: bootstrap} is applied to get a confidence interval for the parameters. The bootstrap time $B$ is chosen to be 100, which ensures a stable result.

\begin{figure}[ht]
        \centering
        \includegraphics[scale = 0.16]{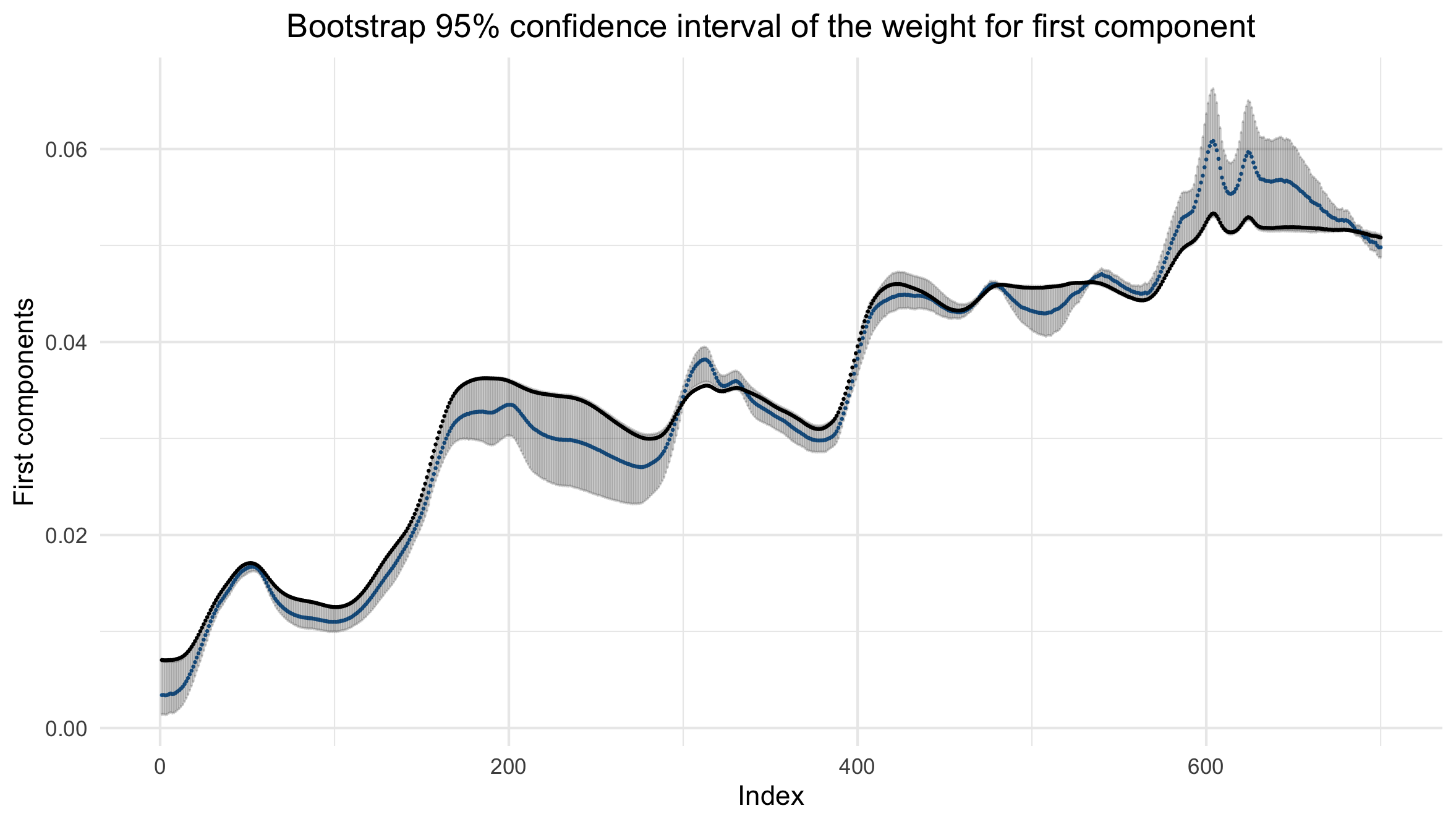}
        \caption{Boostrap 95\% confidence interval for $\hat{\boldsymbol{u}}_1$}
        \label{fig: corn_ci}
\end{figure} 

In the Figure~\ref{fig: corn_ci}, the CI of the weight for the first component is plotted. In the plot, the black line is the $\hat{\boldsymbol{u}}_1$ from applying the algorithm directly. Then the blue line is the middle value of the bootstrapped result, and the grey area is the 95\% confidence interval. From the plot, it's clear that the estimated value lies within the confidence interval, further verifying the proposed bootstrap algorithm.

\section{Conclusion}
\label{sec:conclusion}
This paper extends traditional latent structure models by introducing a unified generative framework, termed Generative Flexible Latent Structure Regression (GFLSR). While a big group of latent structure methods—such as PCA, CCA, and ICA—have well-known algorithms and are widely used in various applications, they still lack a formal statistical model framework. GFLSR addresses this gap via a generative structure that unifies a broad class of linear latent models and establishes a foundation for further model-based inference. We have shown that the proposed structure can be linked to various latent structure methods, including PCA, CCA, ICA, etc.

Then, the special case of the Partial Least Squares (PLS) model is studied in detail. Based on the proposed GFLSR model structure, we proposed two versions of models: Generative PLS-R and Generative (symmetric) PLS-SVD, which correspond to the classical PLS-R and PLS-SVD (or PLS-W2A) methods, respectively. Theoretical contributions include a full asymptotic analysis. Specifically, we proved the convergence of the estimated weights $\hat{\boldsymbol{u}}_h$, and proved that the error between the generated latent components $\boldsymbol{\xi}_h^\bullet$ and their estimates $\hat{\boldsymbol{\xi}}_h$ is bounded by the error term $\boldsymbol{X}_H^\bullet$. To the best of our knowledge, this is the first theoretical work addressing parameter convergence in PLS methods. Furthermore, following the same structure of El Bouhaddani et al. (2018) \cite{el_Bouhaddani_2018}, we demonstrate the identifiability of the proposed model.

The paper also details the model estimation and inference procedure. Then, based on the model structure, a residual bootstrap approach is developed. The model structure, model inference and bootstrap algorithm enabling:
\begin{itemize}
    \item (1) generation of datasets that preserve the PLS structure and reproduce PLS algorithms' results;
    \item (2) estimation of the error term without assuming Gaussianity;
    \item (3) generalisation well to the non-diagonal error covariance scenario;
    \item (4) construction of confidence intervals for model parameters and prediction intervals for unseen data.
\end{itemize}

Overall, the proposed Generative PLS model formalises PLS algorithms into a formal statistical modelling framework, enabling theoretical results and model inference. Simulation studies and real data applications demonstrate that the proposed model achieves satisfactory results without the EM algorithm and maintains good interpretation. We believe that the Generative PLS is a starting point based on the GFLSR model structure. This general model framework sheds light on the potential model structure of all other linear latent structure methods.

On the other hand, the proposed model framework can be further developed. First, more efficient optimisation algorithms are needed to replace the current grid search approach, which, while stable, is computationally intensive and unsuitable for large datasets. A better optimisation algorithm can make the proposed model structure more practical. Then, this framework can be generalised to incorporate non-linear components to involve more latent structure methods. Current work is still based on linear components, which limits the model's representation ability. Finally, further theoretical work could explore the infinite-dimensional setting, enhancing our understanding of inference in high-dimensional PLS contexts.

\section*{Acknowledgments}
This work is supported by the GADI HPC system, Australia. The order of authorship follows alphabetical order.

%Bibliography
\bibliographystyle{unsrt}  
\bibliography{references}  

\newpage
\appendix
\section{Further theoretical results and Proofs}
\label{app: theorem}
Then we list the proof of Lemma~\ref{lemma: star}.
\begin{proof}
\label{proof: star}
In this proof, we first focus on the Generative (symmetric) PLS-SVD case. In this case, we prove all the results by induction. We will start with the case $h = 1$, then under the assumption that the results are correct when $h = h$, prove the case when $h = h+1$. In every step, we will use the result from Lemma~\ref{lemma: bullet}, which is
$$
    ({\boldsymbol{w}^\bullet_h},\boldsymbol{v}_h^\bullet) = \underset{\text{C1}}{\argmax}~\text{Cov}({\boldsymbol{u}_h}^\top\boldsymbol{X}^\bullet_{h-1},{\boldsymbol{v}_h}^\top\widetilde{\boldsymbol{Y}}^\bullet_{h-1}).
$$
While the optimisation problems considered here use the $(\boldsymbol{X}_{h-1}^\star, \boldsymbol{Y}_{h-1}^\star)$ replaced $(\boldsymbol{X}_{h-1}^\bullet, \boldsymbol{Y}_{h-1}^\bullet)$, which is:
$$
({\boldsymbol{u}^\star_h},\boldsymbol{v}_h^\star) \in \underset{\text{C1,C2,C3}^\star}{\argmax}~\text{Cov}({\boldsymbol{u}_h}^\top\boldsymbol{X}^\star_{h-1},{\boldsymbol{v}_h}^\top\widetilde{\boldsymbol{Y}}^\star_{h-1})
$$
We will first prove that, even with the additional constraints and different objective functions, the above two optimisation problems are equivalent and lead to the above results, which is $\boldsymbol{u}_h^\star= \boldsymbol{w}_h^\bullet$ and $\boldsymbol{v}_h^\star= \boldsymbol{v}_h^\bullet$.

Then, based on this, we further prove $\xi_h^\star - \xi_h^\bullet ={\boldsymbol{u}_h^\bullet}^\top \boldsymbol{X}^\bullet_{H}$ and $\boldsymbol{X}^\bullet_{h} - \boldsymbol{X}^\star_{h} = \sum_{i=1}^{h}\boldsymbol{u}_i^\bullet {\boldsymbol{u}_i^\bullet}^\top \boldsymbol{X}^\bullet_{H}$.

Finally, the proof is extended to the Generative PLS-R case.

Starting with $h = 1$.

When $h=1$,  $(\boldsymbol{X}_{0}^\star,\widetilde{\boldsymbol{Y}}_{0}^\star )= (\boldsymbol{X}_{0}^\bullet,\widetilde{\boldsymbol{Y}}_0^\bullet)$, and the condition $\text{C2}^\star$ and $\text{C3}^\star$ is not applied yet. Thus, the two optimisation problems above are identical. From Lemma~\ref{lemma: bullet}, the solution is unique, ensuring that $({\boldsymbol{u}^\star_1},\boldsymbol{v}_1^\star) = ({\boldsymbol{w}^\bullet_1}, \boldsymbol{v}_1^\bullet)$.

Now, from the definition and equation~\ref{eq: pls_generated_xi}, we have
$$
{\xi}_1^\star =  {\boldsymbol{u}^\star_1}^\top{\boldsymbol{X}}_{0}^\star =  {\boldsymbol{w}^\bullet_1}^\top{\boldsymbol{X}}_{0}^\bullet \quad\text{and} \quad
\xi_1^\bullet = {\boldsymbol{w}^\bullet_1}^\top \boldsymbol{X}^\bullet_{0} - {\boldsymbol{w}^\bullet_1}^\top \boldsymbol{X}^\bullet_{H}.
$$
Clearly,
$$
\xi_1^\star - \xi_1^\bullet= {\boldsymbol{w}_1^\bullet}^\top \boldsymbol{X}^\bullet_{H}.
$$
Then we have,
$$
{\boldsymbol{X}}^\star_{1} = {\boldsymbol{X}}_{0}^\star-{\boldsymbol{u}}^\star_1{\xi_1^\star} \quad \textrm{and}\quad
{\boldsymbol{X}}_{1}^\bullet = {\boldsymbol{X}}_{0}^\bullet-{\boldsymbol{w}}_1^\bullet\xi_1^\bullet.
$$
So,
$$\boldsymbol{X}_{1}^\bullet -{\boldsymbol{X}}_{1}^\star =  {\boldsymbol{X}}_{0}^\bullet-{\boldsymbol{w}}^\bullet_1\boldsymbol{\xi}_1^\bullet -{\boldsymbol{X}}_{0}^\star+{\boldsymbol{u}}^\star_1\boldsymbol{\xi}_1^\star=-{\boldsymbol{w}}^\bullet_1({\boldsymbol{w}^\bullet_1}^\top\boldsymbol{X}_{0}^\bullet -{\boldsymbol{u}^\bullet_1}^\top\boldsymbol{X}_{H}^\bullet)+{\boldsymbol{u}}_1^\star{\boldsymbol{u}_1^\star}^\top{\boldsymbol{X}}_{0}^\star =  {\boldsymbol{w}}^\bullet_1{\boldsymbol{w}^\bullet_1}^\top\boldsymbol{X}_{H}^\bullet.$$

Symmetrically, we have $\omega_1^\star - \omega_1^\bullet= {\boldsymbol{v}_1^\bullet}^\top \tilde{\boldsymbol{Y}}^\bullet_{H}$ and $\widetilde{\boldsymbol{Y}}_{1}^\bullet -\widetilde{\boldsymbol{Y}}_{1}^\star = {\boldsymbol{v}}^\bullet_1{\boldsymbol{v}^\bullet_1}^\top\widetilde{\boldsymbol{Y}}_{H}$

In step $h$, we assume that (1) $({\boldsymbol{u}^\star_h},\boldsymbol{v}_h^\star) = ({\boldsymbol{w}^\bullet_h}, \boldsymbol{v}_h^\bullet)$, (2) $\xi_h^\star - \xi_h^\bullet = {\boldsymbol{w}_h^\bullet}^\top \boldsymbol{X}^\bullet_{H}$, and ${\boldsymbol{X}_{h}^\bullet- \boldsymbol{X}}_{h}^\star = \sum_{i=1}^h {\boldsymbol{w}}^\bullet_i {\boldsymbol{w}^\bullet_i}^\top \boldsymbol{X}_{H}^\bullet$ and (3) $\omega_h^\star - \omega_h^\bullet= {\boldsymbol{v}_h^\bullet}^\top \tilde{\boldsymbol{Y}}^\bullet_{H}$ and $\widetilde{\boldsymbol{Y}}_{h}^\bullet -\widetilde{\boldsymbol{Y}}_{h}^\star = \sum_{i=1}^h {\boldsymbol{v}}^\bullet_i {\boldsymbol{v}^\bullet_i}^\top\widetilde{\boldsymbol{Y}}_{H}^\bullet$

Then at step $h = h+1$.

Since the $\boldsymbol{u}_{h+1}$ and $\boldsymbol{v}_{h+1}$ are searching in the direction that is orthogonal to all the previous ones, so $\boldsymbol{u}_{h+1}^\top \boldsymbol{u_j}^\star = \boldsymbol{u}_{h+1}^\top \boldsymbol{w_j}^\bullet = \boldsymbol{v}_{h+1}^\top \boldsymbol{v_j}^\star = \boldsymbol{v}_{h+1}^\top \boldsymbol{v_j}^\bullet =0$ for all $j \in \{1,\ldots, h \}$. The objection function can be formulated as:
\begin{eqnarray*}
    \text{Cov}({\boldsymbol{u}_{h+1}}^\top\boldsymbol{X}^\star_{h},{\boldsymbol{v}_{h+1}}^\top\tilde{\boldsymbol{Y}}^\star_{h
    })=&\text{Cov}({\boldsymbol{u}_{h+1}}^\top\boldsymbol{X}^\bullet_{h}- \boldsymbol{u}_{h+1}^\top \sum_{i=1}^h {\boldsymbol{w}}^\bullet_i {\boldsymbol{w}^\bullet_i}^\top \boldsymbol{X}_{H}^\bullet,{\boldsymbol{v}_{h+1}}^\top\widetilde{\boldsymbol{Y}}_{h}^\bullet-{\boldsymbol{v}_{h+1}}^\top\sum_{i=1}^h {\boldsymbol{v}}^\bullet_i {\boldsymbol{v}^\bullet_i}^\top\widetilde{\boldsymbol{Y}}_{H}^\bullet)\\
    =& \text{Cov}({\boldsymbol{u}_{h+1}}^\top\boldsymbol{X}^\bullet_{h},{\boldsymbol{v}_{h+1}}^\top\widetilde{\boldsymbol{Y}}_{h}^\bullet)
    \end{eqnarray*}

Again, the objective function in the optimisation problems will be identical to the result in Lemma~\ref{lemma: bullet}. So, if the considered optimisation problem only has constraint C1, the optimal value will be $({\boldsymbol{w}^\bullet_{h+1}}, \boldsymbol{v}_{h+1}^\bullet)$. Then, it's easy to see that this optimal solution also meets C2 and C3. So, we have 
$$
({\boldsymbol{u}^\star_{h+1}}, \boldsymbol{v}_{h+1}^\star) = ({\boldsymbol{w}^\bullet_{h+1}}, \boldsymbol{v}_{h+1}^\bullet).
$$
Inductively,
$$
{\xi}_{h+1}^\star = {\boldsymbol{u}^\star_{h+1}}^\top {\boldsymbol{X}}_{h}^\star = {\boldsymbol{w}^\bullet_{h+1}}^\top \left( {\boldsymbol{X}}_{h}^\bullet - \sum_{i=1}^h {\boldsymbol{w}}^\bullet_i {\boldsymbol{w}^\bullet_i}^\top \boldsymbol{X}_{H} \right) = {\boldsymbol{w}^\bullet_{h+1}}^\top  {\boldsymbol{X}}_{h}^\bullet
$$
From the model,
$$
\xi_{h+1}^\bullet = {\boldsymbol{w}^\bullet_{h+1}}^\top \boldsymbol{X}^\bullet_{h} - {\boldsymbol{w}^\bullet_{h+1}}^\top \boldsymbol{X}^\bullet_{H}
$$
Clearly,
$$
\xi_{h+1}^\star - \xi_{h+1}^\bullet = {\boldsymbol{w}_{h+1}^\bullet}^\top \boldsymbol{X}^\bullet_H
$$

Similarly,
$$
\boldsymbol{X}_{h+1}^\bullet-{\boldsymbol{X}}_{h+1}^\star = {\boldsymbol{X}}_{h}^\bullet-{\boldsymbol{w}}^\bullet_{h+1}{\boldsymbol{\xi}_{h+1}}^\bullet -{\boldsymbol{X}}_{h}^\star+{\boldsymbol{u}}^\star_{h+1}{\boldsymbol{\xi}_{h+1}}^\star =  \sum_{i=1}^h {\boldsymbol{w}}^\bullet_i {\boldsymbol{w}^\bullet_i}^\top \boldsymbol{X}_{H} +  {\boldsymbol{w}}^\bullet_{h+1} {\boldsymbol{w}}^{\bullet\top}_{h+1}\boldsymbol{X}_H= \sum_{i=1}^{h+1} {\boldsymbol{w}}^\bullet_i {\boldsymbol{w}^\bullet_i}^\top \boldsymbol{X}_H
$$
Symmetrically, we have $\omega_{h+1}^\star - \omega_{h+1}^\bullet= {\boldsymbol{v}_{h+1}^\bullet}^\top \tilde{\boldsymbol{Y}}^\bullet_{H}$ and $\widetilde{\boldsymbol{Y}}_{h+1}^\bullet -\widetilde{\boldsymbol{Y}}_{h+1}^\star = \sum_{i=1}^{h+1} {\boldsymbol{v}}^\bullet_i {\boldsymbol{v}^\bullet_i}^\top\widetilde{\boldsymbol{Y}}_{H}$

This completes the proof of the Generative (symmetric) PLS-SVD case.

Then, in the Generative PLS-R case, in step $h = 1$, because $\boldsymbol{Y}_0^\star = \widetilde{\boldsymbol{Y}}_0^\star$, we directly get the same result, except the result for $\boldsymbol{Y}_1^\bullet - \boldsymbol{Y}_1^\star$.

Instead, we have: 
\begin{align*}
    \boldsymbol{Y}_1^\bullet - \boldsymbol{Y}_1^\star &= \boldsymbol{Y}_0^\bullet-b_1^\bullet\boldsymbol{v}_1^\bullet\xi_1^\bullet-\boldsymbol{Y}_0^\star+b_1^\star\boldsymbol{v}_1^\star\xi_1^\star\\
    &= b_1^\star\boldsymbol{v}_1^\star\xi_1^\star-b_1^\bullet\boldsymbol{v}_1^\bullet\xi_1^\bullet\\
    &= b_1^\star\boldsymbol{v}_1^\star(\xi_1^\bullet+\boldsymbol{w}_1^\bullet\boldsymbol{X}_H^\bullet)-b_1^\bullet\boldsymbol{v}_1^\bullet\xi_1^\bullet\\
    &= \boldsymbol{v}_1^\bullet\xi_1^\bullet (b_1^\star-b_1^\bullet)+ \boldsymbol{v}_1^\bullet\boldsymbol{w}_1^\bullet\boldsymbol{X}_H^\bullet b_1^\star
\end{align*}

Then in the step $h+1$, we have 
\begin{align*}
    &\text{Cov}({\boldsymbol{u}_{h+1}}^\top\boldsymbol{X}^\star_{h},{\boldsymbol{v}_{h+1}}^\top\boldsymbol{Y}^\star_{h
    })\\=& \text{Cov}({\boldsymbol{u}_{h+1}}^\top\boldsymbol{X}^\bullet_{h}- \boldsymbol{u}_{h+1}^\top \sum_{i=1}^h {\boldsymbol{w}}^\bullet_i {\boldsymbol{w}^\bullet_i}^\top \boldsymbol{X}_{H},{\boldsymbol{v}_{h+1}}^\top\boldsymbol{Y}_{h}^\bullet-{\boldsymbol{v}_{h+1}}^\top\sum_{i=1}^h \boldsymbol{v}_i^\bullet\xi_i^\bullet (b_i^\star-b_i^\bullet)+ \boldsymbol{v}_i^\bullet\boldsymbol{w}_i^\bullet\boldsymbol{X}_H^\bullet b_i^\star\\
    =& \text{Cov}({\boldsymbol{u}_{h+1}}^\top\boldsymbol{X}^\bullet_{h},{\boldsymbol{v}_{h+1}}^\top\boldsymbol{Y}_{h}^\bullet)
    \end{align*}
From Lemma~\ref{lemma: bullet}, again we have $({\boldsymbol{u}^\star_{h+1}}, \boldsymbol{v}_{h+1}^\star) = ({\boldsymbol{w}^\bullet_{h+1}}, \boldsymbol{v}_{h+1}^\bullet).$
Then, 
\begin{align*}
    \boldsymbol{Y}_{h+1}^\bullet - \boldsymbol{Y}_{h+1}^\star & = \boldsymbol{Y}_0^\bullet-\sum_{i = 1}^{h+1}b_i^\bullet\boldsymbol{v}_i^\bullet\xi_i^\bullet-\boldsymbol{Y}_0^\star+\sum_{i=1}^{h+1}b_i^\star\boldsymbol{v}_i^\star\xi_i^\star\\
    & = \sum_{i = 1}^{h+1}\boldsymbol{v}_i^\bullet\xi_i^\bullet (b_i^\star-b_i^\bullet)+ \boldsymbol{v}_i^\bullet\boldsymbol{w}_i^\bullet\boldsymbol{X}_H^\bullet b_i^\star
\end{align*}
Everything else stays the same with the Generative (Symmetric) PLS-SVD scenario.
\end{proof}
Moreover, the proof of Lemma~\ref{lemma: star} is below:
\begin{proof}
\label{proof: bullet}

We first prove that these two optimisation problems are the same. Then, we focus on the symmetric case. In the symmetric case, we first show that there must exist a solution, then we prove that $(\boldsymbol{w}_h^\bullet, \boldsymbol{v}_h^\bullet)$ is one of the solutions. Finally, we show that the solution is unique.

By definition,  $\widetilde{\boldsymbol{Y}}^\bullet_h = \boldsymbol{Y}^\bullet_h-\sum_{k=1}^h\boldsymbol{v}^\bullet_k\epsilon_{k}^\bullet$ and so  $\boldsymbol{v}^{\bullet\top}_h\boldsymbol{Y}^\bullet_{h-1}={\boldsymbol{v}_h^\bullet}^\top\widetilde{\boldsymbol{Y}}^\bullet_{h-1}+\boldsymbol{v}_h^{\bullet\top} \sum_{k = 1}^{h-1}\boldsymbol{v}^\bullet_k \epsilon_{1,k}^\bullet={\boldsymbol{v}_h^\bullet}^\top\widetilde{\boldsymbol{Y}}^\bullet_{h-1}$ because $V^\bullet$ is an orthonormal matrix. Thus, if $(\boldsymbol{u}_h,\boldsymbol{v}_h^\bullet)$ is a solution of one of the two optimisation problems, it is also a solution of the other.

Now, we focus on the symmetric case, consider the objective function, with $\boldsymbol{\Sigma}^{\bullet({h-1})}_{X,\tilde{Y}}$ to denote the covariance matrix between $\boldsymbol{X}_{h-1}^\bullet$ and $\widetilde{\boldsymbol{Y}}_{h-1}^\bullet$,  we have:
$$
\textrm{Cov}\left({\boldsymbol{X}^{\bullet\top}_{h-1}}\boldsymbol{u}_h, \widetilde{\boldsymbol{Y}}_{h-1}^{\bullet\top}\boldsymbol{v}_h\right) = \boldsymbol{u}_h^\top \boldsymbol{\Sigma}^{\bullet({h-1})}_{X,\tilde{Y}} \boldsymbol{v}_h,
$$
which is a continuous function. The feasible set is $(\boldsymbol{u}_h, \boldsymbol{v}_h) \in \mathbb{S}^{p-1}\times\mathbb{S}^{q-1}$, which is compact. By Weierstrass' extreme value theorem [\cite{rudin2021principles}, Theorem 4.16, Chapter 4], there must exist an optimal value that attains the maximum.

Then, we show that the $(\boldsymbol{w}_h^\bullet,\boldsymbol{v}_h^\bullet)$ is a solution of one of the optimisation problems under constraint C1.
    
From data simulation, for $h=1,\ldots,H$, $\boldsymbol{X}_{h-1}^\bullet =  \boldsymbol{w}_h^\bullet \xi_h^\bullet + \boldsymbol{X}_{h}^\bullet$, we can write for any such $h$:
$$
    \boldsymbol{X}_{h-1}^\bullet = \boldsymbol{X}_H^\bullet + \sum_{k=h}^H \boldsymbol{w}_k^\bullet \xi_k^\bullet,
$$
from which we get
$$
    {\boldsymbol{w}^\bullet_h}^\top \boldsymbol{X}_{h-1}^\bullet = {\boldsymbol{w}^\bullet_h}^\top \boldsymbol{X}_{H}^\bullet + \sum_{k=h}^H ({\boldsymbol{w}^\bullet_h}^\top \boldsymbol{w}_k^\bullet) \xi_k^\bullet= {\boldsymbol{w}^\bullet_h}^\top \boldsymbol{X}_{H}^\bullet + \xi_h^\bullet
    $$
since $\boldsymbol{w}_i^{\bullet\top}\boldsymbol{w}_j^\bullet=\delta_{ij}$. As a result, we have
$$
    \xi_h^\bullet = {\boldsymbol{w}^\bullet_h}^\top \boldsymbol{X}^\bullet_{h-1} - {\boldsymbol{w}^\bullet_h}^\top \boldsymbol{X}^\bullet_{H}.
    $$
Similarly, we have    
    $$
    \omega_h^\bullet = {\boldsymbol{v}^\bullet_h}^\top \widetilde{\boldsymbol{Y}}^\bullet_{h-1}- {\boldsymbol{v}^\bullet_h}^\top \widetilde{\boldsymbol{Y}}_{H}^\bullet = {\boldsymbol{v}^\bullet_h}^\top \boldsymbol{Y}^\bullet_{h-1}- {\boldsymbol{v}^\bullet_h}^\top \boldsymbol{Y}_{H}^\bullet+\epsilon_h^\bullet.
    $$
From the optimization problem, because $\boldsymbol{X}_H$ and $\boldsymbol{Y}_H$ are independent with $\boldsymbol{X}^\bullet_{h-1}$ and $\boldsymbol{Y}^\bullet_{h-1}$, and $\epsilon_{1,h}$ is independent of $\boldsymbol{X}_H^\bullet$ and $\boldsymbol{X}_{h-1}^\bullet$, the objective function can be reformulated as:
    \begin{align*}
        &\text{Cov}({\boldsymbol{u}_h}^\top\boldsymbol{X}^\bullet_{h-1},{\boldsymbol{v}_h}^\top\widetilde{\boldsymbol{Y}}^\bullet_{h-1})\\
        =&\text{Cov}({\boldsymbol{u}_h}^\top\boldsymbol{X}^\bullet_{h-1} -{\boldsymbol{u}_h}^\top\boldsymbol{X}_{H}^\bullet,{\boldsymbol{v}_h}^\top\tilde{\boldsymbol{Y}}_{h-1}^\bullet-{\boldsymbol{v}_h}^\top\tilde{\boldsymbol{Y}}_{H}^\bullet)\\
        =&\text{ Cov}(\boldsymbol{u}_h^\top (\sum_{i = h}^H \boldsymbol{w}^\bullet_i \xi_i^\bullet),\boldsymbol{v}_h^\top (\sum_{i = h}^H \boldsymbol{v}^\bullet_i \omega_i^\bullet))\\
        =& \boldsymbol{u}_h^\top\text{ Cov}( (\sum_{i = h}^H \boldsymbol{w}^\bullet_i \xi_i^\bullet), (\sum_{i = h}^H \boldsymbol{v}^\bullet_i \omega_i^\bullet))\boldsymbol{v}_h\\
        =&\boldsymbol{u}_h^\top(\sum_{i = h}^H \sum_{j = h}^H \text{ Cov}(\xi_i^\bullet,  \omega_j^\bullet)\boldsymbol{w}^\bullet_i{\boldsymbol{v}^\bullet_j}^\top)\boldsymbol{v}_h\\
        =&\sum_{i = h}^H\text{ Cov}(\xi_i^\bullet,  \omega_i^\bullet)(\boldsymbol{u}_h^\top\boldsymbol{w}^\bullet_i)({\boldsymbol{v}^\bullet_i}^\top\boldsymbol{v}_h)\\
    \end{align*}
    From Lemma~\ref{lemma: cov_solution}, without loss of generality, we can consider only $\boldsymbol{u}_h \in \mathcal{W}$ for the above optimisation problem. 

    When $h = 1$, for $\boldsymbol{u}_1\in \mathcal{W}$, it can be uniquely represent as $\boldsymbol{u}_1 = \sum_{i=1}^Ha_i\boldsymbol{w}_i^\bullet$, similarly, $\boldsymbol{v}_1 = \sum_{i=1}^Hc_i\boldsymbol{v}_i^\bullet$. The above optimisation problem becomes:
    $$\underset{\|\boldsymbol{a}\| = \|\boldsymbol{c}\| = 1}{\argmax} \sum_{i = 1}^H\text{ Cov}(\xi_i^\bullet,  \omega_i^\bullet)(a_i)(c_i) = \underset{\|\boldsymbol{a}\| = \|\boldsymbol{c}\| = 1}{\argmax} \sum_{i = 1}^Hb_i^\bullet{\sigma_{\xi_i}^2}^\bullet(a_i)(c_i)$$
    By Cauchy-Schwarz inequality,
    $$\sum_{i = 1}^Hb_i^\bullet{\sigma_{\xi_i}^2}^\bullet(a_i)(c_i) \leq \sum_{i = 1}^Hb_i^\bullet{\sigma_{\xi_i}^2}^\bullet(a_i)^{1/2}$$
    The maximum of the right side is achieved when $a_i^{opti} = \delta_{i,1}$, then $\boldsymbol{u}_1^{opti} = \boldsymbol{w}_1^\bullet$. Symmetrically, $\boldsymbol{v}_1^{opti} = \boldsymbol{v}_1^\bullet$.
    
    Iteratively, in step $h$, the $\boldsymbol{u}_h \perp \{\boldsymbol{w}_j \}_{j<h}$, from lemma~\ref{lemma: cov_solution}, we consider $\boldsymbol{u}_h \in \text{span}\{\boldsymbol{w_j^\bullet} \}_{j=h}^H \subset \mathcal{W}$, similarly we get $({\boldsymbol{w}^\bullet_h},\boldsymbol{v}_h^\bullet)$ is one of the optimal value.

    So, we have
    $$({\boldsymbol{w}^\bullet_h},\boldsymbol{v}_h^\bullet) \in \underset{C}{\argmax}~\text{Cov}({\boldsymbol{u}_h}^\top\boldsymbol{X}^\bullet_{h-1},{\boldsymbol{v}_h}^\top\widetilde{\boldsymbol{Y}}^\bullet_{h-1})$$

    Then, we prove the uniqueness of the solution. The above maximisation problem can also be written as:
    $$
        \underset{C}{\argmax}~\text{Cov}({\boldsymbol{u}_h}^\top\boldsymbol{X}^\bullet_{h-1},{\boldsymbol{v}_h}^\top\widetilde{\boldsymbol{Y}}^\bullet_{h-1})\\
        = \underset{C}{\argmax}~\boldsymbol{u}_h^\top \widetilde{\Sigma}_{X,Y}^{(h-1)\bullet} \boldsymbol{v}_h
    $$
    where $\widetilde{\Sigma}_{X,Y}^{(h-1)\bullet}$ denote the covariance matrix of $\boldsymbol{X}_{h-1}^\bullet$ and $\boldsymbol{Y}_{h-1}^\bullet$.
    
    With constraint C, this optimisation becomes a standard singular value decomposition (SVD) problem. The solution corresponds to the largest left singular and largest right singular vectors. \\
    Again from the data generation procedure $$\boldsymbol{X}^\bullet_{h-1} = \boldsymbol{X}_H^\bullet + \sum_{k=h}^H\boldsymbol{u}_k^\bullet\xi_k^\bullet$$
    $$\widetilde{\boldsymbol{Y}}^\bullet_{h-1} = \widetilde{\boldsymbol{Y}}_H^\bullet + \sum_{k=h}^H\boldsymbol{v}_k^\bullet\omega_{k}^\bullet$$
    Then, the covariance matrix can be calculated as follows:
    \begin{eqnarray*}
    \widetilde{\Sigma}_{X,Y}^{(h-1)\bullet} =& \text{Cov}(\boldsymbol{X}_{h-1}^\bullet, \widetilde{\boldsymbol{Y}}_{h-1}^\bullet)\\
    =& \text{Cov} \left( \boldsymbol{X}_H^\bullet + \sum_{k=h}^{H} \boldsymbol{u}^\bullet_k \xi^\bullet_k, \widetilde{\boldsymbol{Y}}_H^\bullet- \sum_{k=h}^H\boldsymbol{v}_k^\bullet\omega_{k}^\bullet\right)\\
    =& \sum_{k=h}^{H} \sum_{\ell=h}^{H} \text{Cov}(\boldsymbol{u}_k^\bullet \xi_k^\bullet, \boldsymbol{v}^\bullet_\ell \omega^\bullet_\ell)\\
    =& \sum_{k=h}^{H} \boldsymbol{u}_k^\bullet b_k^\bullet {\sigma_\xi^\bullet}^2 {\boldsymbol{v}_k^\bullet}^\top.
    \end{eqnarray*}
    Since $\boldsymbol{u}_k^\bullet$ and $\boldsymbol{v}_k^\bullet$ are orthogonal unit-norm vectors, they already form a natural SVD decomposition of $\boldsymbol{\Sigma}_{XY}^{(h)}$. The singular values are:
    $$
    \lambda_k = b_k^\bullet {\sigma^\bullet_\xi}^2.
    $$
    With the assumption $b_k^\bullet {\sigma^\bullet_\xi}^2$ is strictly decreasing, then the largest singular value $\lambda_1$ is unique, ensuring a unique optimal solution up to sign:
    $$({\pm \boldsymbol{u}^\bullet_h},\pm \boldsymbol{v}_h^\bullet) = \underset{C}{\argmax}~\text{Cov}({\boldsymbol{u}_h}^\top\boldsymbol{X}^\bullet_{h-1},{\boldsymbol{v}_h}^\top\boldsymbol{Y}^\bullet_{h-1})$$
    While in the model, we impose the biggest component of every vector $\boldsymbol{w}_h^\bullet$ to be non-negative, then $\boldsymbol{u}_h^\bullet$ is identifiable, the above optimisation problem is ensured to have a unique solution for the predictive model. Since the two optimisation problems are identical, the unique solution is also the unique solution for the symmetric model
\end{proof}
Then, the proof for Lemma~\ref{lemma: cov_solution}:
\begin{proof}
\label{proof: cov_solution}
    Since $\boldsymbol{X}_0^\bullet \in \mathbb{R}^p$ is generated via the generative PLS-R model or the Generative symmetric PLS-SVD model, we have
    $$\boldsymbol{X}^\bullet_{h-1} = \sum_{i = h}^H \boldsymbol{w}^\bullet_i \xi_i^\bullet+\boldsymbol{X}_H^\bullet.$$
    Then the optimisation problem becomes:
    $$
        \underset{\boldsymbol{w}}{\arg\max}\text{ Cov}(\boldsymbol{w}^\top (\boldsymbol{X}^\bullet_{h-1}-\boldsymbol{X}_H^\bullet),\cdot) = \underset{\boldsymbol{w}}{\arg\max}\text{ Cov}(\boldsymbol{w}^\top (\sum_{i = h}^H \boldsymbol{w}^\bullet_i \xi_i^\bullet),\cdot).
    $$
    If the optimisation problem has a solution $\boldsymbol{w}' \in \mathbb{R}^p$, then an orthogonal decomposition allows us to write:
    $$\boldsymbol{w}' = \mathcal{P}_{\mathcal{W}} \boldsymbol{w}' + \mathcal{P}_{\mathcal{W}}^\perp \boldsymbol{w}'.$$
    Since the vector $\mathcal{P}_{\mathcal{W}}^\perp \boldsymbol{w}'$ is orthogonal to any vector in $\mathcal{W}$, and since $\sum_{i = h}^H \boldsymbol{w}^\bullet_i \xi_i^\bullet \in \mathcal{W}$ almost surely, we have:
    $$
    (\mathcal{P}_{\mathcal{W}}^\perp \boldsymbol{w}')^\top (\sum_{i = h}^H \boldsymbol{w}^\bullet_i \xi_i^\bullet) = 0.
    $$
    As a result
    \begin{align*}
    \text{Cov}(\boldsymbol{w}'^\top (\sum_{i = h}^H \boldsymbol{w}_i^\bullet \xi_i^\bullet), \cdot)
    &= \text{Cov}((\mathcal{P}_{\mathcal{W}} \boldsymbol{w}')^\top (\sum_{i = h}^H \boldsymbol{w}_i^\bullet \xi_i^\bullet) + (\mathcal{P}_{\mathcal{W}}^\perp \boldsymbol{w}')^\top (\sum_{i = h}^H \boldsymbol{w}_i^\bullet \xi_i^\bullet), \cdot) \\
    &= \text{Cov}((\mathcal{P}_{\mathcal{W}} \boldsymbol{w}')^\top (\sum_{i = h}^H \boldsymbol{w}_i^\bullet \xi_i^\bullet), \cdot).
    \end{align*}
    So, there exists $\boldsymbol{w}^{opti} := \mathcal{P}_{\mathcal{W}} \boldsymbol{w}' \in \mathcal{W}$ that reaches the same optimal value as $\boldsymbol{w}'$, also an optimal solution of the optimization problem.

    Moreover, when $\boldsymbol{w}\in \mathcal{W}$is considered and there's additional constraint $\boldsymbol{w} \perp \{\boldsymbol{w}_j \}_{j<h}$, we have:
    $$\boldsymbol{w} \in \{\boldsymbol{w}_j \}_{j<h}^\perp = \{\boldsymbol{w}_j \}_{j=h}^H$$
    
\end{proof}
The following lemma is used when we define the model
\begin{lemma}\label{lemma: UV}
Let $W$ be a $p\times H$ matrix with linearly independent columns $\boldsymbol{w}_1,\ldots,\boldsymbol{w}_H$. Then there exists a (generally non-unique) $p\times H$ matrix $U$ whose columns $\boldsymbol{u}_1,\ldots,\boldsymbol{u}_H$ satisfy $$\boldsymbol{u}_i^\top\boldsymbol{w}_j=\delta_{ij}\quad\text{ for }i,j=1,\ldots,H,$$
or equivalently, $U^\top W=I_H$. 
Furthermore, if the columns of $W$ are orthonormal, then there exists a unique matrix $U$ that satisfies the above condition together with $\|\boldsymbol{u}_h\|_2=1$, for all $h$. This unique matrix is $U = W$.
\end{lemma}
\begin{proof}
Since the columns $\boldsymbol{w}_1,\ldots,\boldsymbol{w}_H$ are linearly independent, the matrix $W$ has full column rank, and the $H\times H$ matrix $W^\top W$ is invertible. Then $U=W(W^\top W)^{-1}$ is one solution as the following equation shows
$$
U^\top W = (W(W^\top W)^{-1})^\top W=(W^\top W)^{-1}W^\top W=I_H.
$$
Moreover, if $\tilde{U}$ is any other solution, then we can write $\tilde{U}=U+(\tilde{U}-U)$ and thus $I_H=\tilde{U}^\top W= U^\top W+(\tilde{U}-U)^\top W=I_H+(\tilde{U}-U)^\top W$, so that we must have $(\tilde{U}-U)^\top W=0$. 
All the solutions can thus be written as $U+Q$ where $Q:p\times H$ is such that $Q^\top W=0$, which shows the non-unicity in the general case.

Now, if we further assume that $W^\top W=I_H$, then $U=W$ and any solution takes the form $\tilde{U}=W+Q$ for some matrix $Q:p\times H$ satisfying $Q^\top W=0$.
If we only consider the $h$th column, we get
$$
\tilde{\boldsymbol{u}}_h = \boldsymbol{w}_h + \boldsymbol{q}_h
$$
but since we impose $\|\tilde{\boldsymbol{u}}_h\|_2=1$, we get
$$
1 = \|\boldsymbol{w}_h + \boldsymbol{q}_h\|_2^2 = \|\boldsymbol{w}_h\|_2^2 + \|\boldsymbol{q}_h\|_2^2 + 2 \boldsymbol{w}_h^\top\boldsymbol{q}_h=\|\boldsymbol{w}_h\|_2^2 + \|\boldsymbol{q}_h\|_2^2
$$
and because $\|\boldsymbol{w}_h\|_2=1$, we necessarily have $\|\boldsymbol{q}_h\|_2^2=0$ and thus $\boldsymbol{q}_h=\boldsymbol{0}$ and thus $\tilde{\boldsymbol{u}}_h = \boldsymbol{w}_h$. The unique solution is thus $U=W$.
\end{proof}

\begin{prop}\label{prop: covariance}
Let $\xi_h$ and $\omega_h$ be two random variables with zero mean and finite variance. Then, their covariance is maximal if and only if there exists a constant $b_h\ne0$ such that
    \begin{equation*}
        \omega_h = b_h \xi_h.
    \end{equation*}
In fact, we have $b_h=\frac{Cov(\omega_h,\xi_h)}{Var(\xi_h)}$.    
\end{prop}
\begin{proof}
This is direct using Cauchy–Schwarz inequality which allows us to write
    $$
    |\text{Cov}(\xi_h, \omega_h)| \leq \sqrt{\text{Var}(\xi_h)\text{Var}(\omega_h)}
    $$
with equality holding if and only if $\xi_h=b_h\omega_h$ (almost surely) for some $b_h\ne0$. 
\end{proof}

\section{Simulation from the proposed model}
\label{app:simulation}
We investigated simulating data from the GFLSR model we defined in Section \ref{sec: latent}. Simulating data from a mathematical model can help define the model and learn the behaviour. Due to the complexity of the general model, four different simple situations are considered to verify the generative property under certain constraints.

Within each situation, the simulation procedure is given, and the simulation is checked via the plot of latent variables $\boldsymbol{\xi}$ and the plot of the response variable and the fitted value. The plot is using instead of the numeric result because the current model defines the X latent variable with equation \eqref{eq: X_deflation}. This equation only makes sure that the $g_h$ function in equation \eqref{eq: xi_g_function} is linear. Although in the model optimisation, the $\boldsymbol{u}_h$ within equation \eqref{eq: xi_g_function} is found within the p-dimensional unit sphere. This setting only ensures the latent variable is identifiable up to a scale. 

\begin{remark}
    Within the model structure defined, the $w_{hj}$ is a free parameter in $\mathbb{R}$. As mentioned above, the $\hat{\boldsymbol{\xi}}$ is identifiable up to a scale. One possible way to have exact $\hat{\boldsymbol{\xi}}$ is to constraint the $w_{hj}$, let $||\frac{1}{\hat{\boldsymbol{w}}_h}|| = 1$ during simulation, and $\sqrt{||\frac{1}{\hat {\boldsymbol{w}_h}}||} = \lambda$, then instead of $X_{h,j}=\mathcal{P}_{\hat{\boldsymbol{\xi}}_{h}^\perp}^\perp X_{h-1,j}$, use $X_{h,j}=(I-\lambda\mathcal{P}_{\hat{\boldsymbol{\xi}}_{h}^\perp}) X_{h-1,j}$.
\end{remark}

\paragraph{Situation 1: The true $\boldsymbol{f}_H$ function is linear and single response}

In this situation, everything is linear; if there's only a single response, this single response will only depend on a single latent variable. It's easy to show that if the response is assumed to depend on more latent variables, it can be written as another linear combination of the response variable; thus, one latent variable is enough.

\begin{eqnarray*}
y &=& \theta_1\times \xi_1+\theta_2\times \xi_2 \\
&=& \theta_1\times \xi_1+\theta_2\times (u_{21}\times x_{11}+u_{22}\times x_{12}) \\
&=& \theta_1\times \xi_1+\theta_2\times (u_{21}\times (x_{01}-w_{11}\times \xi_1)+u_{22}\times (x_{02}-w_{12}\times \xi_1) \\
&=& para\times \xi_1+\theta_2\times (u_{21}\times x_{01}+u_{22}\times x_{02}) \\
&=& para\times (u_{11}\times x_{01}+u_{12}\times x_{02})+\theta_2\times (u_{21}\times x_{01}+u_{22}\times x_{02})\\
&=& para\times x_{01}+para\times x_{02}
\end{eqnarray*}

The data is generated with the below procedure with $p = 2$, $q = 1$, $H = 1$, and $K = 1$. The study is conducted with a sample size of 1000.

The true function is:
$$\mathbb{E}(Y |\xi_1) = 2\times \xi_1$$
$\boldsymbol{X}_H$ and $Y_H$ follow the normal distribution $N(0,0.02)$
then when $h = 1$:
\begin{eqnarray*}
\left(\xi_1, \omega_1\right)^\top &=& 
\begin{cases}
\Psi(U_1)\\
2\times \Psi(U_1)
\end{cases}
~\text{with }U_1\sim\text{Unif}[0,1],\\
X_{0,1} & = & 3\times \xi_1 + N(0,0.02),\\
X_{0,2} & = & 2\times \xi_1 + N(0,0.02),\\
\boldsymbol{Y}_{0} & = &  2\times \xi_1+N(0,0.02)
\end{eqnarray*}
where the $\Psi$ function is the inverse CDF of the standard normal distribution.

\begin{figure}[hbt]
        \centering
        \includegraphics[scale = 0.30]{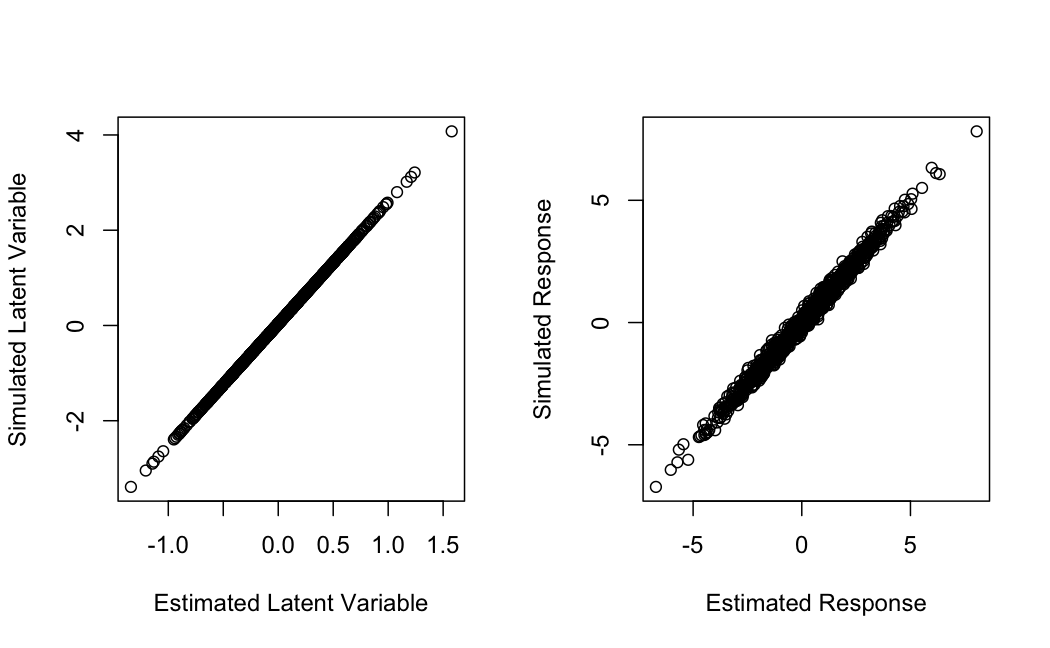}
        \caption{Plot between fitted and simulated values in Situation 1.}
        \label{fig:situation1}
\end{figure}

The $f_H$ function is embedded in \eqref{eq: latent_relationship} in the single variable situation.  The relationship between the latent variables is linear, so $D$ is the Covariance or Pearson will be able to capture the latent variable and fit the model. The plot between the fitted latent variable and the simulated latent variable is the left figure in Figure \ref{fig:situation1}, and the right figure is the plot between the fitted response variable and the simulated response variable. 

From figure \ref{fig:situation1}, the model is able to find the true simulated latent variable, although on a scale, and then the simulated response. In the simulation situation, the $\Psi$ function can be more complicated if a non-linear dependence measure is preferred. For example, the Spearman dependence measure will be able to capture $\psi_1 = qnorm(U_1)$, $\psi_2 = exp(qnorm(U_1))$.

\paragraph{Situation 2: The true $f$ function is nonlinear and single response}
In situation 2, we still assume a single response, but the $f_H$ function is no longer linear. So, if $f_H$ can be represented uniquely by $\xi_h$ and these $\xi_h$ have strictly decreasing correlation with $f_H(\cdot)$, more latent variables can be involved. 

Here we consider a simple situation with p = 2, q = 1, H = 2, and K = 2. So, we have two predictors and two latent variables. The sample size is still set as 1000. With $\boldsymbol{X}_H \sim N(0,0.0001)$ and $Y_H \sim N(0,0.02)$:
$$\mathbb{E}(Y | \xi_1,\xi_2)= exp(\xi_1)+\xi_2^2.$$
When $h = 2$:
\begin{eqnarray*}
\left(\xi_2, \omega_2\right)^\top &=& 
\begin{cases}
\Psi(U_1)\\
\Psi(U_1)^2  + N(0,0.02),
\end{cases}
~\text{with }U_2\sim\text{Unif}[0,1],\\
X_{1,1} & = & 1\times \xi_2 + N(0,0.0001),\\
X_{1,2} & = & 1\times \xi_2 + N(0,0.0001),\\
Y_{1} & = & \xi_2^2 + N(0,0.02),
\end{eqnarray*}
then $h = 1$:
\begin{eqnarray*}
\left(\xi_1, \omega_i\right)^\top &=& 
\begin{cases}
\Psi(U_1)\\
\text{exp}(\Psi(U_1))+y_1 + N(0,0.02),
\end{cases}
~\text{with }U_1\sim\text{Unif}[0,1],\\
X_{0,1} & = & 3\times \xi_1 + X_{1,1},\\
X_{0,2} & = & 1\times \xi_1 + X_{1,2},\\
Y_{0} & = & exp(\xi_1)+\xi_2^2 + + N(0,0.02),
\end{eqnarray*}
where $\Psi$ function is the inverse CDF of the standard normal.

For single-response variable situations, the relationship between $Y_0$ and all the latent variables is embedded in the equation \ref{eq: latent_relationship}. So, a Spearman correlation $D$ and a non-linear $\boldsymbol{f}_H$ function will be able to catch the true data relationship. Here, for simplicity, the GROC algorithm is used.
\begin{figure}[hbt]
        \centering
        \includegraphics[scale = 0.15]{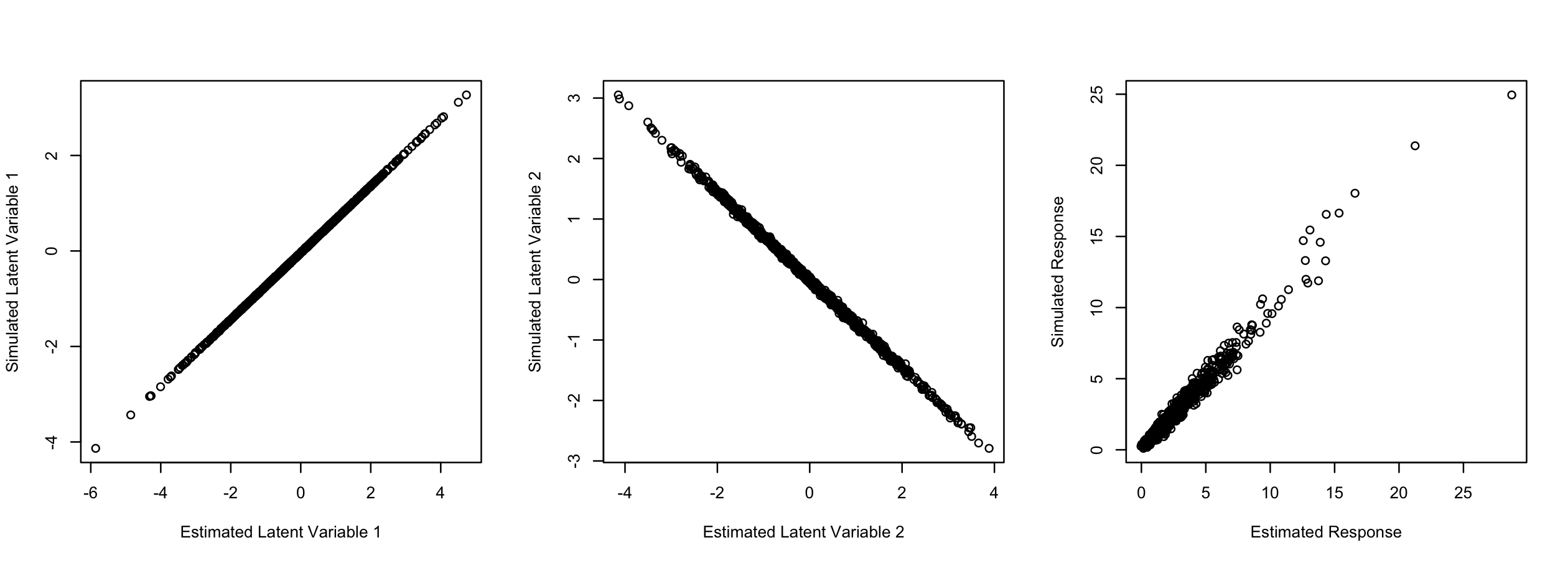}
        \caption{Plot between fitted and simulated values in Situation 2.}
        \label{fig:situation2}
\end{figure}
From figure \ref{fig:situation2}, the model with Spearman correlation is able to capture the non-linearity. In the single response situation, the $\Psi$ function defines both the relationship between $\xi$ and $\boldsymbol{X}$ and $Y$.

\paragraph{Situation 3: The true $\boldsymbol{f}_H$ function is linear, and multiple responses}
The rest two situations are multiple responses. Situation 3 is still assuming the $\boldsymbol{f}_H$ function is linear. So, to let $\boldsymbol{f}_H$ have the unique representation with the latent variables, at most $min(p,q)$ of the latent variables are required. 

We consider the situation when $p = 2$, $q = 2$, $H = 2$ and $K = 2$. The situation is made that the $\boldsymbol{f}_H$ can't be represented with other linear combinations of the predictors. The sample size remains set at 1000. Again, with $\boldsymbol{X}_H \sim N(0,0.0001)$ and $\boldsymbol{Y}_H \sim N(0,0.02)$, $\Psi$ is the inverse CDF of the standard normal distribution. The true function is:
$$\mathbb{E}(Y_{0,1}|\xi_1,\xi_2) = 3\times \xi_2; \quad \mathbb{E}(Y_{0,2}|\xi_1,\xi_2) = 3\times \xi_1+0.9\times \xi_2.$$
then $h = 2$:
\begin{eqnarray*}
\left(\xi_2, \omega_2\right)^\top &=& 
\begin{cases}
\Psi(U_2)\\
\Psi(U_2)^2 + N(0,0.02),
\end{cases}
~\text{with }U_2\sim\text{Unif}[0,1],\\
X_{1,1} & = & 1\times \xi_2 + N(0,0.0001),\\
X_{1,2} & = & 2\times \xi_2 + N(0,0.0001),\\
Y_{1,1} &=& 3 \times\xi_2+N(0,0.02),\\
Y_{1,2} & = & 0.9\times \xi_2+N(0,0.02),
\end{eqnarray*}
then $h = 1$:
\begin{eqnarray*}
\left(\xi_1, \omega_1\right)^\top &=& 
\begin{cases}
\Psi(U_1)\\
3\times \Psi(U_1)+ N(0,0.02),
\end{cases}
~\text{with }U_1\sim\text{Unif}[0,1],\\
X_{0,1} & = & 1\times \xi_1 + X_{1,1},\\
X_{0,2} & = & 1\times \xi_1 + X_{1,2},\\
Y_{0,1} & = & Y_{11}+N(0,0.02),\\
Y_{0,2} & = & 3\times \xi_1+0.9\times \xi_2+N(0,0.02).
\end{eqnarray*}
In this situation, the $Y_{01}$ is only related to $\xi_2$ to ensure a unique function representation. Moreover, the response can not be represented with other linear combinations of X. Pearson correlation is used as $D$, and the non-linear $\boldsymbol{f}_H$ function is used to capture the non-linear relationship.
\begin{figure}[hbt]
        \centering
        \includegraphics[scale = 0.20]{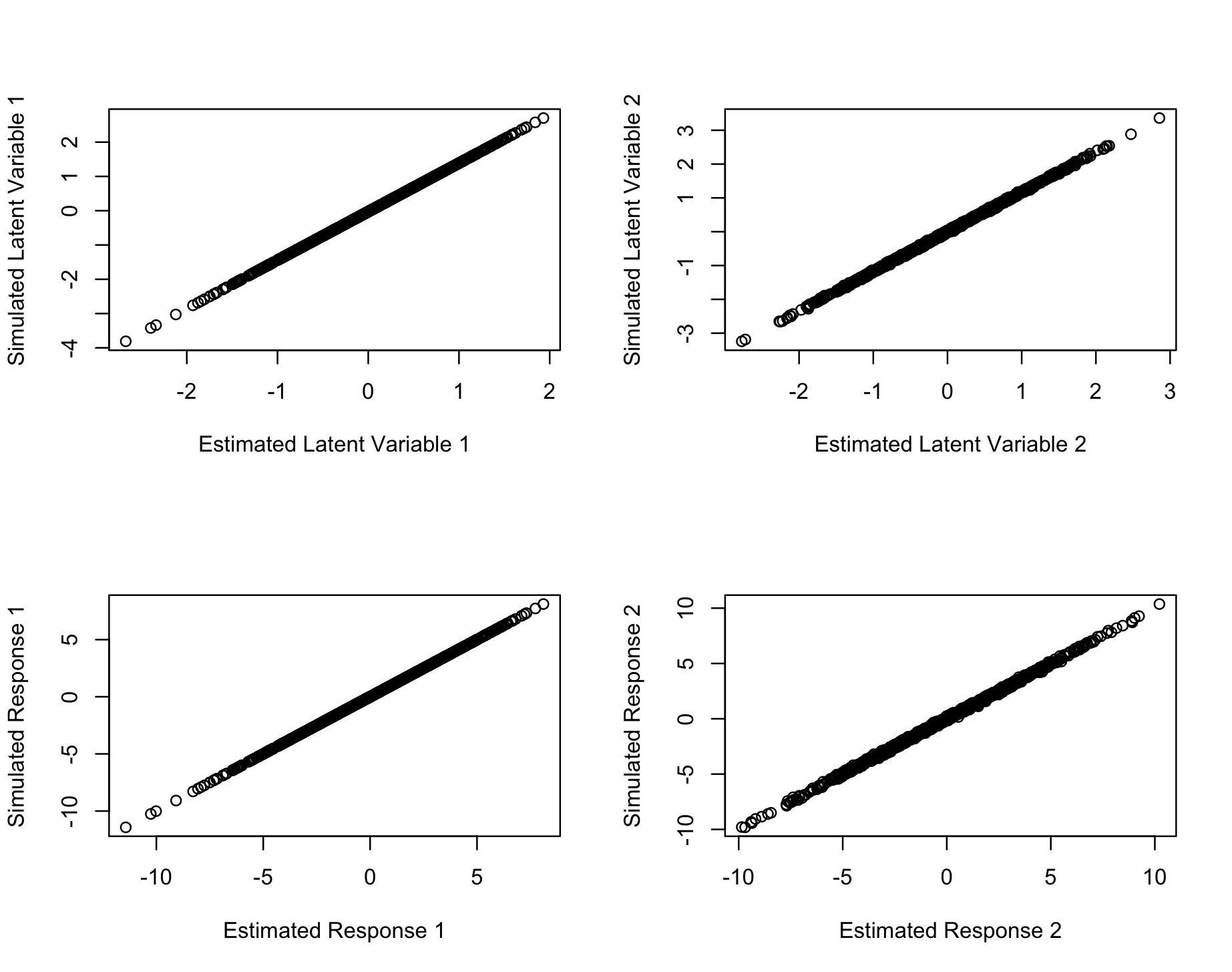}
        \caption{Plot between fitted and simulated values in Situation 3.}
        \label{fig:situation3}
\end{figure}

From figure \ref{fig:situation3}, again, the data structure is captured well.

\paragraph{Situation 4: The true $\boldsymbol{f}_H$ function is non-linear and multiple responses}
The last situation is the most complicated one, with a non-linear structure and multiple responses. 
We consider $p = 3$, $q = 3$, $H = 3$ and $K = 3$. Again, the sample size is 1000.
With a similar setting to all previous scenarios:
$$\mathbb{E}(Y_{0,1}) = \xi_1^3+\xi_2+\xi_3^2; \quad \mathbb{E}(Y_{0,2}) = \xi_1+\xi_2^2+\xi_3^3 ; \quad \mathbb{E}(Y_{0,2}) = \xi_1+0.3\times \mathbb{E}(Y_{0,1})-0.5\mathbb{E}(Y_{0,2}).$$
when h = 3, 
\begin{eqnarray*}
\left(\xi_3, \omega_3\right)^\top &=& 
\begin{cases}
\Psi(U_3)\\
\Psi(U_3) + N(0,0.02)
\end{cases}
~\text{with }U_3\sim\text{Unif}[0,1],\\
X_{2,1} & = & 3\times \xi_3 + N(0,0.0001),\\
X_{2,2} & = & 3\times \xi_3 + N(0,0.0001),\\
X_{2,3} & = & 1\times \xi_3 + N(0,0.0001),\\
Y_{2,1} & = & \xi_3^2 + N(0,0.02),\\
Y_{2,2} & = & \xi_3^3 + N(0,0.02),\\
Y_{2,3} & = & 0.3 \times \xi_3^2-0.5\times \xi_3^3 + N(0,0.02),
\end{eqnarray*}
then $h = 2$:
\begin{eqnarray*}
\left(\xi_2, \omega_2\right)^\top &=& 
\begin{cases}
\Psi(U_2)\\
\Psi(U_2) + N(0,0.02)
\end{cases}
~\text{with }U_2\sim\text{Unif}[0,1],\\
X_{1,1} & = & 2\times \xi_2 + X_{2,1},\\
X_{1,2} & = & 6\times \xi_2 + X_{2,2},\\
X_{1,3} & = & 1\times \xi_2 + X_{2,3},\\
Y_{1,1} & = & \xi_2 + Y_{2,1},\\
Y_{1,2} & = & \xi_2^2 + Y_{2,2},\\
Y_{1,3} & = & 0.3 \times \xi_2-0.5\times \xi_2^2 + Y_{2,3},
\end{eqnarray*}
then $h = 1$:
\begin{eqnarray*}
\left(\xi_1, \omega_i\right)^\top &=& 
\begin{cases}
\Psi(U_1)\\
\Psi(U_1)+ N(0,0.02)
\end{cases}
~\text{with }U_1\sim\text{Unif}[0,1],\\
X_{0,1} & = & 3\times \xi_1 + X_{1,1}\\
X_{0,2} & = & 1\times \xi_1 + X_{1,2}\\
X_{0,3} & = & 2\times \xi_1 + X_{1,3}\\
Y_{0,1} & = & \xi_1^3+Y_{1,1}\\
Y_{0,2} & = & \xi_1+Y_{1,2}\\
Y_{0,3} & = & 1.3 \times \xi_1-0.5 \times \xi_1^3+Y_{1,3})
\end{eqnarray*}
In this case, the true function is more complicated. But, the relationship between the two latent variables is identity; all the dependence measures will be able to capture it. The relationship between the latent variable and response is non-linear; a non-linear $\boldsymbol{f}_H$ is used to capture the relationship.

\begin{figure}[hbt]
        \centering
        \includegraphics[scale = 0.25]{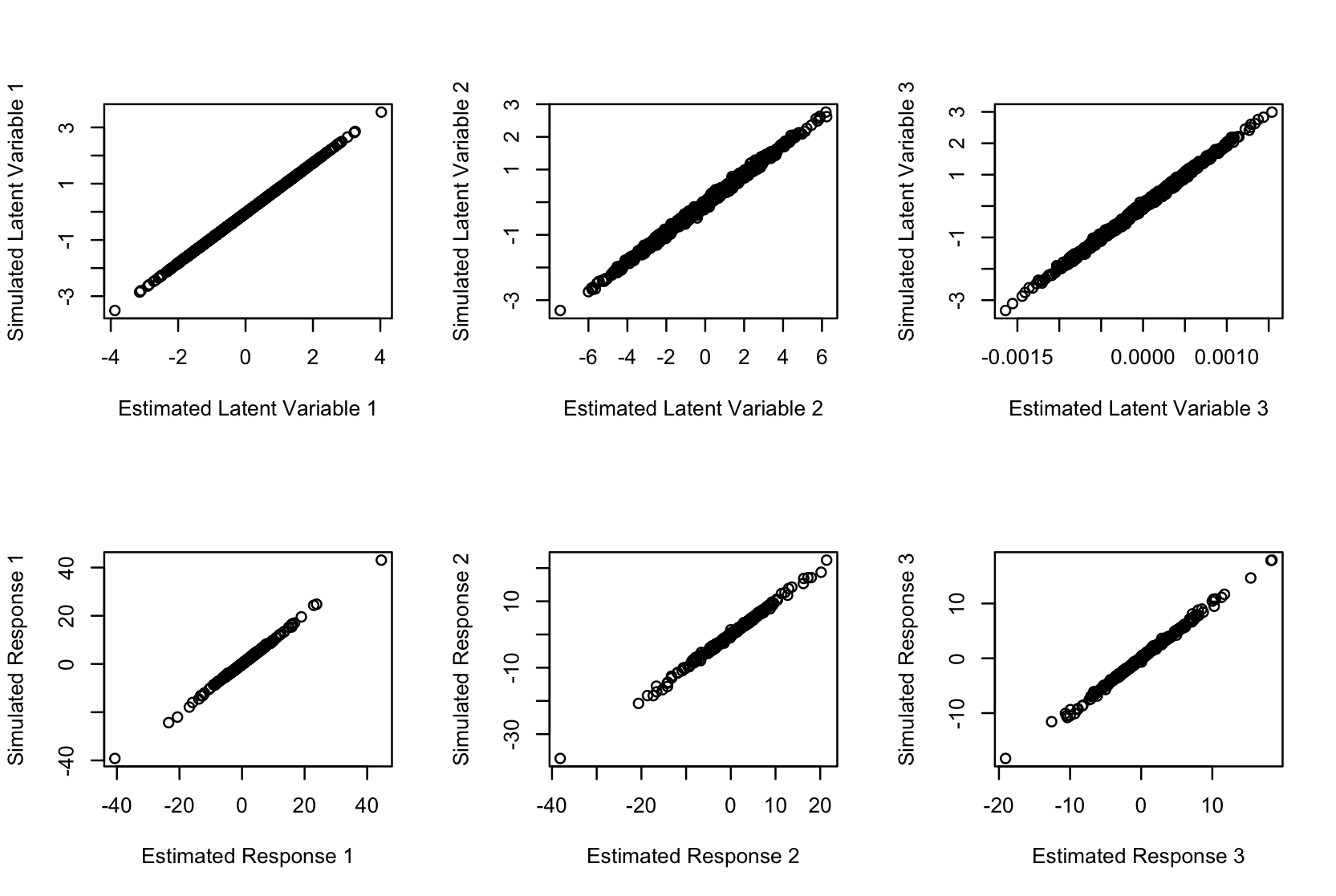}
        \caption{Plot between fitted and simulated values in Situation 4.}
        \label{fig:situation4}
\end{figure}

Again, from figure \ref{fig:situation4}, the proposed model with Pearson correlation is able to capture the data structure.

\end{document}